\documentstyle[aps,epsf,epsfig]{revtex}
\begin{document}

\large
\title {Spectral Statistics of Rectangular Billiards with Localized Perturbations}
\author {Saar Rahav\footnotemark[2] and Shmuel Fishman\footnotemark[2]\footnotemark[3]\footnotemark[4] \\ \dag \hskip.1in Department of Physics, Technion, Haifa 32000, Israel \\ \ddag \hskip.1in  Department of Physics, University of Maryland, College-Park, MD. 20742, USA \\ \S \hskip.1in Institute for Theoretical Physics, University of California at Santa Barbara, Santa Barbara, CA 93106, USA}
\date{28 May 2002}
\maketitle
\begin{abstract}
The form factor $K(\tau)$ is calculated analytically to the
order $\tau^3$ as well as numerically for a rectangular
billiard perturbed by a $\delta$-like scatterer with an angle
independent diffraction constant, $D$. The cases where the
scatterer is at the center and at a typical position in the
billiard are studied. The analytical calculations are
performed in the semiclassical approximation combined with
the geometrical theory of diffraction. Non diagonal
contributions are crucial and are therefore taken into
account. 
 The numerical calculations are performed for a self adjoint extension of a $\delta$ function potential.
 We
calculate the angle dependent diffraction constant for an
arbitrary perturbing potential $U({\bf r})$, that is large in
a finite but small region (compared to the wavelength of the
particles that in turn is small compared to the size of the
 billiard).  The relation to the idealized model of the
$\delta$-like scatterer is formulated. The angle dependent
diffraction constant is used for the analytic calculation of
the form factor to the order $\tau^2$. If the scatterer is at
a typical position, the form factor is found to reduce (in
this order)  to the one found for angle independent
diffraction.  If the scatterer is at the center, the large
degeneracy in the lengths of the orbits involved leads to an
additional small contribution to the form factor, resulting
of the angle dependence of the diffraction constant. The
robustness of the results is discussed.
\end{abstract}


\baselineskip 20pt

\section{Introduction}
\label{intro}

The spectral properties of quantum mechanical systems are closely related to
the ones of their classical counterparts~\cite{haakebook,LH89}. This
relation can be made precise and quantitative in the semiclassical limit. In
the extreme cases this correspondence is quite well understood, although some
important details are subject of research~\cite{AAA95,BK96,bogomolnyschool,marklof98}. The spectral statistics exhibit a
high degree of universality, although
there are 
exceptional systems that do not exhibit this universal behavior, 
due to some special properties.
When a classical system is chaotic the statistical properties
of corresponding quantum spectrum are well
described by Random Matrix Theory (RMT)~\cite{bohigas84} with the symmetry
determined by the underlying Hamiltonian. For integrable systems, on the other
hand, levels are
typically uncorrelated resulting in Poissonian spectral statistics~\cite{BT77}.
There are however intermediate situations. One is of mixed systems, where in
some parts of phase space the motion is chaotic and in other parts it is
regular leading to different statistics~\cite{berry84a,izrailev88}. Another type of systems that exhibit
intermediate
statistics are systems with singularities, that are integrable in absence of
these singularities. These are models of physical situations where an 
integrable
system is strongly perturbed in a region much smaller then
the wavelength of the quantum particle. Examples of such  
systems are billiards
with flux lines, sharp corners and
  $\delta$-like interactions~\cite{seba90,bogomolny99E,bogomolny01b,narevich01,rahav01,bogomolny01a}.
The specific system that will be analyzed in the present work is a billiard
with a $\delta$-like potential perturbation. 

The interest in billiards of various types is primarily theoretical
since it is relatively easy to analyze them analytically and numerically. Billiards were studied also experimentally for electrons~\cite{marcus92,chang97}, microwaves~\cite{stockmann90,sridhar91,richter01}
and for laser cooled atoms~\cite{davidson,raizen01}. We hope that in
the future, perturbations of the type discussed in the present work
will be also introduced in the experimental realizations of billiards.

Semiclassical methods
provide a convenient path to explore how classical dynamical properties affect
the quantum spectral statistics.
The connection between spectral statistics and classical dynamics is
usually made 
using trace formulas which connect the quantum density
of states to a sum over classical periodic orbits.
For classically chaotic systems the trace formula was developed
by Gutzwiller~\cite{gutz67,gutzwiller} while for classically integrable
systems the trace formula was developed by Berry and Tabor~\cite{BT-I}.
Such formulas can be used to compute the energy level-level correlation
function, $R_2(\eta)$, where $\eta$ is the level separation. Its Fourier
transform is the spectral form factor
$K(\tau)$, where the ``time'' $\tau$ is conjugate to  $\eta$. 
For systems with strong perturbations on length scales smaller than the
wavelength, standard semiclassical theory cannot be used and diffraction
effects
have to be taken into account. For billiards with localized perturbations this
can be done in the framework of the Geometrical Theory of
Diffraction (GTD)~\cite{keller62}.
In this approximation, scattering from the perturbation can be described
by a free propagation to the perturbation multiplied by a diffraction
constant, which depends on the incoming and outgoing directions, and then
free propagation away from the perturbation. 

The spectral statistics of the rectangular billiard with a flux line was 
recently studied~\cite{rahav01,bogomolny01a}.
In this system the flux line affects the amplitudes of the contributions
of periodic orbits. When the form factor is computed it is found that
at small times  $\tau$ it tends to a flux dependent constant (instead of $1$ in
absence
of the flux). Certain triangular billiards were shown to have similar 
behavior~\cite{bogomolny01a}. The works mentioned above are based only on the
contributions
from periodic orbits. It is of interest to include the contributions 
of orbits that are diffracted by the flux line in the spectral form factor.   
The contribution of diffracting orbits turns out to be rather complicated.
The contribution of orbits with one diffraction was computed by
Sieber~\cite{sieber99}, while  
Bogomolny, Pavloff and Schmit computed the contributions of diffracting
orbits with several forward diffractions to the density of
states~\cite{bogomolny00}. 
The effect of diffraction on the spectral statistics in these systems was not
determined yet. The difficulty results from the need to know the contributions
of orbits which diffract almost in the forward direction and to include
all of these in the form factor, and this has not been done so far.
Since the diffraction from flux lines and corners is complicated it is  
of special interest to examine a system in which the diffraction is relatively
simple. A class of such systems, that consists of integrable billiards with
localized
perturbations, is the subject of the present work. For  systems
of this type
only diffraction effects are responsible for deviation of the behavior from
the one of
integrable systems, therefore these are most appropriate for the exploration
of such effects. 

The diagonal approximation, where the cross terms between the contributions of
classical orbits are ignored because of rapidly varying phases associated with
them, is used for a large variety of systems.  
Berry has
shown that using the diagonal approximation leads to agreement with the
form factor of random matrices, at small times~\cite{berry85}.
Only recently Sieber and Richter~\cite{sieber01} were able to go beyond the
diagonal
approximation and to compute the next power of $\tau$ for chaotic  
systems with time reversal symmetry. This non diagonal contribution
was found to originate from orbits with small angle self intersections and
orbits which are close to them in coordinate space, except near the
intersection.
Such configurations are not possible if there is no time reversal
symmetry, thus this is an example where the difference in the spectral
statistics
is connected to a difference in the dynamical properties of the classical
counterpart, and the difference can be obtained directly from a formula
involving the classical periodic orbits. 
For chaotic systems with singular perturbations, where effects of diffraction
are important, also the non diagonal contributions to the form factor   
$K(\tau)$ are important. The final result, however, is that such perturbations
do not
affect the spectral statistics as is clear from the works of
Sieber~\cite{sieber99b,sieber00} and of Bogomolny, Leboeuf and
Schmit~\cite{bogomolny00c}.

In contrast to the case of chaotic systems, addition of a localized
scatterer,
is expected to change qualitatively the spectral properties of integrable
systems.
Moreover, for such systems the diagonal approximation is not sufficient to
obtain
any interesting physics and one should go beyond this approximation. Non
diagonal contributions can be calculated following a method that was developed
by Bogomolny~\cite{bogomolny00b}. By this method contributions of some
manifolds consisting of orbits that are almost parallel are calculated.  
If a point interaction,
that is a generalization of a $\delta$-function potential, is added to
a rectangular billiard, one obtains the so called \v{S}eba
billiard~\cite{seba90}.
The spectral statistics of this system were computed 
exactly, subject to the assumption that the underlying
integrable system exhibits Poissonian statistics~\cite{bogomolny01b} and
it was shown
not to be those of integrable or chaotic systems.
The spectral statistics of the \v{S}eba billiard with periodic boundary
conditions were shown by Berkolaiko, Bogomolny and Keating~\cite{berkolaiko01} to be identical
to those of certain star
graphs, where the form factor was calculated by Berkolaiko
and Keating~\cite{berkolaiko99}. 
 
After the work was completed a paper by Bogomolny and Giraud~\cite{BG}, where
the expansion of $K(\tau)$, was calculated to all orders in $\tau$ was published
on the Web. It is more general than the present paper, but in our paper more of
the physical properties are calculated and discussed in detail. The methodology
of the
derivation is different. A comparison between various results of these two
papers is presented in App. \ref{appbg}. 
 
Generalized $\delta$-functions were extensively studied in mathematical
physics in the framework of the self adjoint extension theory~\cite{rsbooks,pointbook,jackiw,seba91,albeverio91,shigehara94,weaver95,cheon96,shigehara97,legrand97,zorbas80}. The calculations 
in~\cite{bogomolny01b,berkolaiko01} were performed within this framework. In
the present work also a potential of finite small extension $a$ is studied,    
under the assumption that $a$ is much smaller than the wavelength of the
quantum
particle, that in turn is much smaller than the dimensions of the billiard. The
relation of the $\delta$-like scatterers to well defined potentials is
discussed in the paper. 

The aim of this work is to use periodic and diffracting orbits to compute
the form factor of rectangular billiards with localized perturbations.
The calculation includes non diagonal contributions and leads to a
power expansion of $K(\tau)$ where only the first three powers of $\tau$
are calculated.
The structure of this paper is as follows.
In Sec.~\ref{noangle} the form factor is calculated for angle
independent diffraction. In Sec.~\ref{point} the model of point
$\delta$-like interactions is 
presented, in order to introduce a simple example of angle independent
scattering. For this model the results of Sec.~\ref{noangle} are compared to numerical
results in Sec.~\ref{numerics}.
The scattering from localized potentials that are physically well defined is
studied in Sec.~\ref{scatter},
where it is shown how to compute the diffraction constant in the limit
where the size of the perturbation is smaller than the wavelength.
In Sec.~\ref{angles} the form factor for angle dependent scattering, with the
diffraction constant that was calculated Sec. ~\ref{scatter}, 
is computed.
The results are summarized and discussed in Sec.~\ref{discussion}.

\section{The Form Factor of a point-like S wave Scatterer}
\label{noangle}

In this section the form factor will be calculated for a rectangular
billiard with a scatterer such that
its diffraction constant $D$, is angle independent.
The form factor is
\begin{equation}
\label{formfactor}
K(\tau)=  \int_{-\infty}^{\infty} d \eta R_2 (\eta) e^{2 \pi i \eta \tau},
\end{equation}
where
$R_2 (\eta)$ is the energy levels 
correlation function:
\begin{equation}
\label{correlation}
R_2 (\eta)= \left\langle d_{osc} \left( E-\frac{\eta \Delta}{2} \right) d_{osc} \left( E+\frac{\eta \Delta}{2} \right) \right\rangle \Delta^2
\end{equation}
and $\Delta$ denotes the mean level spacing of the system.
The brackets denote averaging over an energy window which is large compared to the mean level spacing $\Delta$ but small when compared to the semiclassical energy $E$. In the semiclassical limit, the oscillatory part of the
density of states, $d_{osc} (E) \equiv d(E) - \langle d(E) \rangle $, is of the form
\begin{equation}
 d_{osc}(E) \simeq \sum_p A_p e^{i \frac{S_P}{\hbar}} + c.c.,
\end{equation}
where $c.c.$ denotes the complex conjugate
and the sum is over periodic orbits. Using this density
of states, the form factor is given by:
\begin{equation}
\label{nondiagfactor}
K(\tau) = 2 \pi \hbar \Delta \left\langle \sum_{pp'} A_p A_{p'}^{*} e^{\frac{i}{\hbar}(S_p - S_{p'})} \delta \left(\frac{2 \pi \hbar}{\Delta} \tau - \frac{t_p + t_{p'}}{2} \right) \right\rangle 
\end{equation}
where $S_p$ is the action of the orbit, $t_p$ is its period, and $\tau > 0$ is assumed.
The sum is over the leading semiclassical contributions to the density of states.
For smooth potentials $d_{osc}$ is given by the Berry-Tabor formula for 
integrable systems~\cite{BT-I} and by the Gutzwiller trace formula
for chaotic systems~\cite{gutz67,gutzwiller}.
In these formulas the sum is over the periodic orbits of the
system. In the present work, periodic orbits and diffracting orbits
will be included in the sum.

Due to the averaging only action differences that are of the order of $\hbar$
or less
can contribute to the form factor. As a result, the diagonal approximation can be used, where only pairs of orbits with the same action contribute. This 
approximation is valid for integrable systems~\cite{berry85} and thus also
valid when there are only once diffracting orbits (since their density is
proportional to that of periodic orbits as will be explained in the following). When orbits with more than one diffraction are
included this approximation is no longer valid and non diagonal terms should
be included as is done in what follows.
Since the motion in the billiard is free, the periods of the orbits are
related to the lengths by $t_p=\frac{l_p}{2k}$ where $k=\sqrt{E}$ is the wavenumber (the units $\hbar=1$, $2m=1$ are used from now). The mean level spacing
, in the semiclassical limit, is given by $\Delta=\frac{4 \pi}{\cal A}$. 
In the diagonal approximation the form factor is expressed as a sum over orbits of identical lengths
\begin{equation} 
\label{diagfactor}
K(\tau) = \sum_p \frac{16 \pi^2 k}{\cal A} \left\langle |A_p|^2 \delta (\l_p - {\cal A} k \tau) \right\rangle.
\end{equation}
The averaging widens the delta function and enables to replace the sum over orbits by an integral using the density of orbits. Every factor of length, $l_p$, is replaced by a factor of ${\cal A} k \tau$, and as will be seen, every diffraction order eventually contributes a power of $\tau$
 to the from factor. 
The form factor $K(\tau)$ will be calculated to the third order in $\tau$.

The spectral statistics are very sensitive to the location of the scatterer,
since they are influenced by degeneracies in the lengths of the diffracting orbits. Therefore we choose to treat first the simplest case when the scatterer is in the center of the rectangle.
 Then length degeneracies are maximal. Later we treat the case in which the scatterer is at a typical location, such that its coordinates (relative to the rectangle sides) are typical irrational numbers. 

\subsection{Scatterer at the center}

The periodic and diffracting orbits are classified by the number of
bounces from the boundaries. We will denote by ${\bf N}_p=(N_p,M_p)$ a
periodic orbit with $2N_p$ and $2M_p$ bounces from the boundaries and by
${\bf N}_j=(N_j,M_j)$ a diffracting orbit with $N_j$ and $M_j$ bounces
from the boundary.
When the scatterer is at the center of the rectangle the length of the diffracting segments is
given by
\begin{equation} 
\label{lengthper}
l_j=\sqrt{a_x^2 N_j^2 + a_y^2 M_j^2}
\end{equation} 
while the length of the periodic orbits is 
\begin{equation}
\label{lengthdif}
l_p=2\sqrt{a_x^2 N_p^2 +a_y^2 M_p^2}
\end{equation}
where  $a_x$ and $a_y$ are the lengths of the sides of the rectangle.

\begin{figure}[p]
\begin{flushright}
\vspace{.5cm}
 \leavevmode
\epsfig{file=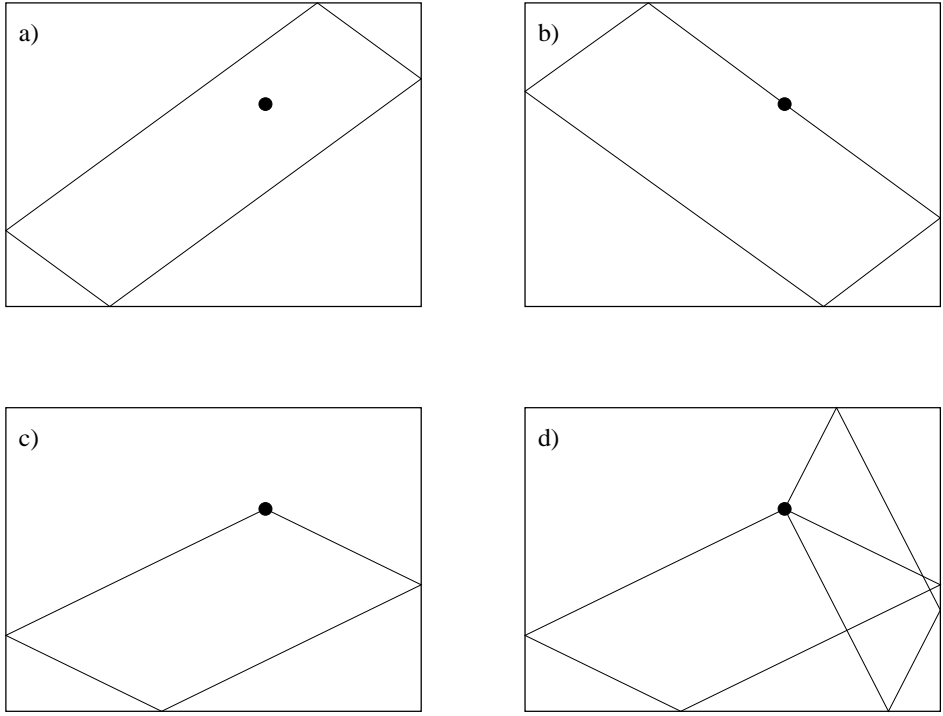,height=9.5cm,width=12.5cm,angle=0}
\vspace{.5cm}
\centering{\caption{Several orbits contributing to the density of states: a) a periodic orbit (the ($1,1$) orbit), b) a forward once diffracting orbit ($2,2$), c) a once diffracting orbit ($2,1$), and d) a twice diffracting orbit. \label{orbitsfig}}}
\end{flushright}
\vspace{0cm}
\end{figure}

The contributions to the oscillatory part of the density of states
were calculated in the framework of geometrical theory of diffraction
in App.~\ref{diffcont}. For an angle independent scatterer
the contributions~(\ref{finaldos}) to $d_{osc}$ can be simplified since $D(\theta,\theta')$ reduces to a constant $D$ and if the scatterer is at the center the 
terms do not depend on $\mu$ and $\nu$.
 The contribution of them
with no more than three diffractions is
\begin{equation}
\label{dosc2}
d_{osc} (E)= \sum_p A_p^{(0)} e^{ikl_p} + \sum_j A_j^{(1)} e^{i k l_j}
+\sum_{j_1,j_2} A_{j_1,j_2}^{(2)} e^{ik(l_{j_1}+l_{j_2})} + \sum_{j_1,j_2,j_3} A_{j_1,j_2,j_3}^{(3)} e^{ik(l_{j_1}+l_{j_2}+ l_{j_3})}+c.c.+\cdots
\end{equation}
where
\begin{equation}
\label{apo}
A_p^{(0)} = \frac{2 {\cal A}}{\pi \sqrt{8 \pi k l_p}} e^{-i \frac{\pi}{4}},
\end{equation}
\begin{equation}
\label{a1dif}
A_j^{(1)} = \frac{(-1)^{N_j+M_j}\sqrt{l_j}}{\pi k \sqrt{8 \pi k}} D e^{-i \frac{3}{4} \pi},
\end{equation}
\begin{equation}
\label{a2dif}
A_{j_1,j_2}^{(2)} = (-1)^{N_{j_1}+M_{j_1}+N_{j_2}+M_{j_2}} \frac{l_{j_1}+l_{j_2}}{4 \pi^2 k^2 \sqrt{l_{j_1} l_{j_2}}} D^2 e^{-i \frac{3}{2} \pi} 
\end{equation}
and
\begin{equation}
\label{a3dif}
A_{j_1,j_2,j_3}^{(3)} = (-1)^{N_{j_1}+M_{j_1}+N_{j_2}+M_{j_2}+N_{j_3}+M_{j_3}} \frac{16 D^3}{3 \pi k (8 \pi k)^{\frac{3}{2}}} \frac{l_{j_1}+l_{j_2}+l_{j_3}}{\sqrt{l_{j_1} l_{j_2} l_{j_3}}} e^{-i \frac{\pi}{4}}.
\end{equation}
We omitted the factors $\beta_j$  defined in~(\ref{perioddosc}) since the fraction of orbits where these
differ from unity is negligible in the calculations of the present section.
Examples of the various orbits contributing to (\ref{dosc2}-\ref{a2dif}) are
presented in Fig.~\ref{orbitsfig}. 

First the contribution of terms that involve no more than one diffraction,
resulting from the first two terms in the expansion~(\ref{dosc2}) of
$d_{osc}$, will be calculated. Following~(\ref{inangle}) the contributions
of the diffracting terms can be classified according the parity of $N_j$
and $M_j$. If both $N_j$ and $M_j$ are even (even-even orbits) the orbit
$j$ coincides with a periodic orbit, since in the notation of~(\ref{inangle}),
at the scatterer $\theta_{in} = \theta_{out}$ 
 (compare Fig.~\ref{orbitsfig} a) and b)) 
and the contribution
of these orbits will be calculated first. There is no such relation
for the other parities, (even-odd, odd-even, odd-odd, referring to the parities
of $N_j$ and $M_j$) and actually for such parities the diffracting orbits
are halves of periodic orbits
 (this is only true if the scatterer is at the center, therefore for example the orbit
in Fig.~\ref{orbitsfig} c) is not a half of a ($2,1$) periodic orbit).
For even-even diffracting orbits the first two sums in the expression~(\ref{dosc2}) can be grouped as
\begin{equation}
\label{1difftod}
d_{osc}^{(1)} (E)= \sum_p \tilde{A}_P e^{i k l_p} + \sum_j A_j^{(1)} e^{i k l_j} + c.c.
\end{equation}
where the first sum runs over periodic orbits while the second sum
runs over diffracting orbits, omitting orbits where both $N_j$ and
$M_j$ are even and
\begin{equation}
\label{tildea}
\tilde{A}_p = \left( \frac{2{\cal A}}{\pi \sqrt{8 \pi k l_p}}+\frac{\sqrt{l_p}}{\pi k \sqrt{8 \pi k}} D e^{-i \frac{\pi}{2}} \right) e^{-i \frac{\pi}{4}}.
\end{equation}
The diagonal approximation is now applied to various terms in~(\ref{1difftod})
and the sum in~(\ref{diagfactor}) is replaced by an integral with the
density
\begin{equation}
\label{density}
\rho (l) = \frac{\pi l}{8 {\cal A}}
\end{equation}
of orbits of length $l$. This is the density of periodic orbits as well as 
the density of diffracting orbits with a given parity of $N_j$ and $M_j$
as can be easily seen from~(\ref{lengthper}) and~(\ref{lengthdif}).
The resulting contribution of the periodic and even-even diffracting orbits
to the form factor is
\begin{eqnarray}
\label{ktau1}
K^{(1,e)}(\tau) & = \frac{16 \pi^2 k}{{\cal A}} \int_0^{\infty} dl \frac{\pi l}{8 {\cal A}} \left( \frac{{\cal A}^2}{2 \pi^3 k l} + \frac{{\cal A}}{2 \pi^3 k^2} \Im D + \frac{l}{8 \pi^3 k^3} |D|^2 \right) \delta ( l - {\cal A} k \tau ) \nonumber \\ &  = 1+\Im D \tau +\frac{|D|^2}{4} \tau^2.
\end{eqnarray}
The contribution of the other diffracting orbits (odd-odd, odd-even, even-odd) with
the amplitudes $A_j^{(1)}$ of~(\ref{a1dif}) are
\begin{equation}
\label{formparts}
K^{(1,o-o)} (\tau) = K^{(1,o-e)} (\tau) =K^{(1,e-o)}(\tau) = \frac{|D|^2 \tau^2}{4}
\end{equation}
and their combined contribution is
\begin{equation}
\label{oncedifandpr}
K^{(1,o)} (\tau) = \frac{3 |D|^2 \tau^2}{4}.
\end{equation}
Thus, the contribution of the once-diffracting orbits and the periodic orbits to
the form factor is
\begin{equation}
\label{k1tau}
K^{(1)}(\tau)  = K^{(1,e)}(\tau)+K^{(1,o)}(\tau)= 1+\Im D \tau +|D|^2 \tau^2.
\end{equation}

From~(\ref{apo}-\ref{a2dif}) we note that the powers of $\frac{1}{\sqrt{k}}$
and $\sqrt{l_j}$ (the powers in~(\ref{apo}-\ref{a2dif}) combined with the 
increasing density of multiple diffracting orbits resulting
from powers of~(\ref{density})) increase with the powers of $D$ and by~(\ref{nondiagfactor})
and~(\ref{diagfactor}) these lead to increasing orders of $\tau$.
We are interested first in all contributions to the order $\tau^2$. The additional terms that may contribute to this order result from non diagonal terms
in~(\ref{nondiagfactor}) where $p$ is a periodic orbit and $j'$ is an
orbit with two diffractions, namely terms of the form
\begin{equation}
\label{orbitcomb}
A_p A_{j'}^{*} e^{\frac{i}{\hbar}(S_p-S_{j'})}=A_p A_{j_1,j_2}^{(2) *} e^{ik(l_p-l_{j_1}-l_{j_2})}.
\end{equation}
The contribution of such terms will be analyzed in the following. All other
contributions are of higher order in $\tau$.

We turn to analyze the contribution of non diagonal terms involving periodic and
twice diffracting orbits. These oscillate as $e^{ik(l_p-l_{j_1}-l_{j_2})}$
that may be potentially fast.
Naively, one can guess that after smoothing,
the contributions relevant to the form factor are those for which $l_p=l_{j_1}+l_{j_2}$.
However, it turns out that there are not enough orbits that satisfy this
equality in order to compensate for the extra factor of $1/k$ due to
the additional diffraction. The contribution comes from combinations
of orbits for which the lengths are almost equal. Such combinations survive
since there are enough orbits for which the phase of the exponent is of the
order of unity in the semiclassical limit.
This is caused by the fact that these orbits have hidden saddle manifolds
on which their action differences are slowly varying. These manifolds
were identified by Bogomolny~\cite{bogomolny00b}, and were used to 
compute the density-density correlation function and mean values of powers of Green's functions for integrable systems.
The diffraction problem discussed here has lengths (or actions) that are related to the length of the periodic orbits so that saddle manifolds exist.
We use Bogomolny's method to compute this contribution.
It will be convenient to expand the
length of the diffracting orbits around the length of orbits that 
satisfy $l_p=l_{j_1}+l_{j_2}$. This expansion is given by
\begin{eqnarray}
\label{expandlength}
l_j & = &\sqrt{a_x^2 (N_j+ \delta N_j)^2 + a_y^2 (M_j+\delta M_j)^2}  \simeq \sqrt{a_x^2 N_j^2 + a_y^2 M_j^2} \nonumber \\ & & +\frac{1}{\sqrt{a_x^2 N_j^2 + a_y^2 M_j^2}} (a_x^2 N_j \delta N_j +a_y^2 M_j \delta M_j) +\frac{a_x^2 a_y^2}{2 (a_x^2 N_j^2 + a_y^2 M_j^2)^{\frac{3}{2}}} (M_j \delta N_j -N_j \delta M_j )^2.
\end{eqnarray}
The length difference in the exponent is 
\begin{equation}
\delta l_p =  l_{j_1} + l_{j_2} -l_p
\end{equation}
The periodic orbit is not expanded because for each periodic orbit there
will be a saddle manifold composed of different twice-diffracting orbits.
The summation over the orbits in the manifold will be also
combined with summation over all of the periodic orbits.
Equivalently, it is possible to fix any of the lengths, expand the other lengths, and to reach
the same contribution to the form factor.

The condition that the orbits are on the saddle manifold is that the length
difference, $\delta l_p$, is stationary leading to:
\begin{equation}
\label{saddle1}
\frac{1}{\sqrt{a_x^2 N_{j_1}^2 + a_y^2 M_{j_1}^2}} (a_x^2 N_{j_1} \delta N_{j_1} +a_y^2 M_{j_1} \delta M_{j_1}) +\frac{1}{\sqrt{a_x^2 N_{j_2}^2 + a_y^2 M_{j_2}^2}} (a_x^2 N_{j_2} \delta N_{j_2} +a_y^2 M_{j_2} \delta M_{j_2}) =0.
\end{equation}
The orbits on the manifold are the {\em integer} solutions of this equation. Since
the aspect ratio of the rectangle is assumed to be a typical irrational number,
such solutions can be found only if the orbits are related by
\begin{equation}
\label{precond1}
\frac{\sqrt{a_x^2 N_{j_1}^2 + a_y^2 M_{j_1}^2}}{\sqrt{a_x^2 N_{j_2}^2 + a_y^2 M_{j_2}^2}}=\frac{r_{j_1}}{r_{j_2}}.
\end{equation}
This means that the orbits are repetitions of some simpler orbit, for example
\begin{equation}
\sqrt{a_x^2 N_{j_1}^2 + a_y^2 M_{j_1}^2}=r_{j_1} \sqrt{a_x^2 n^2 + a_y^2 m^2} = r_{j_1} l_0.
\end{equation}
Substituting this condition in~(\ref{saddle1}) leads to
\begin{equation}
a_x^2 n \delta N_{j_1} + a_x^2 n \delta N_{j_2} + a_y^2 m \delta M_{j_1} + a_y^2 m \delta M_{j_2} =0.
\end{equation}
Since the aspect ratio is a typical irrational number an integer solution can be found only if
\begin{eqnarray}
\label{cond1}
\delta M_{j_1} & = & - \delta M_{j_2} \nonumber \\
\delta N_{j_1} & = & - \delta N_{j_2}.
\end{eqnarray}
We are interested in the contribution that survives the energy averaging, 
required for contributions to the form factor. Therefore we will consider the saddle manifold for which
\begin{equation}
\label{dif+dif=pr}
 l_p=2\sqrt{a_x^2 N_p^2 + a_y^2 M_p^2} =(r_{j_1} + r_{j_2}) l_0 .
\end{equation}
This condition states that the periodic orbit (which will contribute) is also proportional to $(n,m)$ and $l_p= 2 r_p l_0$ with
\begin{equation}
\label{cond2}
2r_p=r_{j_1}+r_{j_2}
\end{equation}
is satisfied.

\begin{figure}[p]
\begin{flushright}
\vspace{.5cm}
 \leavevmode
\epsfig{file=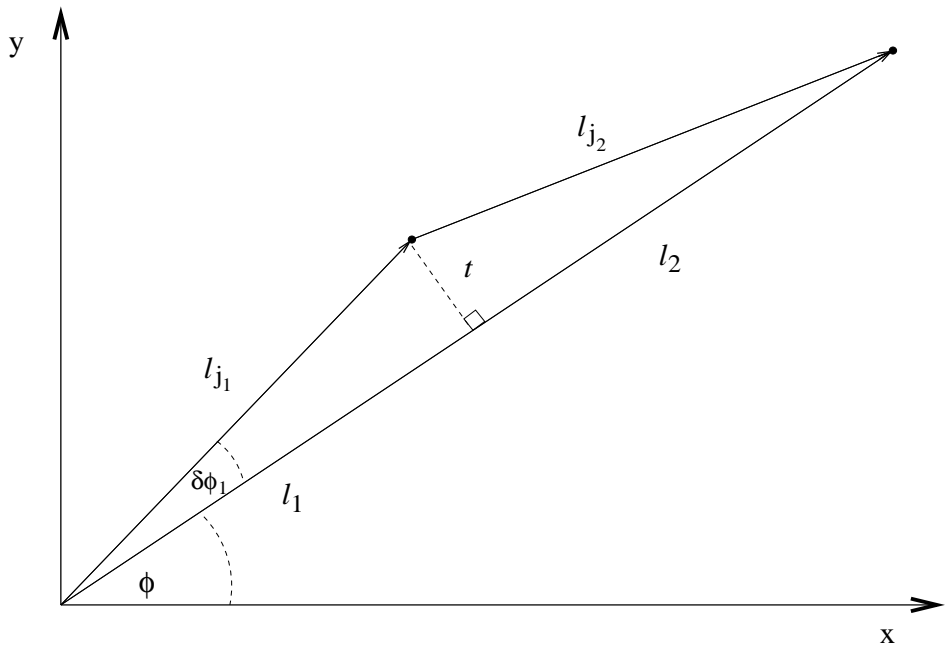,height=8.5cm,width=12.5cm,angle=0}
\centering{\caption{The configuration of lengths on the saddle manifold \label{lengths}}}
\end{flushright}
\vspace{0cm}
\end{figure}

The contribution of the terms of this saddle manifold to the level-level 
correlation function $R_2(\eta)$ is:
\begin{eqnarray}
\label{summm}
{R'}_2^{(2)} (\eta) = \left\langle \sum_{p,j_1,j_2} \frac{2{\cal A}}{\pi \sqrt{8 \pi l_p k}} e^{-i\frac{\pi}{4}} \frac{l_{j_1}+l_{j_2}}{4 \pi^2 k^2 \sqrt{l_{j_1} l_{j_2}}}e^{i \frac{3\pi}{2}} e^{ik(l_p-l_{j_1}-l_{j_2})} D^2 e^{i\frac{\eta \Delta}{4 k}(l_{j_1}+l_{j_2}+l_p)} \right\rangle \Delta^2 \nonumber \\ +c.c. +\left( \eta \rightarrow -\eta \right).
\end{eqnarray} 
The symbol $\left( \eta \rightarrow -\eta \right)$ denotes two terms similar to
the first ones but with the opposite sign of $\eta$.
This is actually a sum over $(N_p,M_p)$, $(N_{j_1},M_{j_1})$ and $(N_{j_2},M_{j_2})$ that define the orbits $p$, $j_1$ and $j_2$ respectively. The
contributions with similar phases, that do not cancel each other,
and therefore survive the averaging over energy involved in the
calculation of the form factor are those that satisfy~(\ref{cond1}) and~(\ref{cond2}). The sum will be calculated and then Fourier
transformed in order to obtain the corresponding contribution to the form factor.
The dominant contributions are these in which the
length difference $\delta l_p$ is much smaller than unity. The configuration of these
contribution is shown in Fig.~\ref{lengths}.
The two diffracting segments are almost
parallel to the periodic orbit. Such a configuration is allowed since
the difference between any lattice points is also a lattice point, moreover,
the conditions~(\ref{cond1}), (\ref{cond2}) force the orbit to have this configuration. Note that $l_1, l_2$ are the projections of the lengths of the diffracting segments on the direction of the periodic orbit (to the leading order in $\delta \phi_j$). Since they are almost parallel they are
a good approximation to the lengths of the diffracting segments.

Since the contributing orbits have the configuration of Fig.~\ref{lengths},
the variables $l_1$ and $t$ are convenient coordinates. This choice also
takes into account the conditions~(\ref{cond1}) and (\ref{cond2}). 
The length difference is given by
\begin{equation}
\label{ldif}
\delta l_p = l_{j_1}+l_{j_2}-l_p \simeq \frac{1}{2} t^2 \left( \frac{1}{l_1} +\frac{1}{l_2}\right) - \frac{1}{8} t^4 \left( \frac{1}{l_1^3} + \frac{1}{l_2^3} \right)
\end{equation}
where $l_2=l_p-l_1$. The length difference $\delta l_p$ appears as a phase in an exponent multiplied
by $k$. Thus only length differences which scale as $\frac{1}{k}$ can
contribute to the correlation function. The lengths of orbits
that will contribute to the form factor scale as $k$ and thus the corresponding values of
$t$ are of the order of unity or less.
Since the sum is dominated by terms in which $t \sim 1$ or less, the expansion
(\ref{ldif}) is justified. This expansion may break down when 
either $l_1 \rightarrow 0$ or $l_2 \rightarrow 0$. However, there is a minimal
length for the diffracting orbits and such lengths do not occur in the sum.

\begin{figure}[p]
\begin{flushright}
\vspace{.5cm}
 \leavevmode
\epsfig{file=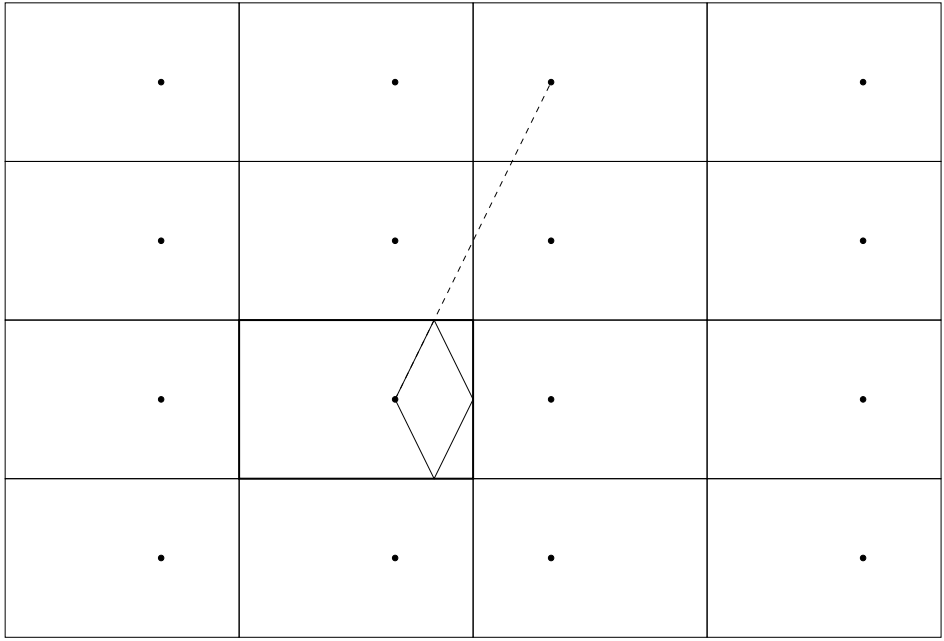,height=8.5cm,width=12.5cm,angle=0}
\vspace{.5cm}
\centering{\caption{Unfolding of orbits using reflections around the rectangle boundaries. \label{lattice}}}
\end{flushright}
\vspace{0cm}
\end{figure}

The correlation function is computed by replacing the sum over orbits
by an integral over $l_1$ and $t$. The slowly varying terms are replaced
by their values on the saddle (when $t=0$) and the phase is replaced by the 
leading term of~(\ref{ldif}). To transform the sum into an integral the density of contributions is needed.
The configuration in Fig.~\ref{lengths} is obtained when the plane is tiled
with rectangles using reflections with respect to 
 the hard walls. The scatterer is also
reflected and its images form a lattice (since there is a scatterer in the
center of each rectangle).
An example for this unfolding is presented in Fig.~\ref{lattice} (for a 
scatterer which is not at the center).
The ($1,2$) orbit in the original rectangle (solid line) is equivalent to the
one connecting the scatterer with its image (dashed line).
 The density of points in this lattice is $\frac{1}{\cal A}$. After these manipulations the correlation function is given by
\begin{eqnarray}
\label{newcalc1}
R^{'(2)}_2 (\eta) = & \left\langle \sum_p \int_0^{l_p} d l_1 \int d t \; \frac{2}{\pi \sqrt{8 \pi l_p k}} e^{-i\frac{\pi}{4}} \frac{l_1+l_2}{4 \pi^2 k^2 \sqrt{l_1 l_2}} e^{i\frac{3\pi}{2}} e^{i \frac{\eta \Delta}{4k} ( l_1+l_2+l_p)} D^2 e^{-i \frac{kt^2}{2} \left( \frac{1}{l_1} + \frac{1}{l_2} \right)} \right\rangle \Delta^2 \nonumber \\ &+ c.c + (\eta \rightarrow -\eta).
\end{eqnarray}
Calculation of the Gaussian integral leads to
\begin{equation}
 R^{'(2)}_2 (\eta) = \left\langle \sum_p \frac{l_p}{2 \pi^3 k^3} (-\Re D^2) e^{i \frac{\eta \Delta}{2 k} l_p} \right\rangle \Delta^2 + (\eta \rightarrow -\eta).
\end{equation}
The Fourier transform leads to the form factor
\begin{equation}
\label{newcalc3}
K^{'(2)} (\tau)= \left\langle \sum_p \frac{2 \Delta l_p}{\pi^2 k^2} \delta \left( l_p - \frac{4 \pi k \tau}{\Delta}\right) (-\Re D^2) \right\rangle
\end{equation}
where only positive times were taken into account. The sum over periodic
orbits is replaced by an integral using the density~(\ref{density})
leading to 
\begin{equation}
\label{form2diff}
{K'}^{(2)}(\tau) = \tau^2 \Re (-D^2),
\end{equation}
which is the non diagonal contribution to the form factor.  
This contribution is indeed of order $\tau^2$ as was expected.

After computing all of the contributions of order $\tau^2$ we turn to compute
the contributions to the form factor which are of order $\tau^3$.
There are three possible contributions of this order: diagonal
contributions from twice diffracting orbits that will be denoted by $K^{(3)} (\tau)$,
non diagonal contributions resulting from combinations of once and twice
diffracting orbits, denoted by $K^{'(3a)} (\tau)$, and
from combinations of periodic orbits and orbits
with three diffractions, denoted by $K^{'(3b)} (\tau)$.
These contributions are computed in the following.

To obtain the diagonal contribution of twice diffracting orbits
(\ref{dosc2}) is substituted in~(\ref{correlation}), and only
the terms with two diffractions are kept. A Fourier transform
leads to
\begin{equation}
K^{(3)} (\tau) = \left\langle \sum_{j_1 j_2 j_3 j_4} {A^{(2)}_{j_1 j_2}}^* A_{j_3 j_4}^{(2)} e^{-ik(l_{j_1}+l_{j_2}-l_{j_3}-l_{j_4})} \delta \left( \frac{l_{j_1}+l_{j_2}+l_{j_3}+l_{j_4}}{2} - \frac{4 \pi k \tau}{\Delta} \right) \right\rangle 4 \pi k \Delta
\end{equation}
where contributions from negative times were omitted.
The diagonal approximation requires that $l_{j_1}+l_{j_2}=l_{j_3}+l_{j_4}$,
this means that $l_{j_1}=l_{j_3}$ and $l_{j_2}=l_{j_4}$ or that $l_{j_1}=l_{j_4}$ and $l_{j_2}=l_{j_3}$. Both cases result in identical contributions  (the
fraction of orbits
where $l_{j_1}=l_{j_2}$ is negligible).
Using~(\ref{a2dif}) and replacing the sums by integrals with the density
of diffracting segments $\frac{\pi l}{2 {\cal A}}$ resulting of 
multiplication of~(\ref{density}) by $4$, to take account of the
various possible parities, leads to
\begin{equation}
\label{k3tau}
K^{(3)} (\tau) = \int_0^{\infty} d l_1 d l_2 \; \frac{\pi^2 l_1 l_2}{4 {\cal A}^2} \frac{(l_1+l_2)^2 \Delta}{2 \pi^3 k^3 l_1 l_2} |D|^4 \delta \left( l_1 + l_2-\frac{4 \pi k \tau}{\Delta}\right) = \frac{1}{2} |D|^4 \tau^3.
\end{equation}
One should note that there is a non diagonal contribution from combinations
of two twice diffracting orbits but this contribution is of higher order in $\tau$.

\begin{figure}[p]
\begin{flushright}
\vspace{.5cm}
 \leavevmode
\epsfig{file=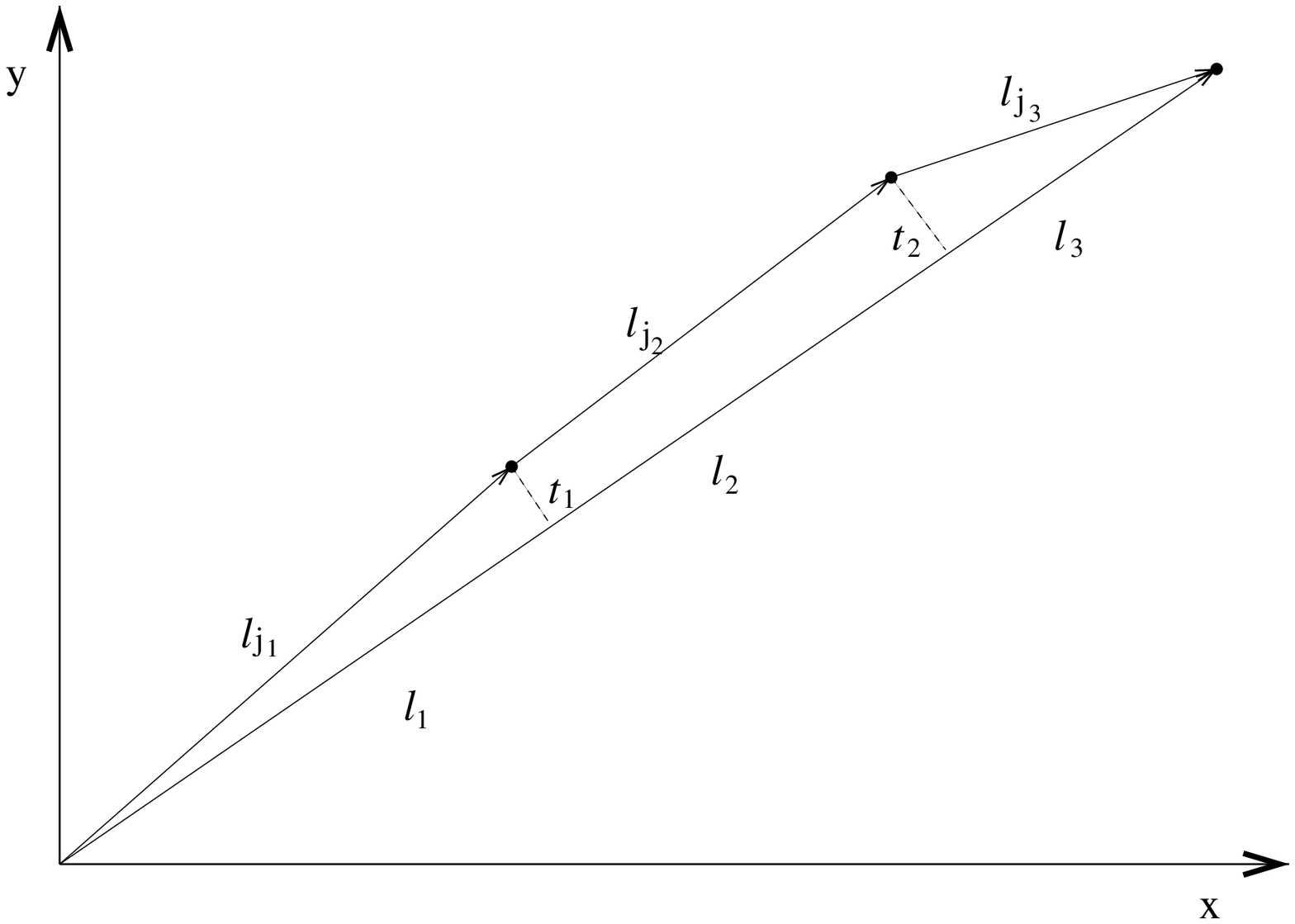,height=8.5cm,width=12.5cm,angle=0}
\centering{\caption{The configuration of lengths on the saddle manifold composed of periodic orbits and orbits with three diffractions\label{saddle3}}}.
\end{flushright}
\vspace{0cm}
\end{figure}

The contribution which results from combinations of once and twice diffracting
orbits is computed along the same lines as that from the combinations
from twice diffracting and periodic orbits which was computed earlier.
Using~(\ref{a1dif}) and (\ref{a2dif}), 
the relevant contribution to the correlation function is found to be given by 
\begin{eqnarray}
R_2^{'(3a)} (\eta)& = & \left\langle \sum_{j_1 j_2 j_3} (-1)^{N_{j_1} + M_{j_1}+N_{j_2} + M_{j_2}+N_{j_3} + M_{j_3}} \frac{\sqrt{l_{j_1}}(l_{j_2}+l_{j_3})}{4 \pi^3 k^3 \sqrt{8 \pi k l_{j_2} l_{j_3}}} \left[ |D|^2 D^* e^{ik (l_{j_1} - l_{j_2} -l_{j_3})} \right. \right. \nonumber \\
& \times & \left. \left. e^{-i \frac{\eta \Delta}{4k} (l_{j_1} + l_{j_2} +l_{j_3})} e^{i \frac{3 \pi}{4}} + c.c. + \left( \eta \rightarrow - \eta \right) \right] \rule{0mm}{6mm} \right\rangle \Delta^2
\end{eqnarray} 
where $j_1$ denotes the once diffracting orbit with length $l_{j_1}$ and $j_2,j_3$ denotes the twice diffracting orbit with length $l_{j_2}+l_{j_3}$.
 Only combinations of orbits for which $l_{j_1} \simeq l_{j_2}+ l_{j_3}$
survive the smoothing. This leads to a saddle manifold of orbits that
satisfy $N_{j_1}= N_{j_2}+N_{j_3}$ and $M_{j_1}= M_{j_2}+M_{j_3}$
(note that on this saddle manifold $N_{j_1}+ N_{j_2}+N_{j_3}+M_{j_1}+ M_{j_2}+M_{j_3}$
is even). To compute this contribution the sum over $j_2,j_3$ is restricted
to the saddle manifold, on which the length difference is
\begin{equation}
l_{j_1} - l_{j_2} -l_{j_3} \simeq -\frac{t^2}{2 l_2} -\frac{t^2}{2 l_3}
\end{equation}
where $l_2 + l_3 = l_{j_1}$, in analogy with~(\ref{ldif}).
The summation on the saddle manifold is then replaced by integration over
$t,l_2$ with density $\frac{1}{{\cal A}}$
\begin{eqnarray}
\label{r2four}
R_2^{'(3a)} (\tau) & = & \left\langle \sum_{j_1} \int_0^{l_{j_1}} \! dl_2 \! \int \! dt  \frac{l_{j_1}^{\frac{3}{2}}}{4 \pi^3 k^3{\cal A} \sqrt{8 \pi k l_2 l_3}} \left[ |D|^2 D^* e^{-ik \left( \frac{t^2}{2l_2}+\frac{t^2}{2l_3} \right)} e^{-i \frac{\eta \Delta}{2k}l_{j_1}+i\frac{3\pi}{4}} +\! c.c.\! +\left( \eta \! \rightarrow \! -\eta \right)\right] \right\rangle \Delta^2 \nonumber \\
& = & \left\langle \sum_{j_1} \frac{l_{j_1}^2}{4 \pi^3 k^4 {\cal A}} |D|^2 (\Im D) e^{-i\frac{\eta \Delta}{2k}} + \left( \eta \rightarrow -\eta \right)\right\rangle \Delta^2.
\end{eqnarray}
The contribution to the form factor is obtained by Fourier transforming~(\ref{r2four}) and replacing the sum over $j_1$ by an integral with the density
$\frac{\pi l_1}{2{\cal A}}$, resulting in
\begin{equation}
\label{k4tau}
K^{'(3a)} (\tau) = \int_0^{\infty} d l_1 \; \frac{l_1^3 \Delta}{2 \pi k^3 {\cal A}^2} |D|^2 \Im D \delta \left( l_1 -\frac{4\pi k \tau}{\Delta} \right) = 2 |D|^2 \Im D \tau^3.
\end{equation}
The last contribution to be computed is from combinations of periodic orbits
and orbits with three diffractions.
The calculation is very similar to the one just performed. 
Using~(\ref{apo}) and (\ref{a3dif}) the contribution
to the correlation function from combinations of periodic orbits and
orbits with three diffractions is found to be
\begin{equation}
\label{r2no5}
R_2^{'(3b)} (\eta) = \left\langle \sum_{p j_1 j_2 j_3} \frac{{\cal A}(l_{j_1}+l_{j_2}+l_{j_3})}{6 \pi^4 k^3 \sqrt{l_p l_{j_1} l_{j_2} l_{j_3}}} \left[ (D^3)^* e^{ik(l_p-l_{j_1}-l_{j_2}-l_{j_3})} e^{-i\frac{\eta \Delta}{4 k}(l_p+l_{j_1}+l_{j_2}+l_{j_3})} + c.c. + \left( \eta \rightarrow -\eta \right) \right] \right\rangle \Delta^2
\end{equation}
where the factor $(-1)^{N_{j_1}+N_{j_2}+N_{j_3}+M_{j_1}+M_{j_2}+M_{j_3}}$ was omitted since
on the saddle manifold it reduces to unity.
The saddle manifold is composed of orbits which satisfy
\begin{eqnarray}
N_{j_1} + N_{j_2} + N_{j_3} & = & N_p \nonumber \\
M_{j_1} + M_{j_2} + M_{j_3} & = & M_p.
\end{eqnarray}
 The dominant contributions to the correlation function 
are from diffracting segments which are almost parallel to the periodic orbit
as is shown in Fig.~\ref{saddle3}.
The length difference is given by
\begin{equation}
l_p - l_{j_1} -l_{j_2} - l_{j_3} \simeq - \frac{1}{2} \frac{t_1^2}{l_1} - \frac{1}{2} \frac{t_2^2}{l_3}-\frac{1}{2} \frac{(t_1-t_2)^2}{l_2}.
\end{equation}
The summation over $j_1,j_2,j_3$ in~(\ref{r2no5}) is restricted to the saddle
manifold and then replaced by integration over $l_1,t_1,l_2,t_2$ where $l_1$ varies
between $0$ and $l_p$ and $l_2$ between $0$ and $l_p-l_1$.
When the sums over $j_1,j_2$ are replaced by integrals a factor
of $\frac{1}{{\cal A}^2}$, resulting from the density of contributions
is introduced.
The integral over $t_1,t_2$ is Gaussian and can be evaluated
\begin{equation}
\int d t_1 d t_2 \; e^{-i \frac{k}{2}\left[ \frac{t_1^2}{l_1} +\frac{t_2^2}{l_3} +\frac{(t_1-t_2)^2}{l_2}\right]} = \frac{2 \pi}{i k} \sqrt{\frac{l_1 l_2 l_3}{l_p}}.
\end{equation}
The resulting contribution to the correlation function is given by
\begin{equation}
\label{r25final}
R_2^{'(3b)} (\tau) = \left\langle \sum_p \int_0^{l_p} dl_1 \int_0^{l_p-l_1} dl_2 \; \frac{1}{3 \pi^3 k^4 {\cal A}} \left[ -i (D^3)^* e^{-i \frac{\eta \Delta}{2 k}l_p} + c.c+ \left( \eta \rightarrow -\eta \right) \right] \right\rangle \Delta^2.
\end{equation}
The integrals can be computed and then~(\ref{r25final}) is Fourier transformed
leading to the following contribution to the form factor 
\begin{equation}
K^{'(3b)} (\tau) = - \left\langle \sum_p \frac{4 \Delta l_p^2}{3 \pi^2 k^3 {\cal A}} \Im (D^3) \delta \left(l_p - \frac{4 \pi k \tau}{\Delta} \right) \right\rangle.
\end{equation}
The sum over periodic orbits can be replaced by an integral using the density
(\ref{density}), leading to
\begin{equation}
\label{k5tau}
K^{'(3b)} (\tau) = - \frac{2}{3} \Im (D^3) \tau^3.
\end{equation}

Combining all the different contributions, that are given by~(\ref{k1tau}), (\ref{form2diff}), (\ref{k3tau}), (\ref{k4tau}) and~(\ref{k5tau}), the form factor is 
\begin{equation}
\label{centerform}
K (\tau) = 1 + \Im D \tau +\left( |D|^2 -\Re D^2 \right) \tau^2 + \frac{|D|^4 \tau^3}{2}+2|D|^2 \Im D \tau^3 -\frac{2}{3} \Im D^3 \tau^3 + O(\tau^4).  
\end{equation}
The diffraction is assumed to conserve probability and therefore the diffraction constant has to satisfy the optical theorem. For an angle
independent scatterer in two dimensions it takes the form:
\begin{equation}
\label{opth}
\Im D =  - \frac{|D|^2}{4}, 
\end{equation}
leading to
\begin{equation}
\label{opth2}
|D|^4 =  8 ( |D|^2 - \Re D^2). 
\end{equation}
With the help of the last equations~(\ref{centerform}) can be simplified to
\begin{equation}
\label{centerf}
K (\tau) = 1 -\frac{|D|^2}{4} \tau +\frac{1}{8} |D|^4 \tau^2 + \left(\frac{1}{2} |D|^4 - \frac{1}{24} |D|^6 \right) \tau^3 + O(\tau^4).  
\end{equation}
The term linear in time results from the combination of (forward) diffracting orbits and periodic orbits and is related, by the optical theorem, to the total
cross section of the scatterer and thus its sign is always negative. 
The resulting form factor is $1$ at $\tau=0$ and decreases with $\tau$
due to the scatterer. At larger $\tau$ the form factor starts to rise, and will
get back to $1$ for large $\tau$ since the spectrum is discrete~\cite{berry85}.
Qualitatively, this type of behavior should be observed for integrable
systems with point perturbations, since the forward diffracting orbits are
always on a family of periodic orbits. From the terms to
order $\tau^2$ one may get the wrong impression that the form factor
depends on $D$ and $\tau$ only via the combination $|D|^2 \tau$
and in particular its value at the minimum is independent of $|D|$.
This is {\em not} correct as one finds from the term of order $\tau^3$.
In particular the expansion exhibits dependence on $|D|^2$ of the minimum
that is found reasonable compared to the numerical calculations presented
in Sec.~\ref{numerics}. For this reason the calculation was terminated
at the order $\tau^3$.  
We now turn to the case where the scatterer is located at some typical (irrational) position. Only slight modifications of the calculation are needed, and
these are related to the lifting of length degeneracies.

\subsection{Scatterer at a typical position}

\begin{figure}[p]
\begin{flushright}
\vspace{.5cm}
 \leavevmode
\epsfig{file=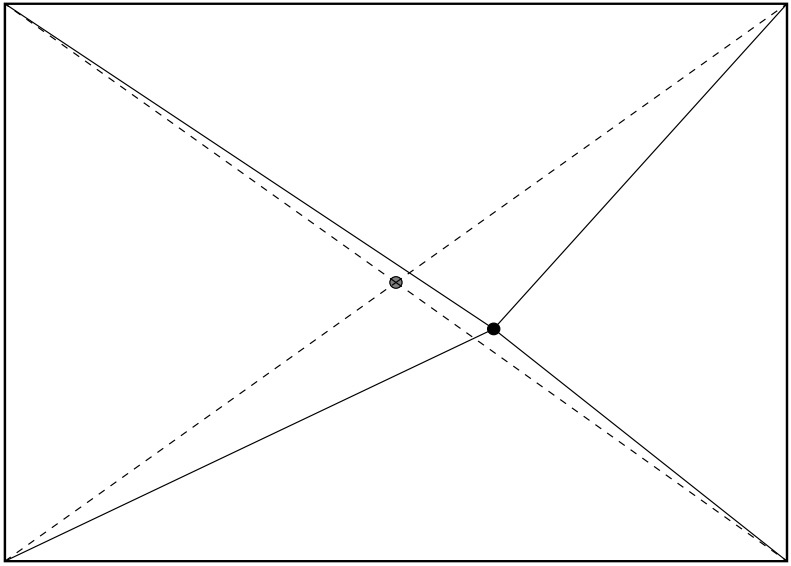,height=7.5cm,width=10.5cm,angle=0}
\centering{\caption{All $(1,1)$ diffracting orbits for various scatterer positions \label{degen}}}
\end{flushright}
\vspace{0cm}
\end{figure}

\begin{figure}[p]
\begin{flushright}
\vspace{.5cm}
 \leavevmode
\epsfig{file=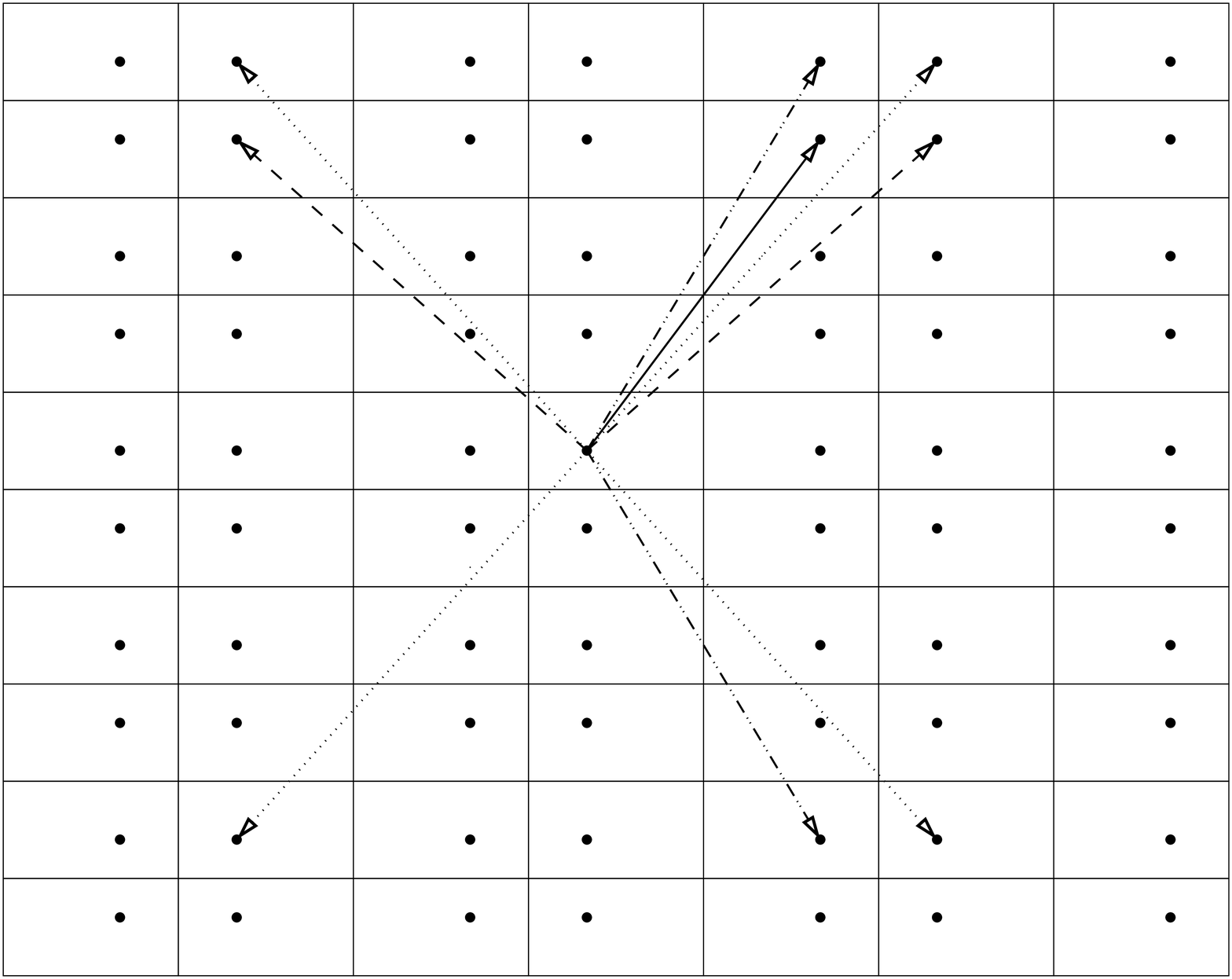,height=8.5cm,width=12.5cm,angle=0}
\vspace{.5cm}
\centering{\caption{ Once diffracting orbits for a typical scatterer location. The solid line represents the ($1,3$) odd-odd orbit, the dashed lines represent two of the even-odd ($2,3$) orbits, the dashed-dotted lines represent the odd-even ($1,4$) orbits and the dotted lines represents the even-even ($2,4$) orbits.  \label{moreorbits}}}
\end{flushright}
\vspace{0cm}
\end{figure}

When the scatterer is in the center of the rectangle the lengths of all
the diffracting orbits with the same indices ${\bf N}_1$ were the same.
If the scatterer is not in the center this is no longer true. An example
is given in Fig.~\ref{degen}.
The four dashed lines in Fig.~\ref{degen} represent the four $(1,1)$ once 
diffracting orbits which go from the scatterer to a corner and then back, they all have equal lengths. When the scatterer is not in the center the orbits with the same index $(N_j,M_j)$ (solid lines) have different lengths. This situation is typical for odd-odd
orbits (orbits where both $N_j$ and $M_j$ are odd). If the scatterer is shifted by $(a_x \delta_x,a_y \delta_y)$ from
the center the lengths of the odd-odd $(N_j,M_j)$ orbits take
all four values $l_j^{(o,o)}=\sqrt{a_x^2 (N_j  \pm 2 \delta_x)^2 + a_y^2 (M_j  \pm 2\delta_y)^2}$. It will be assumed in what follows that $\delta_x$ and $\delta_y$
are typical irrational numbers, therefore the various lengths are not related
in any simple way. For even-odd orbits the lengths take the two 
values $l_j^{(e,o)} =\sqrt{a_x^2 N_j^2 +a_y^2 (M_j  \pm 2 \delta_y)^2}$ while
for odd-even orbits the lengths take the two values
$l_j^{(o,e)}=\sqrt{a_x^2 (N_j \pm 2 \delta_x)^2+a_y^2 M_j^2}$. 
From~(\ref{inangle}) one sees that these are related by time reversal 
symmetry. For even-even orbits all lengths take the same value
$l_j^{(e,e)}=\sqrt{N_j^2a_x^2+N_j^2a_y^2}$ that is identical to the 
value found for the periodic orbit $(N_p,M_p)=(N_j/2,M_j/2)$.
Examples of such (unfolded) orbits are presented in Fig.~\ref{moreorbits}.
The four orbits in Fig.~\ref{moreorbits} that are in the first quadrant have all $\mu=1$, this
notation is introduced since orbits in different quadrants generally have different lengths. Alternatively one can denote orbits using negative values of their index. The value of $\mu$ (see~(\ref{eqa18}) and~(\ref{finaldos})) thus determines the signs (of $\delta_x$ and $\delta_y$) in the formulas for the lengths
of the orbits. The dashed line orbit in the second quadrant in Fig.~\ref{moreorbits}
is also a ($2,3$) orbits but with $\mu=2$. Its length is equal to the length
of the ($2,3$) orbit with $\mu=1$. This results from the fact these orbits
are related by time reversal. The other ($2,3$) orbits ($\mu=3,4$) which are not shown
have the same length which differs from that of the $\mu=1,2$ orbits. This situation is typical to all even-odd and odd-even orbits (the latter have $\mu=1,4$ and are marked by dotted dashed lines in Fig.~\ref{moreorbits}). The lengths
of odd-odd orbits are different for different $\mu$ (these orbits are their own time reversals). The lengths of even-even orbits, marked by the dotted lines in Fig.~\ref{moreorbits}, do not depend on $\mu$ and are the same as the lengths of periodic orbits.

For this reason the contribution of the even-even diffracting orbits
with a single diffraction combined with one of the periodic orbits
is identical to the one obtained when the scatterer is in the center
and reduces to $K^{(1,e)} (\tau)$ of~(\ref{ktau1}).
Also the other contributions to the form factor are simply related
to the ones found if the scatterer is at the center.
For odd-even and even-odd $(N_j,M_j)$ the lengths are divided into
two degenerate lengths and therefore their amplitude is $A_j^{(1)}/2$
(the sums over $\mu$ in~(\ref{finaldos}) consists of two pairs
of identical terms)
and their density is $2 \rho (l_j)$ (see~(\ref{a1dif}) and~(\ref{density})).
The reason is that the number of orbits of each length is reduced by a factor
of $2$ and the number of values of lengths in some interval increases
by a factor of $2$. Since the form factor is proportional to $|A_j|^2$
and to $\rho(l_j)$ the resulting contributions to the form factor
are $\frac{1}{2}K^{(1,o-e)}(\tau)=\frac{1}{2} K^{(1,e-o)} (\tau)=\frac{|D|^2 \tau^2}{8}$ (see~(\ref{formparts})).
For similar considerations (all terms in the sum
over $\mu$ in~(\ref{finaldos}) are different) for the odd-odd orbits 
the amplitude is $\frac{1}{4} A_j^{(1)}$
and the density is $4 \rho(l_j)$ leading to the contribution
$\frac{1}{4} K^{(1,o-o)} (\tau) = \frac{1}{16} |D|^2 \tau^2$ to the form factor.
The contribution to the form factor corresponding to~(\ref{oncedifandpr}) is
\begin{equation}
K^{(1,o)} (\tau) = \frac{5}{16} |D|^2 \tau^2.
\end{equation}

Therefore if the scatterer is located at a typical point the total
contribution to the form factor, resulting from periodic orbits and
once diffracting orbits corresponding to~(\ref{k1tau}) is
\begin{equation}
\label{1diffnocen}
K^{(1)} (\tau) = 1 +\Im D \tau + \frac{9}{16} |D|^2 \tau^2.
\end{equation}

We turn now to look for the saddle manifold when the scatterer is at a 
typical position $(a_x \delta_x,a_y \delta_y)$. 
In order to find a saddle manifold one expands the length of the diffracting
segments as was done in~(\ref{expandlength}). Since the shifts are of order
of unity they are combined with $\delta N$ or $\delta M$ by defining
\begin{eqnarray}
\delta \tilde{N_j} & = \delta N_j \pm \delta_x \nonumber \\
\delta \tilde{M_j} & = \delta M_j \pm \delta_y.
\end{eqnarray}
The appropriate sign is the sign that appears in the length that is expanded.
Using these definitions the rest of the calculation is identical to the
one performed for the scatterer at the center
except that the expansion is done with respect to
$\delta \tilde{N},\delta \tilde{M}$. The condition~(\ref{cond1})
is replaced by
\begin{eqnarray}
\label{nonccond1}
\delta \tilde{N_{j_1}} & = -\delta \tilde{N_{j_2}} \nonumber \\
\delta \tilde{M_{j_1}} & = -\delta \tilde{M_{j_2}}.
\end{eqnarray}
Since $(\delta_x,\delta_y)$ are irrational this condition can be
satisfied only if the signs of the shifts of both orbits are
opposite and also if the condition~(\ref{cond1}) is satisfied.
Note that the conditions for two diffracting orbits to be on the saddle
manifold force these orbits to have the same parity. The contribution of
each type of orbits to the form factor is calculated in what follows.
For orbits of the odd-odd type both the shifts appear in the length, thus
if the orbit $j_1$ has $\mu=1$ (the shifts have negative signs) then to satisfy~(\ref{nonccond1}) the
$j_2$ orbit has to have $\nu=3$. Similarly, if one orbit has $\mu=2$ then the
other one has to be of the $\nu=4$ type. Therefore, from the $16$ combinations
that appear in~(\ref{finaldos}) only four will satisfy~(\ref{nonccond1}).
The density of these contribution is $\frac{1}{4 {\cal A}}$
since the odd-odd orbits are a quarter of all the orbits. The calculation
of the contribution is exactly as was done in~(\ref{newcalc1}-\ref{newcalc3}),
however the two factors of $\frac{1}{4}$ result in 
\begin{equation}
K^{'(2,o-o)} (\tau) = -\frac{1}{16} \tau^2 \Re D^2.
\end{equation} 
The contribution from the odd-even and even-odd orbits is similar, the only
difference results from the fact that one of the shifts does not appear
in the length of the orbit. This results in a degeneracy of lengths for 
orbits with different $\mu$. Thus if one orbit has $\mu=1$ then not only
the orbits with $\nu=3$ will contribute but also the orbit with $\nu=2$ (or
$\nu=4$) marked by the dashed line (dashed-dotted line) in Fig.~\ref{moreorbits} which has identical length. Therefore, there are $8$ combinations
of $\mu,\nu$ that will contribute to the saddle manifold leading to
\begin{equation}
K^{'(2,e-o)} (\tau) = K^{'(2,o-e)} = -\frac{1}{8} \tau^2 \Re D^2.
\end{equation}
The contribution of the even-even orbits is
\begin{equation}
K^{'(2,e-e)} (\tau) = -\frac{1}{4} \tau^2 \Re D^2.
\end{equation} 
which is exactly as it was for a scatterer at the center. This is a result
of the fact that the length of these orbits does not change when the
scatterer position is changed.
The sum of the non diagonal contributions of order $\tau^2$ is
\begin{equation}
\label{ndk2tau}
K^{'(2)} (\tau) = \frac{9}{16} \Re (-D^2) \tau^2.
\end{equation}

The computation of the $\tau^3$ terms is fairly similar to the case in
which the scatterer is at the center of the rectangle.
Therefore, only the relative factors resulting from the lifting of degeneracies
of lengths will be calculated in what follows.
The twice diffracting orbits are composed of two segments
${j_1}$, ${j_2}$. In the diagonal approximation each segment is summed 
independently. For the diagonal terms resulting from once diffracting orbits
it was shown that a relative factor of $\frac{9}{16}$ is obtained in
~(\ref{1diffnocen}) compared to (\ref{k1tau}).
Since the sums over ${j_1}$ and ${j_2}$ are independent a relative factor
of $\left( \frac{9}{16} \right)^2=\frac{81}{256}$ is obtained.
The diagonal contribution of twice diffracting orbits for a scatterer at some typical location, corresponding to~(\ref{k3tau}), is therefore,
\begin{equation}
\label{nck3tau}
K^{(3)} (\tau) = \frac{81}{512} |D|^4 \tau^3.
\end{equation}

We turn to compute the relative factor that results from relocating the scatterer from the center 
to a typical location for non diagonal terms of once and twice diffracting
orbits. For this contribution, the saddle 
condition~(\ref{nonccond1}) is replaced by
\begin{eqnarray}
\label{cond2-1}
\delta \tilde{N}_{j_1} & = & \delta \tilde{N}_{j_2}+\delta \tilde{N}_{j_3} \nonumber \\
\delta \tilde{M}_{j_1} & = & \delta \tilde{M}_{j_2}+\delta \tilde{M}_{j_3}.
\end{eqnarray} 
To understand the effect of the location of the scatterer the non integer
part of~(\ref{cond2-1}) is studied. This non integer part depends on the 
displacements
of the scatterer from the center $\delta_x$, $\delta_y$.
It is convenient to discuss different parities of ${j_1}$ separately,
giving a factor of $\frac{1}{4}$ to each case (since each case contributes a $\frac{1}{4}$ of the total contribution when the scatterer is at the center).
If the orbit ${j_1}$ is even-even then the parity of
$j_2$ should be equal to the one of ${j_3}$. This is exactly identical
to the calculation of the 
non diagonal contribution of order $\tau^2$ which resulted
in a relative factor of $\frac{9}{16}$ in~(\ref{ndk2tau}) compared to~(\ref{form2diff}). Therefore, the contribution
when ${j_1}$ is even-even, is $\frac{9}{64}$ of the total term
for the scatterer at the center, given by~(\ref{k4tau}). 

If the orbit ${j_1}$ is of the even-odd parity then the combinations of
${j_2}$ and ${j_3}$ that satisfy~(\ref{cond2-1}) are:
(a) ${j_2}$ even-even; ${j_3}$ even-odd and (b)  ${j_2}$ odd-odd; ${j_3}$ odd-even and also two combinations where the roles of ${j_2}$ and ${j_3}$
are interchanged.
In case (a) there
are $8$ combinations of $\mu_{j_2}$, $\mu_{j_3}$ that will satisfy~(\ref{cond2-1})
for each $\mu_{j_1}$ (compared to $16$ when the scatterer is at the center).
Since this case (with ${j_1}$  even-odd) accounts for a $\frac{1}{4}$ of all contributions  it results in a relative factor of $\frac{1}{4} \frac{8}{16} = \frac{1}{8}$.
In case (b) there are
$4$ combinations of $\mu_{j_2}$ and $\mu_{j_3}$ that satisfy~(\ref{cond2-1})
for each~$\mu_{j_1}$. This results in a relative factor of $\frac{1}{16}$.
The relative contribution of terms where ${j_1}$ is even-odd is thus
given by $\frac{1}{4} \left( 2\frac{1}{8}+2\frac{1}{16} \right)=\frac{3}{32}$,
where the factors of $2$ result from the cases where the parities of
${j_2}$ and ${j_3}$ are interchanged.
The case where ${j_1}$ is odd-even gives an identical contribution
to the case in which ${j_1}$ is even-odd.

When ${j_1}$ is odd-odd~(\ref{cond2-1}) is solved if (c) ${j_2}$ is
even-even and ${j_3}$ is odd-odd or if (d) ${j_2}$ is even-odd and
${j_3}$ is odd-even. There are also two additional solutions that result
from interchanging the parities of ${j_2}$ and ${j_3}$.
One can verify that for all the parities of ${j_2}$ there are $4$
such solutions for each value of $\mu_{j_1}$ compared to $16$ solutions
when the scatterer is at the center. This results in a relative
factor of $\frac{1}{4} \frac{1}{4}=\frac{1}{16}$ compared to~(\ref{k4tau}).

Combining all the contributions from different parities of ${j_1}$ one
finds that for a scatterer at a typical location the contribution of non diagonal
terms from once and twice diffracting orbits is $\frac{9}{64}+2\frac{3}{32}+\frac{1}{16}=\frac{25}{64}$
relative to the case where the scatterer is at the center. Therefore using~(\ref{k4tau}) one finds
\begin{equation}
\label{nck4tau}
K^{'(3a)} (\tau) = \frac{25}{32} |D|^2 \Im D \tau^3.
\end{equation}

The contributions from non diagonal terms of periodic orbits and orbits
with three diffractions should satisfy the condition
\begin{eqnarray}
\sum_{i=1}^3 \delta \tilde{N}_{j_i} & = & 0 \nonumber \\
\sum_{i=1}^3 \delta \tilde{M}_{j_i} & = & 0
\end{eqnarray}
corresponding to~(\ref{nonccond1}).
The non integer part of this condition behaves exactly in the same
way as the non integer part of~(\ref{cond2-1}). This is true since
both signs of the displacements always appear. This results in the fact
that the factors relative to the case where the scatterer is at the
center are equal for both of these terms. Therefore using~(\ref{k5tau}) one
finds
\begin{equation}
\label{nck5tau}
K^{'(3b)} (\tau) = -\frac{25}{96} \Im D^3 \tau^3.
\end{equation}

Adding all the contributions~(\ref{1diffnocen}), (\ref{ndk2tau}), (\ref{nck3tau}), (\ref{nck4tau}) and (\ref{nck5tau}) leads to
the form factor for a typical position of the scatterer 
\begin{equation}
\label{noncenterform}
K(\tau)= 1+\Im D \tau +\frac{9}{16} |D|^2 \tau^2 + \frac{9}{16}\Re(-D^2) \tau^2 + \frac{81}{512} |D|^4 \tau^3 + \frac{25}{32} |D|^2 \Im D \tau^3 - \frac{25}{96} \Im D^3 \tau^3+ O(\tau^4).  
\end{equation}
The optical theorem~(\ref{opth}) and~(\ref{opth2}) can be used to obtain 
\begin{equation}
\label{typacalf}
K(\tau)= 1-\frac{|D|^2}{4} \tau + \frac{9}{128} |D|^4 \tau^2 + \frac{81}{512} |D|^4 \tau^3-\frac{25}{1536} |D|^6 \tau^3+O(\tau^4).  
\end{equation}
This form factor has the same qualitative features of the form factor~(\ref{centerf}), it is $1$ for $\tau=0$ and decreases linearly for small $\tau$.
See the comment following~(\ref{centerf}).

\section{A Model of an Integrable Billiard with a Point Scatterer}
\label{point}

An easily solvable model is that of a point interaction~\cite{pointbook,jackiw}.
This interaction is the self-adjoint extension of a Hamiltonian where
a point is removed from the domain. It can be viewed as formally
represented by a $\delta$-function potential.
The influence of such an interaction on the spectral statistics was
first investigated by \v{S}eba~\cite{seba90}. The system that was investigated
was the rectangular billiard with a point interaction, and the spectral statistics were found to differ from Poisson statistics when the point scatterer is added.
The spectral statistics of this system and of similar systems were the 
subject of many works~\cite{bogomolny01b,berkolaiko01,BG,seba91,albeverio91,shigehara94,weaver95,cheon96,shigehara97,legrand97}. Since the energy levels of this system are obtained from the zeroes of a function (as will be explained in the following) it is possible to compute a large number of levels and thus also the form factor with
considerable accuracy. Therefore, we use the rectangular billiard with a point
scatterer to obtain numerically the form factor and compare it to the predictions of the analytical theory, presented in the previous section.

Point interactions are useful since the Green's function of the problem
$G_{\xi}(z; {\bf x},{\bf x'})$ can be expressed by the Green's functions of the problem in absence of the
scatterer $G(z; {\bf x},{\bf x'})$~\cite{zorbas80}
\begin{equation}
\label{greenpoint}
G_{\xi} (z; {\bf x},{\bf x'})= G(z; {\bf x},{\bf x'}) + T(z,\xi)   G(z; {\bf x},{\bf x}_0) G(z; {\bf x}_0,{\bf x'})
\end{equation}
where $\xi$ is the self adjoint parameter that is related to the interaction strength, ${\bf x}_0$ is the scatterer location and $\Im \sqrt{z} > 0$.
$T$ is the on shell transition matrix and it is given by
\begin{equation}
\label{zorbast}
 T(z,\xi) = \left(e^{i \xi}\!-\!1\right) \! \left[(\Lambda i \! - \! z) \int \! dy G(z; {\bf y},{\bf x}_0)G(\Lambda i; {\bf y},{\bf x}_0) + e^{i\xi}(\Lambda i \! + \! z)\int \! dy G(z; {\bf y},{\bf x}_0)G(-\Lambda i; {\bf y}, { \bf x}_0) \right]^{-1}.
\end{equation}
$\Lambda$ is an energy scale usually set to be unity, since there is only one parameter in the problem which is the interaction strength.
The point scatterer is characterized by the self-adjoint parameter $\xi$,
while $T$ is independent of angles, resulting in s-wave scattering.
The energy levels
of the perturbed system are the poles of the resolvent, and if they do not coincide with the eigenvalues of the unperturbed system they are also the poles
of the transition matrix $T$. The unperturbed Green's function of the rectangular
billiard
is given by
\begin{equation}
\label{recgreen}
G (z; {\bf x},{\bf x'}) = \sum_{n=1}^{\infty} \frac{\psi_n({\bf x})\psi_n({\bf x'})}{z-E_n}
\end{equation}
where $\psi_n$ and $E_n$ are the eigenfunctions and eigenvalues of the billiard. 
Substituting ~(\ref{recgreen}) into~(\ref{zorbast}) leads to the
equation for the energy levels of the perturbed problem
\begin{equation}
\label{numeq}
 \left( \frac{\sin \xi}{1-\cos \xi} \right) \sum_n \psi_n^2 ({\bf x}_0) \frac{\Lambda}{\Lambda^2 + E_n^2}  - \sum_n \psi_n^2 ({\bf x}_0) \left[ \frac{1}{z-E_n} + \frac{E_n}{E_n^2+\Lambda^2}\right] = 0.
\end{equation}
In the geometrical theory of diffraction used in the previous section
the effect of the scatterer that is located in the billiard was expressed
by the diffraction constant $D$ that determines the scattering in
free space. The diffraction constant in free space is the $T$ matrix~(\ref{zorbast})
if $G (z; {\bf x},{\bf x'})$ is the Green's function in free space.
It was calculated in App.~\ref{pointd} leading to
\begin{equation}
\label{dfree}
D(z)=\frac{2 \pi}{-\frac{1}{2} \ln \left( \frac{z}{\Lambda}\right) + \frac{\pi}{4}\frac{\sin \xi_1}{1-\cos \xi_1} + i \frac{\pi}{2}}.
\end{equation}
The self adjoint extension parameter in free space is $\xi_1$ and it need not to be the same 
as $\xi$ that was used
for the bound problem~(\ref{numeq}). The reason is that the self adjoint extensions
depend not only on the scatterer but also on the boundary conditions.
The general relation between extensions of different problems is unexplored to the
best of our knowledge, however, a connection can be made in the case that
is considered here.

The energy scale associated with the scatterer is $\Lambda$. For $\Lambda\gg \Delta$
the spectrum is effectively continuous, therefore the $T$ matrix of the bound
problem is approximately the diffraction constant of the unbound problem,
and also the self adjoint extension parameters $\xi$ and $\xi_1$ should be
approximately equal. Another way to establish this correspondence
is to compare length scales. The length scale associated with the scatterer
is $1/\sqrt{\Lambda}$ and the condition $\Lambda \gg \Delta$ is
that $1/\sqrt{\Lambda} \ll \sqrt{\cal A}$ where ${\cal A}$ is the
area of the billiard. The condition is that the length associated
with the scatterer is much smaller then the dimensions of the
billiard. This is also the condition for the validity of the
semiclassical geometrical theory of diffraction used in the
previous section. The scattering is strongest ($|D|^2$ maximal) when
$D=-4i$. The variation of $D(z)$ with $z$ is weak and therefore its
value at the maximum is expected to be a good approximation
for a whole energy region around the maximum. 
Substitution of $D=-4 i$ in~(\ref{centerf}) and~(\ref{typacalf}) yields:
\begin{equation}
\label{maxdcen}
K(\tau) = 1 - 4 \tau + 32 \tau^2 -\frac{128}{3} \tau^3 + O(\tau^4)
\end{equation}
when the scatterer is at the center of the rectangle and
\begin{equation}
\label{maxdnonc}
K(\tau) = 1 - 4 \tau + 18 \tau^2 -\frac{157}{6} \tau^3+ O(\tau^4)
\end{equation}
when the scatterer is at some typical location. 

When the scatterer is at the center of the rectangle the system has a symmetry. The (unperturbed) eigenfunctions that vanish on the scatterer remain eigenfunctions of the perturbed problem, and so are their eigenvalues. To obtain
the other eigenvalues one has to use only the non-vanishing eigenfunctions in~(\ref{numeq}).
These are the eigenfunctions that are even functions with respect to reflection with
respect to both the $X$ and $Y$ axes that pass through the center of the rectangle. Since the value of these eigenfunctions for all eigenvalues is the same
($\psi_n ({\bf x}=0)=\frac{2}{\sqrt{\cal A}}$)
 at the center of the
rectangle, the resulting equation~(\ref{numeq}) is the same as for the \v{S}eba billiard
with periodic boundary conditions~\cite{bogomolny01b}.
The spectral statistics of the \v{S}eba billiard
with periodic boundary conditions were also shown to be the same as the spectral statistics of some star graphs~\cite{berkolaiko01,berkolaiko99}. 
However,~(\ref{maxdcen}) describes the form factor of the full spectrum.
Therefore, in addition to the eigenvalues that are perturbed due to the scatterer, all of the eigenvalues that are unaffected are included in the calculation of the form factor.

The form factor of the perturbed levels is related to the form factor of the
full spectrum. It is composed of the even-even eigenfunctions of the
rectangle. The eigenvalues of the four different symmetry classes of the
rectangle can be assumed to be uncorrelated, therefore their combined form factor
is the sum of the form factors (weighted using their density) of the different symmetry classes, with $\tau$ rescaled as required by~(\ref{formfactor}) and~(\ref{correlation}). Since the unperturbed levels have
Poisson level statistics and since levels belonging to the four
symmetry classes have the same mean density (in the semiclassical limit)
the relation between the full form factor and the form factor obtained from only
perturbed levels is given by
\begin{equation}
\label{scale}
K_{full} (\tau) = \frac{3}{4} + \frac{1}{4} K_{per} (4 \tau).
\end{equation}
The perturbed form factor is exactly the form factor that was computed
for the \v{S}eba billiard with periodic boundary conditions and for star
graphs. The first powers in $\tau$ of the form factor for the star
graphs are given by
\begin{equation}
K_{per}(\tau) = 1 -4 \tau + 8\tau^2 -\frac{8}{3} \tau^3 +O(\tau^4),
\end{equation}
as can be seen, for example, from equation (14) of~\cite{berkolaiko01} where the form factor is given up to
order of $\tau^8$. Using~(\ref{scale}) gives exactly the form factor~(\ref{maxdcen}).

\section{Numerical Results}
\label{numerics}

The spectral form factor is not easy to compute numerically. The form factor is
a very oscillatory function of time and much averaging is needed in order
to obtain its mean behavior. This requires computation of many levels, which
is not always possible numerically. However,
such calculations are relatively easy for the point scatterer defined in
the previous section by equations~(\ref{greenpoint}) and~(\ref{zorbast}).
The eigenvalues can be calculated from~(\ref{numeq}).
This equation in particularly convenient since $\sum_n \frac{1}{z-E_n}$ is monotonically decreasing when $z$ is varied between any adjacent unperturbed levels, ensuring that every level must be in an interval between two adjacent unperturbed levels $E_n$ and $E_{n+1}$ (unless the unperturbed wavefunctions vanish at
the scatterer). 
To compute the energy levels numerically equation~(\ref{numeq}) is solved, for example by truncating
 the sums at some large $n$ and replacing the tails by integrals.

This equation is solved in some energy window. The energy scale parameter
is chosen so that $\Lambda \gg \Delta$, and therefore $D$ and $\xi$ can be 
approximated by their values for a scatterer in free space, as discussed
in the previous section. In calculation presented here $\xi$ was chosen so 
that $D=-4i$, resulting in maximal scattering, for $z$ in the center
of the window. The other values of $D$ that were also examined, are $D=2-2i$ and $D=\frac{4}{5} (2-i)$. Since $D$ does not change much with $z$ the results can be
compared with the analytical predictions obtained within
the semiclassical geometrical theory of diffraction.

\begin{figure}[p]
\begin{flushright}
\vspace{.5cm}
 \leavevmode
\epsfig{file=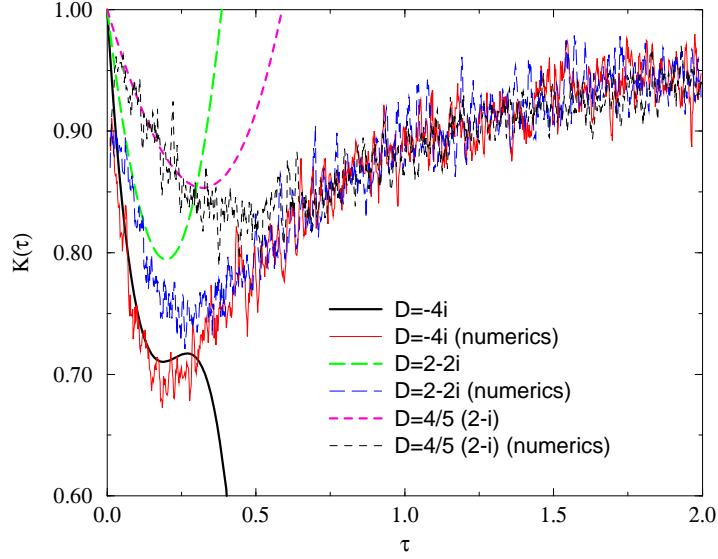,height=8.5cm,width=10.5cm,angle=0}
\centering{\caption{The form factor of a scatterer at a typical position, for some diffraction constants, thin lines, compared to the analytical result~({\protect{\ref{typacalf}}}), heavy lines.\label{noncenter}}}
\end{flushright}
\vspace{0cm}
\end{figure}

For a scatterer at a typical location the form factor is presented
in Fig.~\ref{noncenter}.
The form factor is computed from energy levels between $20000$ and $40000$. The mean level spacing is $\Delta=1$ while $\Lambda=1000$. To smooth the form factor an
ensemble average over $20$ systems was performed.  The aspect-ratio of each systems was taken as a random number between $1/2$ and $1$.
Additional smoothing was performed by averaging the results found for $4$
consecutive values of $\tau$.
 The position of the scatterer was also chosen at random, but it was the same position for all ensembles 
of aspect ratios. Similar results were found for several other scatterer positions.
The agreement between the numerical results and the theoretical prediction~(\ref{typacalf})
is good at short times, where the expansion truncated at a low
order is a good approximation. 

\begin{figure}[p]
\begin{flushright}
\vspace{.5cm}
 \leavevmode
\epsfig{file=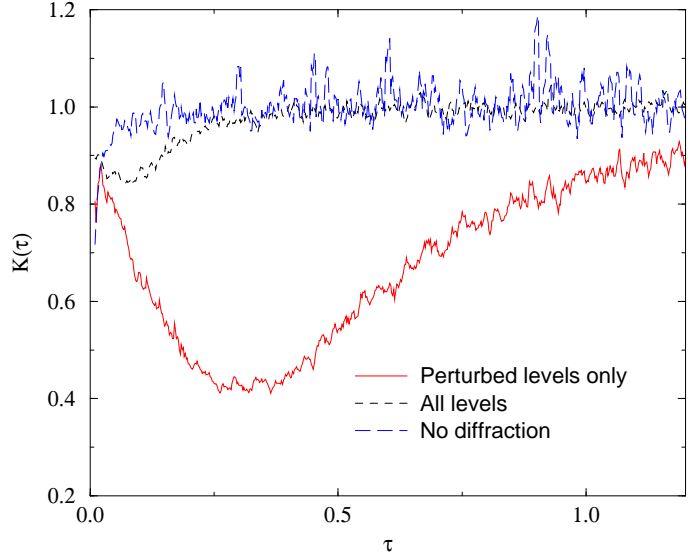,height=8.5cm,width=10.5cm,angle=0}
\centering{\caption{The form factor of a scatterer at the center (${\bf x}_0={\bf 0}$), for only perturbed levels as well as the full form factor\label{cent}}}
\end{flushright}
\vspace{0cm}
\end{figure}

\begin{figure}[p]
\begin{flushright}
\vspace{.5cm}
 \leavevmode
\epsfig{file=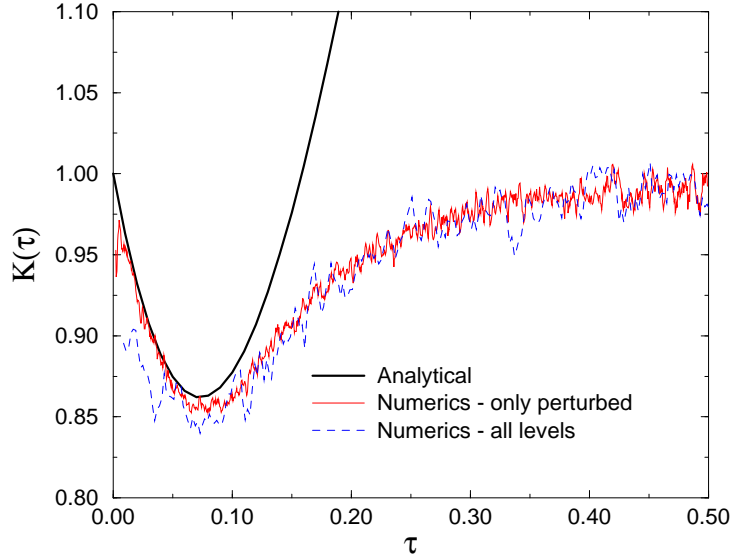,height=8.5cm,width=10.5cm,angle=0}
\centering{\caption{The form factor for the scatterer at the center (${\bf x}_0={\bf 0}$) compared to the scaled form factor calculated from perturbed levels by~({\protect{\ref{scale}}}) and the analytical result~({\protect{\ref{maxdcen}}})\label{centerth}}}
\end{flushright}
\vspace{0cm}
\end{figure}

For the scatterer at the center (${\bf x}_0={\bf 0}$) only levels with
eigenfunctions that are symmetric with respect to the $X$ and $Y$ axes
are perturbed by the scatterer as is clear from~(\ref{numeq}).
Based on the assumption that the spectra of the various symmetry
classes of levels are uncorrelated~(\ref{scale}) was derived.
To test it the form factor of the full spectrum and of the
levels which are affected by the scatterer were
calculated. These are shown in Fig.~\ref{cent}.
The calculation is done for parameters similar to the ones used in the 
case where the scatterer is not at the center, and an averaging over
$70$ systems is performed here, while the energy range considered
was $60000-80000$.
The form factor of the rectangular billiard was included for comparison, 
demonstrating
that at short times all form factors deviate from $1$. This deviation is due
to the fact that one examines times of the order of the period of the
shortest orbits and there are not enough orbits that contribute
to the form factor for such times. In order to demonstrate it,
the form factor of the rectangle was calculated from the exact energy
levels in the same way as the form factors with diffraction.
The result is shown in Fig.~\ref{cent} and this form factor  
is also found to deviate
from $1$ at short times.
The form factor calculated from all levels and the scaled form factor obtained from the perturbed levels with the help of~(\ref{scale}) are
compared to the analytical result~(\ref{maxdcen}) in Fig.~\ref{centerth}.
It is clear that the symmetry argument~(\ref{scale}) is valid. The scaled form factor is indeed the same as the one that is obtained from the full spectrum. Both slightly deviate from the analytical prediction~(\ref{maxdcen}). The deviation
at short times of the full form factor is due to 
the fact that there are not enough relevant
short orbits as was
discussed earlier, and at these times the scaled form factor is 
probably more accurate.  Since the periodic
orbits prediction is in agreement with the form factor studied in~\cite{bogomolny01b,berkolaiko01,berkolaiko99} (that is exact) the deviation probably
results from the numerical solution, or of the fact that the energy is not high enough.

The agreement between the analytical results obtained in Sec.~\ref{noangle}
and the numerical results of the model of the point scatterer outlined
in Sec.~\ref{point} was found for $\Lambda \gg \Delta$, where for the point
scatterer the values of $\xi$ and $D$ of a scatterer in free space were used. The
requirement $\Lambda \gg \Delta$ is essential for this agreement.
For $\Lambda \approx \Delta$ if one uses $D=-4i$ and the corresponding
values of the self adjoint extension parameter $\xi$, found for the
scatterer in free space, the effect of the point scatterer on the spectrum
of the bounded system is even not maximal (as is the case when $\Lambda \gg \Delta$) and clearly there is no agreement
with predictions of Sec.~\ref{noangle} for these values of $\Lambda$ and $D$.

\section{Scattering From a Localized Potential}
\label{scatter}

So far it was assumed that the diffraction constant is given. In this section the diffraction constant for
a given potential, which is localized around some point, is calculated.
Consider a potential which is assumed to have a length scale $a$ such that
\begin{equation}
\label{potscale}
U({\bf r}) = \frac{1}{a^2} f \left( \frac{{\bf r}}{a} \right)
\end{equation}
where $f({\bf y})$ is small for ${\bf y}$ that is large compared to unity.
The scattering from such a potential at energies for which the wave number
$k$ satisfies, $ka \ll 1$,
is studied in what follows.

The diffraction constant is the on shell matrix element of the ${\bf T}$
matrix
\begin{equation}
\label{defd}
D( \theta',\theta) = \langle {\bf k} | {\bf T}(E) | {\bf q} \rangle 
\end{equation}
where ${\bf k}$ is the outgoing momentum (in direction $\theta'$) and
${\bf q}$ is the incoming momentum (in the direction $\theta$). 
The energies of the incoming and outgoing waves are equal, that is
\begin{equation}
k=q=\sqrt{E}
\end{equation}
(in units $\hbar=1$ and $m=\frac{1}{2}$ used in the paper).
In order to compute the diffraction constant it is convenient to start from the
Lippman-Schwinger equation for the ${\bf T}$ matrix
\begin{equation}
\label{LSeq}
{\bf T} (E) = g {\bf U} + g {\bf U G} (E) {\bf T} (E)
\end{equation}
where ${\bf G}(E) $ is the Green's function for free
propagation at energy $E$ and 
$g$ is an expansion parameter which will be set to unity at the end of the
calculation.
One might be tempted to compute the on shell matrix elements of ${\bf T}$
using the Born approximation which is obtained by iterating equation~(\ref{LSeq})
\begin{equation}
\label{born}
\langle {\bf k} | {\bf T}(E) | {\bf q} \rangle = g \langle {\bf k} | {\bf U} | {\bf q} \rangle + g^2  \langle {\bf k} | {\bf UGU} | {\bf q} \rangle + g^3  \langle {\bf k} | {\bf UGUGU} | {\bf q}\rangle + \cdots
\end{equation}
However, in the limit $ka \rightarrow 0$ the Green's function diverges as $\ln ka$ 
and therefore the Born series diverges.

A method which behaves regularly at low energies ($ka \ll 1$) 
was developed by Noyce~\cite{noyce65}. In this method the scattering
in the forward direction (another chosen direction can also be used)
is in some sense resummed. In two dimensions this method was used to describe low energy bound
states~\cite{patil80} and low energy scattering~\cite{averbuch86}.
The Noyce method is derived in the following.
The ${\bf T}$
matrix is written as a product of the diagonal matrix element times a function of the 
incoming and outgoing momenta,
\begin{equation}
\label{dunno}
\langle {\bf k} | {\bf T} | {\bf q} \rangle = h({\bf k}, {\bf q}) \langle {\bf q} | {\bf T} | {\bf q} \rangle.
\end{equation}
Taking the diagonal matrix elements of the Lippmann-Schwinger equation~(\ref{LSeq}) leads to
\begin{equation}
\label{diagpart}
\label{noyce1}
\langle{\bf q}|{\bf T}|{\bf q}\rangle \left( 1 - g \frac{1}{(2 \pi)^2} \int d^2 k_1 \frac{\langle{\bf q}|{\bf U}|{\bf k}_1\rangle h({\bf k}_1,{\bf q})}{E+ i \epsilon - k_1^2}\right)= g \langle {\bf q}| {\bf U}|{ \bf q}\rangle,
\end{equation}
where the Green's function in the momentum representation is
\begin{equation}
G ( E; {\bf k}_1, {\bf k}_2) = \frac{(2 \pi)^2 \delta \left({\bf k}_1-{\bf k}_2 \right)}{E-k_1^2+i \epsilon}.
\end{equation}
Note that ${\bf k}_1$ is not on the energy shell.

Using (\ref{diagpart}) in the non diagonal matrix elements of (\ref{LSeq}) leads to an integral equation for
$h$:
\begin{eqnarray}
\label{noyce2}
h({\bf k}_1,{\bf q}) & = & \frac{\langle {\bf k}_1 | {\bf U}|{ \bf q}\rangle}{\langle {\bf q}| {\bf U}|{ \bf q}\rangle} + g \frac{1}{(2 \pi)^2} \int d^2 k_2 \left(\langle {\bf k}_1 | {\bf U}|{ \bf k}_2 \rangle -
\frac{\langle {\bf k}_1 | {\bf U}|{ \bf q}\rangle \langle {\bf q}| {\bf U}|{ \bf k}_2 \rangle}{\langle {\bf q}| {\bf U}|{ \bf q}\rangle}\right) \nonumber \\ 
& & \times \frac{1}{E + i \epsilon -k_2^2} h({\bf k}_2 ,{\bf q}).
\end{eqnarray}
This equation can be iterated resulting in an expansion of $h$ in powers of $g$. Therefore with the help of (\ref{dunno}) and (\ref{noyce1}) the
${\bf T} $ matrix can be written in the form
\begin{equation}
\label{tnoyce}
\langle {\bf k} | {\bf T} | {\bf q} \rangle=\frac{g h({\bf k}, {\bf q})\langle {\bf q}| {\bf U}|{ \bf q}\rangle }{D_g({\bf q})}.
\end{equation}
Both $h({\bf k},{\bf q})$ and the denominator $D_g ({\bf q})$ can be expanded in powers of $g$ as
\begin{eqnarray}
h({\bf k}, {\bf q})& = & \frac{\langle {\bf k}| {\bf U}|{ \bf q}\rangle}{\langle {\bf q}| {\bf U}|{ \bf q}\rangle}+ g \left( \frac{\langle {\bf k}| {\bf UGU}|{ \bf q}\rangle}{\langle {\bf q}| {\bf U}|{ \bf q}\rangle} - \frac{\langle {\bf k}| {\bf U}|{ \bf q}\rangle\langle {\bf q}| {\bf UGU}|{ \bf q}\rangle }{\langle {\bf q}| {\bf U}|{ \bf q}\rangle^2}\right) \nonumber \\ & & + g^2 \left( 
\frac{\langle{\bf k}| {\bf UGUGU}|{ \bf q}\rangle}{\langle {\bf q}| {\bf U}|{ \bf q}\rangle} -
\frac{\langle {\bf k}| {\bf UGU}|{ \bf q}\rangle \langle {\bf q}| {\bf UGU}|{ \bf q}\rangle}{\langle {\bf q}| {\bf U}|{ \bf q}\rangle^2} \right. \nonumber \\
& & \left. - \frac{\langle {\bf k}| {\bf U}|{ \bf q}\rangle \langle{\bf q}| {\bf UGUGU}|{ \bf q}\rangle}{\langle {\bf q}| {\bf U}|{ \bf q}\rangle^2} +\frac{\langle {\bf k}| {\bf U}|{ \bf q}\rangle \langle {\bf q}| {\bf UGU}|{ \bf q}\rangle^2}{\langle {\bf q}| {\bf U}|{ \bf q}\rangle^3}\right) + \cdots
\end{eqnarray}
and
\begin{eqnarray}
\label{dg}
D_g ({\bf q}) & = & 1 - g \frac{\langle {\bf q}| {\bf UGU}|{ \bf q}\rangle}{\langle {\bf q}| {\bf U}|{ \bf q}\rangle} - g^2 \left( \frac{\langle {\bf q}| {\bf UGUGU}|{ \bf q}\rangle}{\langle {\bf q}| {\bf U}|{ \bf q}\rangle}-\frac{\langle {\bf q}| {\bf UGU}|{ \bf q}\rangle^2}{\langle {\bf q}| {\bf U}|{ \bf q}\rangle^2} \right) \nonumber \\ & &
- g^3 \left( \frac{\langle {\bf q}| {\bf UGUGUGU}|{ \bf q}\rangle}{\langle {\bf q}| {\bf U}|{ \bf q}\rangle} - 2 \frac{\langle {\bf q}| {\bf UGU}|{ \bf q}\rangle \langle {\bf q}| {\bf UGUGU}|{ \bf q}\rangle}{\langle {\bf q}| {\bf U}|{ \bf q}\rangle^2} \right. \nonumber \\ & & \left. + \frac{\langle {\bf q}| {\bf UGU}|{ \bf q}\rangle^3}{\langle {\bf q}| {\bf U}|{ \bf q}\rangle^3} \right) + \cdots
\end{eqnarray}
While the expansions for $h({\bf k},{\bf q})$ and $D_g ({\bf q})$ seem to be complicated they are actually 
composed of all the ways to break the matrix elements, containing chains of the form ${\bf UGUG \cdots UGU}$ with several Green's
functions, to products of smaller elements with less Green functions.
Each order of $g$ consists of terms with the same number of Green's functions in it, and when an element is broken into a product an extra factor
of $\frac{-1}{\langle {\bf q}| {\bf U}|{ \bf q} \rangle} $ is added. 
In the expansion for $D_g({\bf q})$ all factors contain at least one Green's function
but in the expansion for $h({\bf k},{\bf q})$ factors of the type $\langle {\bf k}| {\bf U}|{ \bf q} \rangle$ 
appear. Consequently the $g^N$ term for $h({\bf k},{\bf q})$ can be explicitly written.  It
includes $2^N$ terms (and $2^{N-1}$ terms for $D_g({\bf q})$, some of which may be identical).
Note that $\frac{1}{D_g({\bf q})}$ can be expanded in powers of $g$ leading back
to the Born series~(\ref{born}).

To compute the diffraction constant the matrix elements are computed 
and the length scale $a$ of the potential is scaled out. For example
\begin{eqnarray}
\langle{\bf k}| {\bf UGU} | {\bf q}\rangle & = & \int d^2 r_1 d^2 r_2 e^{-i {\bf k \cdot r}_1} U({\bf r}_1 ) G_0 ({\bf r}_1,{\bf r}_2) U({\bf r}_2) e^{i {\bf q \cdot r}_2} \nonumber \\
&  = & \int d^2 y_1 d^2 y_2 e^{-i a{\bf k \cdot y}_1} f({\bf y}_1 ) G_0 (ka; {\bf y}_1,{\bf y}_2) f({\bf y}_2) e^{i a {\bf q \cdot y}_2}
\end{eqnarray}
where ${\bf y}_i = \frac{ {\bf r}_i }{a}$ and~(\ref{potscale}) was used
and $G_0$ is the free Green's function in the position representation,
namely $G_0 ({\bf r}_1,{\bf r}_2)\equiv \langle{\bf r}_1 |{\bf G}(E)|{\bf r}_2 \rangle\equiv G_0 (ka ; {\bf y}_1 ,{\bf y}_2 )$, where $k=\sqrt{E}$.
These matrix elements are expanded for $ka \ll 1$ keeping terms of order $k^2 a^2$. Then the series in the parameter $g$ is summed. 

Both the exponentials  $\langle {\bf k}| {\bf r}\rangle=e^{-i {\bf k \cdot r}}$
and the Green's function should be expanded to the same order.
In the limit of small $ka$ one finds
\begin{equation}
\label{greenexp}
 G_0 (ka ; {\bf y}_1 ,{\bf y}_2 ) =- \frac{i}{4} H_0^{(1)} (ka | {\bf y}_  - {\bf y}_2 |) \simeq d_1 + \phi ( {\bf y}_1 ,{\bf y}_2 ) - \frac{k^2a^2}{4}| {\bf y}_1 - {\bf y}_2  |^2 \left( d_1 - \frac{1}{2\pi}   + \phi ( {\bf y}_1 ,{\bf y}_2 ) \right) 
\end{equation}
where 
\begin{equation}
\label{defd1}
d_1=\frac{1}{2\pi} \left( \gamma + \ln \left( \frac{ka}{2} \right) \right)-\frac{i}{4} 
\end{equation}
and 
\begin{equation}
\phi ( {\bf y}_1 ,{\bf y}_2 )=\frac{1}{2\pi} \ln | {\bf y}_1 - {\bf y}_2 |
\end{equation}
 while $\gamma$ is Euler's constant. The leading order part of the Green's function which is
coordinate independent is $d_1$ and, as was mentioned earlier, it diverges logarithmically when $ka \rightarrow 0$. In App.~\ref{definitions} it is shown that in many of
the terms to be computed this constant part will be canceled. This is the advantage of the Noyce method~\cite{noyce65} compared to the Born expansion for the present problem.

Calculations presented in App.~\ref{definitions} lead to the asymptotic expression
for the angular dependence of the diffraction constant
\begin{eqnarray}
\label{desired}
D(\theta ', \theta) & \simeq & \frac{g V_0}{V(g) - g V_0 d_1 + k^2 a^2 Q(g)} \left[ 1 - k^2 a^2 M_0 \right. +  i \frac{ka}{2} V(g) \left\{ \left( e^{i \theta} - e^{i \theta '} \right) \frac{\alpha (g)}{V_0} + c.c \right\} \nonumber \\
& - & \frac{k^2 a^2}{8} \left\{  V(g) \left( e^{2 i \theta} + e^{2 i \theta'}\right)\frac{\sigma(g)}{ V_0} + c.c. \right\} +\frac{k^2a^2}{4} \left[ M_{++}e^{i(\theta + \theta')}+ M_{+-}e^{i(-\theta + \theta')} \right. \nonumber \\
& +& \left. \left. M_{-+}e^{i(\theta - \theta')}+ M_{--}e^{-i(\theta + \theta')}\right] \right]
\end{eqnarray}
where $c.c.$ stands for complex conjugate,
\begin{equation}
\label{defqg}
Q(g) \equiv \frac{1}{4} V^2 (g) \sum_{k=1}^{\infty} g^k \frac{{{\cal L}_k}}{V_0}+\frac{1}{2} \left(V(g) -\frac{g V_0}{2 \pi} \right)V(g)\left( \frac{\delta (g)}{V_0} - V(g) \frac{|\alpha (g)|^2}{V_0^2}\right),
\end{equation}
\begin{equation}
M_0=\frac{V^2 (g)}{2} \frac{|\alpha (g)|^2}{V_0^2}
\end{equation}
and
\begin{eqnarray}
M_{++} &= & \left( V(g) - g d_1 V_0 \right) \frac{\mu^{(++)}(g)}{V_0} + V(g) g d_1 V_0 \frac{\alpha^2 (g)}{V_0^2} \\
M_{+-}& = & \left( V(g) - g d_1 V_0 \right) \frac{\mu^{(+-)}(g)}{V_0} + V(g) g d_1 V_0 \frac{|\alpha (g)|^2}{V_0^2}\\
M_{-+}& = & \left( V(g) - g d_1 V_0 \right) \frac{\mu^{(-+)}(g)}{V_0} + V(g) g d_1 V_0 \frac{|\alpha (g)|^2}{V_0^2} \\
M_{--}& = & \left( V(g) - g d_1 V_0 \right) \frac{\mu^{(--)}(g)}{V_0} + V(g) g d_1 V_0 \frac{(\alpha^* (g))^2}{V_0^2}.
\end{eqnarray}
The constants $V_0,V (g),\mu^{(a,b)} (g),\alpha (g),\sigma (g), \delta (g)$, and ${\cal L}_k$ are angle and energy independent and are given by a series of integrals over the potential. They are defined in~(\ref{defv0}),(\ref{defvg}), and (\ref{lkdef})-(\ref{sumsdef}).

Equation~(\ref{desired}) is the main result of this section. It provides
the functional form of the diffraction constant for $ka \ll 1$. It also gives a recipe for the computation of the coefficients of the
angle dependent factors as sums of certain integrals involving the potential~(\ref{potscale}). The physical scattering constant is obtained when $g=1$ is substituted in~(\ref{desired}). It is of interest to note that
the simple dependence on the angles $\theta$ and $\theta '$ is expected since
formally it should be the same dependence as in the Born series~(\ref{born})
which is also simple. This diffraction constant will be used in Sec.~\ref{angles} for the calculation of the effects of angular dependence on the form factor.

Finally, one can verify that the
diffraction constant~(\ref{desired}) satisfies the optical theorem
\begin{equation}
\Im D(\theta,\theta) = -\frac{1}{8 \pi} \int_0^{2 \pi} d \theta' \; \left| D (\theta' , \theta) \right|^2 
\end{equation}
up to order $k^2 a^2$.

\section{The effect of Angular Dependence on the Form Factor}
\label{angles}

In this section the form factor resulting of angle dependent diffraction
is calculated to order $\tau^2$. The complicated parts of the calculation
are not related to the dependence on the angles but rather to the contribution
from non diagonal terms that were calculated in Sec.~\ref{noangle},
therefore this calculation will not be repeated here, and only the modifications
needed to describe angle dependent diffraction constants will be presented.

Since the diffraction constant depends on angles, the density
of diffracting orbits of length $l$ and outgoing direction $\theta$
is needed (the incoming direction depends on the outgoing direction, via~(\ref{inangle})). For simplicity consider first the case where the
 scatterer is at the center.
Instead of orbits that leave the scatterer, hit the walls, and return to the scatterer it is possible to examine orbits which connect the scatterer
to one of its images, when the plane is tiled with rectangles, as described in detail in App.~\ref{diffcont}.
 These orbits
have the same length and the same outgoing directions as the diffracting orbits. To compute the (averaged) density of such orbits the number of orbits
with lengths in the range $(l-\frac{\delta l}{2} ,l+\frac{\delta l}{2}) $ and
outgoing directions in the range $(\theta-\frac{\delta \theta}{2},\theta+\frac{\delta \theta}{2} )$ is estimated. The area of this domain is $l \delta \theta \delta l$. When computing the form factor, in the semiclassical limit, the lengths of orbits that contribute scales as $k$. There is a natural width $\delta l$ which
results from the averaging in~(\ref{correlation}) and is taken such that
$1 \ll \delta l \ll l$. Therefore, for any small but finite $\delta \theta$
the area $l \delta \theta \delta l $ is large in the semiclassical limit.
The number of orbits (of a each type) in such area is given by $\frac{l}{4{\cal A}} \delta \theta \delta l$. If the diffraction constant depends slowly on the angles, that is,
its value changes considerably only when the angles change by $\delta \theta \sim 1$ (namely $O(k^0)$) then the sums over diffracting orbits can be replaced by integrals
with the density
\begin{equation}
\label{goint}
\tilde{\rho} (\l,\theta) = \frac{l}{4{\cal A}}.
\end{equation}
On this scale the angles of diffracting orbits are distributed uniformly.
Note that if the diffraction constant is angle independent one can integrate
(\ref{goint}) over $\theta$ in the interval $(0,\frac{\pi}{2})$ and obtain~(\ref{density}).
 As a result
the form factor for an angle dependent diffraction constant can be
computed if instead of the constant $D$ of Sec.~\ref{noangle}
one uses $D(\theta',\theta)$ of~(\ref{desired}) with $g=1$ with appropriate 
averages on angles. These averages depend on the type of the
diffracting orbits that contribute and are computed in the following.

\subsection{Scatterer at the center}

To compute the contributions to the form factor each type of orbits should
be considered separately. This results from the fact that the incoming
direction is related to the outgoing direction
in a different way characterized by the parity and by the
indices $\mu$ and $\nu$ of App.~\ref{diffcont}, where it is 
explicitly given by~(\ref{inangle}). First the diagonal contribution
from diffracting orbits is computed and later the non diagonal contributions
from twice diffracting orbits and periodic orbits will be computed.

First consider the contribution of the even-even orbits following the lines
of the calculation leading to (\ref{ktau1}) of Sec.~\ref{noangle}. This is
the contribution of
forward diffracting orbits. For these orbits the incoming
direction is the same as the outgoing direction. The index $\mu$ of~(\ref{finaldos}) specifies the quadrant of the angle. In Sec.~\ref{noangle} the 
factor resulting of summation over $\mu$ was included in the amplitude
but for the angle dependent $D(\theta',\theta)$ the averaging
over angles is performed with the help of the density~(\ref{goint}).
The summation
over $\mu$ results in extension of the integration interval from $(0,\frac{\pi}{2})$ to $(0,2\pi)$. When the scatterer is at the center orbits with
different $\mu$ (but with the same index ${\bf N}_1,{\bf N}_2$) have identical lengths.
Since
\begin{equation}
\label{forwardD}
\Im \frac{2}{\pi} \int_0^{\frac{\pi}{2}} \! d \theta \; \frac{1}{4}\left[ {D(\theta,\theta)+ D(\pi \! - \! \theta,\pi \! - \! \theta)+D(\pi \!+\! \theta,\pi\! + \! \theta)+D(2\pi\! -\! \theta,2\pi\! -\! \theta)}\right] = \frac{1}{2 \pi} \Im \int_0^{2 \pi} d \theta \; D(\theta,\theta)
\end{equation}
and
\begin{eqnarray}
\frac{2}{\pi} \int_0^{\frac{\pi}{2}} d \theta \; \frac{1}{16}\left| {D(\theta,\theta)+ D(\pi - \theta,\pi-\theta)+D(\pi+\theta,\pi+\theta)+D(2\pi-\theta,2\pi-\theta)}\right|^2 \nonumber \\ = \frac{1}{8 \pi} \int_0^{2 \pi} d \theta \; D^{*} (\theta,\theta) \left[ D(\theta,\theta)+ D(\pi-\theta,\pi-\theta)+ D(\pi+\theta,\pi+\theta)+ D(2\pi-\theta,2\pi-\theta) \right]
\end{eqnarray}
the contribution of even-even orbits, that are the forward diffracting orbits, to the form factor is given by
\begin{eqnarray}
\label{k1first}
K^{(1,e)} (\tau)& = & 1 + \frac{\tau}{2 \pi} \Im \int_0^{2 \pi} d \theta \; D(\theta,\theta)  \\ & + & \frac{\tau^2}{32 \pi} \int_0^{2 \pi} d \theta \; D^{*} (\theta,\theta) \left[ D(\theta,\theta)+ D(\pi-\theta,\pi-\theta)+ D(\pi+\theta,\pi+\theta)+ D(2\pi-\theta,2\pi-\theta) \right] \nonumber .
\end{eqnarray}
This contribution corresponds to~(\ref{ktau1}) of Sec.~\ref{noangle}.
The diagonal contributions from the other types of diffracting orbits are
computed in a similar manner 
and presented in App.~\ref{appda}.
Substitution of the diffraction constant~(\ref{desired}) in the diagonal contribution~(\ref{centdiagang}) leads to
\begin{equation}
\label{centshort}
K^{(1)} (\tau) = 1 - \frac{1}{4} |C|^2 \tau + |C|^2 C' \tau^2
\end{equation}
where
\begin{equation}
\label{defc}
C \equiv \frac{V_0}{V(1)-V_0 d_1 + k^2 a^2 Q(1)}
\end{equation}
and 
\begin{equation}
\label{defc'}
C' \equiv 1 - k^2 a^2 V^2(1) \left| \frac{\alpha(1)}{V_0} \right|^2.
\end{equation}
See~(\ref{defd1}), (\ref{defqg}), (\ref{defv0}), (\ref{partitution}), and (\ref{sumsdef})  for the definitions of the various quantities used
in~(\ref{defc}) and~(\ref{defc'}).

An additional non diagonal
contribution results from orbits which have almost identical lengths. At order $\tau^2$ such a contribution results from combinations of twice diffracting orbits and periodic 
orbits. For angle independent scattering this contribution was computed in Sec.~\ref{noangle} (see~(\ref{orbitcomb}-\ref{form2diff}) there). The modifications which result from the angle dependence are computed in App.~\ref{appdb}
leading to (\ref{centnondangle}) that yields
\begin{equation}
\label{nondshort}
K^{'(2)} (\tau) = - (\Re C^2) C' \tau^2
\end{equation}
where $C$ and $C'$ are defined by~(\ref{defc}) and~(\ref{defc'}).
To obtain the form factor of an angle dependent scatterer located at the center
of a rectangular billiard, to order $\tau^2$, equations (\ref{centshort}) and (\ref{nondshort}) are summed resulting in
\begin{equation}
\label{totalshort}
K(\tau) = 1 -\frac{1}{4} |C|^2 \tau + \frac{1}{8} |C|^4 C' \tau^2.
\end{equation}
This form factor resembles the form factor~(\ref{centerf}) that was obtained for angle independent scattering. It starts at $1$ and decreases for small times until at some point it has a minimum. At larger times it rises and approaches $1$ for long times.
The angle dependence just modifies the constants in front the powers of $\tau$, but these changes are of the order $k^2a^2$ and typically cannot change the sign of these coefficients.

\subsection{Scatterer at a typical position}

When the scatterer is not at the center the lengths of the diffracting segments
of the same parity vary and some length degeneracies are lifted, as discussed at the
end of Sec.~\ref{noangle}. The form factor for an angle dependent scatterer
in some typical position is calculated
along the lines of the calculation of Sec.~\ref{noangle}. Only the modifications
due to the angle dependence are presented. First the diagonal contribution
from once diffracting orbits is computed and then the non diagonal contributions
from combinations of twice diffracting orbits and periodic orbits are computed.

To compute the diagonal contribution it is convenient to separate the orbits 
into types determined by the number of bounces from the walls (or the parity of
their index ${\bf N}_j$). For orbits of the even-even type the lengths of
all $4$ possible segments (with different $\mu$) are identical and do not change
when the scatterer changes position. Therefore their diagonal contribution 
is identical to the contribution that was obtained for a scatterer at the 
center and is given by~(\ref{k1first}). The other contributions are calculated
in App.~\ref{appdc} leading to
\begin{equation}
\label{ncdiagshort}
K^{(1)} (\tau) = 1 - \frac{1}{4} |C|^2 + \frac{3}{16} |C|^2 (\Re C_1') \tau^2
\end{equation}
where
\begin{equation}
\label{defc1}
C_1' \equiv 3 + k^2 a^2 \left[ V(1) - V_0 d_1 \right] \left[ \frac{\Re \mu^{(+-)}(1)}{V_0} - \frac{V(1)|\alpha(1)|^2}{V^2_0} \right]
\end{equation}
and the various quantities are defined by~(\ref{defd1}), (\ref{defc}), (\ref{defv0}), (\ref{partitution}), and (\ref{sumsdef}).
This is not the total contribution to the form factor (in order $\tau^2$).
There is a non diagonal contribution from twice diffracting orbits
and periodic orbits which is computed in App.~\ref{appdd} and 
is described in what follows.

The non diagonal contributions are from twice diffracting orbits and periodic
orbits with almost identical lengths. The condition for such combinations
to contribute is given by~(\ref{nonccond1}). This condition ensures that
the two segments of the diffracting orbit are of the same type, as it was
for the scatterer at the center.
The fact that the lengths of the segments might depend on the index $\mu$
was discussed following the condition~(\ref{nonccond1}). This discussion is 
also valid for angle depended scattering. As for the case where the scatterer
is at the center the integral over angles can be factored out of the saddle
manifold integral.

For even-even orbits the lengths do not depend upon $\mu$ and the non
diagonal contribution from these orbits and periodic orbits is 
given by~(\ref{nondee}).
The other contributions are calculated in App.~\ref{appdd} resulting in
(\ref{nonck2tauang}) leading to
\begin{equation}
\label{ncndshort}
K^{'(2)} (\tau) = - \frac{3}{16} \Re ( C^2 C_1') \tau^2,
\end{equation}
where $C_1'$ is given by~(\ref{defc1}). Finally summing~(\ref{ncdiagshort}) and~(\ref{ncndshort}) results in an expression for the form factor,
correct to order $\tau^2$, for the case where the scatterer is at a typical position,
\begin{equation}
\label{typicalendf}
K (\tau) = 1 - \frac{1}{4} |C|^2 \tau + \frac{9}{128} |C|^4 \tau^2.
\end{equation}
This form factor is rather similar to the one obtained for angle independent
scattering~(\ref{typacalf}). The angle dependence modifies the coefficients only slightly. The coefficient $C$ satisfies the optical theorem~(\ref{opth}),
therefore this form factor is identical to one obtained for an angle
independent potential, with diffraction constant $C$.

\section{Summary and Discussion}
\label{discussion}

In this paper the form factor $K(\tau)$ was calculated for a rectangular
billiard 
perturbed by a strong potential, that is confined to a region  which is much
smaller
than the wavelength. In the first part (Secs.~\ref{noangle},~\ref{point} and~\ref{numerics}) a highly
idealized situation
of a scatterer that is confined to a point and its action is represented by the
diffraction constant $D$,
that is a complex number like $D=-4i$ or $D=2-2i$, was explored. The form factor
$K(\tau)$
was calculated to $O(\tau^3)$ and the results are given by~(\ref{centerf}), if the
scatterer is at the
center, and by~(\ref{typacalf}) if it is at a typical position. These results 
were compared
to the ones
obtained numerically for various values of $D$ in Figs. \ref{noncenter}
and \ref{centerth}. Reasonable agreement was found in the small
$\tau$ regime, up to the minimum. The numerical calculations were performed for
the
\v{S}eba billiard, that is a rectangular billiard with a $\delta$-like
scatterer inside.
Its $T$ matrix is (\ref{zorbast}) and the energy levels, used in the numerical calculation
of the
form factor, were obtained from the numerical solution of (\ref{numeq}). This is an
idealized model
approximating a
rectangular billiard with a strong potential that is confined to a very small
region. The
model is defined by the self-adjoint parameter $\xi$ and the energy scale is
set by
$\Lambda$. The connection with the analytical results of Sec.~\ref{noangle} is via the
diffraction
constant $D$ of (\ref{dfree}), that is defined in free space. If
$\Lambda \gg \Delta$, where $\Delta$ is the mean level spacing (or
$1/\sqrt{\Lambda}$ is much smaller then the dimensions of the billiard), then
the
self-adjoint parameter $\xi_1$ of (\ref{dfree}) is approximately equal to $\xi$, that is
the
corresponding parameter in the presence of the boundary.
This condition was found to be essential.
 The variation of $D$
with energy
is slow. For these reasons the numerical results of Figs. \ref{noncenter} and \ref{centerth} are expected
to agree
with the
analytical predictions of Sec. \ref{noangle}, for small $\tau$, and indeed such agreement
is found.

In the second part of the paper (Secs. \ref{scatter} and \ref{angles}) a problem with an arbitrary
perturbing potential $U({\bf r})$  of (\ref{potscale}), that is large in a small region of
extension
$a$, was
studied. For rectangles with such potentials the diffraction constant was
calculated to
the order
$(ka)^2$, where $k=2\pi/\lambda=\sqrt{E}$ is the wavenumber. The result is
given by
(\ref{desired}) with $g=1$. The angle dependent terms are of order $ka$ and $(ka)^2$. In
the limit
$(ka)^2 \ll 1$ or $a \ll \lambda$ the diffraction constant $D(\theta',\theta)$
is
approximately independent of angles and reduces to
\begin{equation}
\label{defc0}
C_0 = \frac{1}{\frac{V(1)}{V_0}-\frac{1}{2\pi} \left( \gamma + \ln \left(
\frac{ka}{2}
\right) \right)+\frac{i}{4} },
\end{equation}
where $V_0$ and $V(g)$ are given by~(\ref{defv0}) and (\ref{defvg}) while $\gamma$ is Euler's constant.
One can easily see that $C_0$ satisfies the optical theorem. We do not know
much about the
convergence of the series (\ref{defvg}) and (\ref{partitution}) for $V(g)$. Therefore $V(1)$ is
assumed to be the
analytic continuation from the small $g$ region. Exploration of these series
for various
potentials is left for further research. If the potential is such that in the
$a \rightarrow 0$ limit $C_0$ is well defined, then in this limit it approaches
the
diffraction constant $D$ that was studied in the first part of the paper,
taking the
constant value $D$ in Sec. \ref{noangle} and (\ref{dfree}) in Secs. \ref{point} and \ref{numerics}. For this to hold it
is
required
that in the limit  $a \rightarrow 0$ also
\begin{equation}
\label{limit}
\frac{V(1)}{V_0}-\frac{1}{2\pi} \left( \gamma + \ln \left( \frac{ka}{2}
\right) \right) \longrightarrow -\frac{1}{4\pi} \ln \left(
\frac{k^2}{\Lambda}\right)
+\frac{1}{8}\frac{\sin \xi_1}{1-\cos \xi_1}.
\end{equation}
If the function $f({\bf y})$ of~(\ref{potscale}) is independent of $a$,
then also $V(1)/V_0$ is independent of $a$, as is clear from~(\ref{defv0}) and (\ref{defvg}).
The calculations of Sec.~\ref{scatter} and App.~\ref{definitions}, leading to~(\ref{desired}),
do {\em not} require that $f$ is independent of $a$.
Assume for example that $f$ factors in the form
\begin{equation}
\label{newscale}
f (a,{\bf y}) = \bar{f} \left( \frac{a}{l_{\Lambda}} \right) f_1 ({\bf y})
\end{equation}
where $f_1 ({\bf y})$ is independent of $a$, while $\bar{f}$ is independent
of ${\bf y}$ and $l_{\Lambda}$ is some characteristic length scale.
If $\lim_{a \rightarrow 0} \bar{f} (a/l_{\Lambda}) = 0$
then $\lim_{a \rightarrow 0} V(1) =1$.
The factorization~(\ref{newscale}) implies
\begin{equation}
V_0 = \bar{f} \left( \frac{a}{l_{\Lambda}} \right) \int d^2 y  f_1 ({\bf y}).
\end{equation}
Existence of the limit~(\ref{limit}) requires that
\begin{equation}
V_0 \sim 2\pi \left[ \ln \left( \frac{a}{l_{\Lambda}}\right)+\bar{f}_0 \right]^{-1}
\end{equation}
where $\bar{f}_0$ is a constant independent of $a$,
in agreement with the scaling used in the mathematical
literature (\cite{pointbook} p. 103). It leads
to the natural identification $l_{\Lambda}^2 = \frac{1}{\Lambda}$.
Therefore $l_{\Lambda}$ is the length scale associated with the energy
scale $\Lambda$ that was discussed after~(\ref{dfree}).
The limit is characterized by the two parameters $\xi$ and $\Lambda$.
If the factorization~(\ref{newscale}) exists, then
 for $a \ll \lambda$ the results obtained in the first part of the
paper
provide a good approximation for the ones obtained in the second one, for
arbitrary but
well defined potentials $U$ of (\ref{potscale}). For any given potential $U({\bf r})$ that
is
concentrated in a small region, of extension $a$, one can use (\ref{desired}) to compute
the form
factor. The condition for the applicability of the semiclassical approximation
combined
with the geometrical theory of diffraction (GTD) is
\begin{equation}
\label{apcond}
a \ll \lambda \ll a_x,~a_y,
\end{equation}
where $a_x$ and $a_y$ are the sides of the rectangle. 

If the scatterer is at a
typical
position then the form factor is given by (\ref{typicalendf}). Note that $C$ of Eq.~(\ref{typicalendf}) (defined by~(\ref{defc})) satisfies the optical theorem, since $Q(1)$
is real. Therefore the form factor of the angle dependent scatterer reduces
(to the order $\tau^2$) to~(\ref{typacalf}). That is, the angle dependent
scattering is affecting the form factor in the same way (to the order computed here) as angle independent scattering. 
Therefore the angle dependence plays no role up to this order.
For the scatterer at the center the situation is somewhat different. In this
case, the
form factor (\ref{totalshort}) should be compared with (\ref{centerf}). There is a correction $C'$
resulting of
the angular dependence of $D(\theta',\theta)$ given by (\ref{desired}). This difference
is a
consequence of the increased number of length degeneracies of the diffracting
orbits when
the scatterer is at the center compared to the situation when it is located at
a typical
position.

For the case where the scatterer is at a typical position, using angle
dependent terms to
order
$(ka)^2$ does not change the result for the form factor, compared to the one found
for the
angle independent
leading order.
Therefore
it is reasonable to assume that the results are robust and the limit $a
\rightarrow 0$
describes correctly the physics of the regime $a < \lambda$. This is so
although the
classical dynamics (in the long time limit) are expected to be chaotic in
nearly all of
phase space and similar to the ones of the Sinai billiard. This improves the
chances for the experimental realization of the results of the present work.
Note
that for $ a \gg \lambda$, semiclassical theory works and the system should
behave as a
Sinai billiard, with  level statistics given by Random Matrix Theory (RMT), in
some range
\cite{bohigas84,berry85}.

The spectral statistics found in the present work differ from the ones of the
known
universality classes. It is characterized by the form factor of the type
presented in
Figs. \ref{noncenter}
and \ref{centerth}.
A characteristic feature of the form factor is that it is equal to 1 at
$\tau=0$,
resulting of the fact that for small $\tau$ the number of classical orbits that
are
scattered is small. The contribution that is first order in $\tau$ originates
from the
term (\ref{tildea}), that leads to the contribution $\Im D \tau$ in (\ref{k1tau}), (\ref{centerform}) and (\ref{noncenterform}).
By the
optical theorem (\ref{opth}) it is always negative. This is expected to hold for a
larger class of
systems, where there is forward diffraction along periodic orbits. For $\tau
\gg 1$ the
form factor approaches unity because of the discreteness of the spectrum~\cite{berry85}.
These
are the physical reasons for the qualitative form of $K(\tau)$ depicted in Figs
\ref{noncenter} and~\ref{centerth}.

The form factor was computed when the scatterer is at the center and when it is
at a
typical position shifted by $(a_x \delta_x,a_y \delta_y)$ from the center (with
all
numbers
in this expression being irrational). The results are different since for the
scatterer at
the center there is a high degree of the degeneracy of the orbits involved. In
this case
the form factor is related by the symmetry argument (\ref{scale}) to the one found for
periodic
boundary
conditions, that is known exactly. 
The validity of (\ref{scale}) is demonstrated in Fig.~\ref{centerth}.
Off diagonal contributions of orbits
belonging to saddle manifolds turn out to be of great importance (see (\ref{orbitcomb}) and
the
calculations that
follow). In view of the work of Bogomolny~\cite{bogomolny00b} this should be generic for
integrable
systems in presence of localized perturbations.

A natural question that should be explored is whether the problem of the
rectangular
billiard
with a point scatterer, that was studied here, represents a larger universality
class and
whether it can be related to some RMT models.

\begin{acknowledgments}
It is our great pleasure to thank Eugene Bogomolny and Martin Sieber for
inspiring,
stimulating, detailed and informative discussions and for informing us about
their
results prior publication. SF would like to thank Richard Prange for
stimulating and
critical discussions related to this work and for his hospitality at the
University
of Maryland, to thank Michael Aizenman for illuminating
discussions about the mathematical background
and to thank Peter \v{S}eba for discussion about the \v{S}eba billiard. 
We also  thank Eric Akkermans for bringing references~\cite{noyce65,averbuch86} to our attention.
This research was supported in part by the US-Israel Binational
Science
Foundation (BSF), by the US National Science Foundation under Grant No.
PHY99-07949,
by the Minerva Center of Nonlinear Physics of Complex Systems and by the fund
for
Promotion of Research at the Technion.
\end{acknowledgments}

\begin{appendix}
\section{Diffracting Orbits Contributions}
\label{diffcont}

In this Appendix the contributions of diffracting orbits to the
density of states are calculated in the semiclassical approximation.
The corrections to the trace formula
due to diffraction have been extensively investigated in recent years.
Most of the work was done within the Geometrical Theory of Diffraction (GTD)~\cite{keller62}.
This is an approximation in the spirit of the semiclassical approximation.
In this approximation the Green's function is composed of free propagation from the source to
the diffraction point multiplied by an (angle dependent) diffraction constant and followed by
propagation to the end point. Using the GTD approximation the contributions
of diffracting orbits were calculated in a number of papers~\cite{vattay94,pavloff95,whelan96}.
These contributions as well as the Berry-Tabor trace formula for the 
rectangular billiard
are derived in this Appendix, for completeness.

The GTD approximation fails in a number of cases, typically when a diffracting orbit is close to a classically allowed one. In these cases the contributions
of the periodic orbits are given by uniform approximations which are more
complicated, especially in the case of multiple diffraction. 
Such contributions have been examined for the penumbra diffraction~\cite{primack96}, for wedge diffraction~\cite{sieber97}, for the diffraction from a flux line~\cite{sieber99} and for multiple diffraction from wedges or flux lines~\cite{bogomolny00}. 
Since the scattering from a point-like perturbation is accurately described
using the GTD approximation, uniform approximations are not needed in this work. 

A convenient starting point is the Boundary Integral Method (BIM).  
In the case of Dirichlet boundary conditions this is a Fredholm equation of the second kind for the normal derivative
(with respect to the boundary of the billiard) of the wave function. The kernel
is the normal derivative of the Green's function of the problem with some arbitrary boundary conditions.
That is,
\begin{equation}
\label{beq}
u(s)=-2 \oint_{\partial {\cal D}} d {s'} \; \partial_{n} G(k; {\bf r},{\bf r'}) u(s'),
\end{equation}
where ${\bf r}(s)$ is a parameterization of the boundary of the billiard by arc length, $\partial_{n}$ is the derivative in the (outward) normal direction of the boundary, $u(s)$ denotes the normal derivative of the eigenfunctions and
the integral is taken over the boundary of the billiard.  The
Green's function of the system is $ G(k; {\bf r},{\bf r'}) $.
The units $\hbar=1$, $2m=1$ are used. For the derivation
of this equation and some applications see~\cite{berry84,bogomolny92,boasman94,alonso94,li95,pisani96,kosztin97}. This integral equation has nontrivial solutions only if
\begin{equation}
\det ( \hat{I} - \hat{Q} (k_i))=0,
\end{equation}
where $\hat{Q} (k)$ is the integral operator which gives the right hand side
of~(\ref{beq}) when it is applied. The solutions $k_i$, of this equation
are the exact eigenvalues of the problem. The oscillatory part of the density 
of states is thus given by~\cite{alonso94,sieber97}
\begin{equation}
\label{dexp}
 d_{osc}(k)=\frac{1}{\pi} \Im \sum^{\infty}_{n=1} \frac{1}{n} \frac{d}{dk} (-2)^n \oint_{\partial {\cal D}} ds_1 ds_2 \cdots  ds_n \; \partial_{\hat{n}_n} G (k; {\bf r}_1,{\bf r}_n) \partial_{\hat{n}_{n-1}} G (k; {\bf r}_n,{\bf r}_{n-1}) \cdots \partial_{\hat{n}_1 } G (k; {\bf r}_2,{\bf r}_1).
\end{equation}
In the semiclassical limit these integrals can be approximated using the
stationary phase approximation as long as the boundary is smooth. 

The system that is investigated in this work is a rectangular billiard with
a localized perturbation. The Green's function for the localized perturbation without the boundary is approximately given by
\begin{equation}
\label{green}
G(k; {\bf r},{\bf r'}) \simeq G_0 (k; {\bf r},{\bf r'}) + G_0 (k; {\bf r},{\bf r}_0) D (\theta,\theta')  G_0 (k; {\bf r}_0,{\bf r'}) ,
\end{equation}
where $\theta$ and $\theta'$ are the angles of the vectors ${\bf r}-{\bf r}_0$
and ${\bf r}_0-{\bf r'}$ respectively.
This approximation is exactly the GTD approximation and is valid far from the perturbation. The free Green's function in two dimensions is  $ G_0(k; {\bf r},{\bf r'}) =-\frac{i}{4} H_0^{(1)}(k | {\bf r}- {\bf r'} |) $ ( $H_0^{(1)}(x)$ is the Hankel function of the first kind). The Green's function~(\ref{green}) together with the expansion~(\ref{dexp}) lead to the periodic and diffracting orbits
contributions to the density of states. In this Appendix it will be calculated
to the third order in $D$.

\begin{figure}[p]
\begin{flushright}
\vspace{.5cm}
 \leavevmode
\epsfig{file=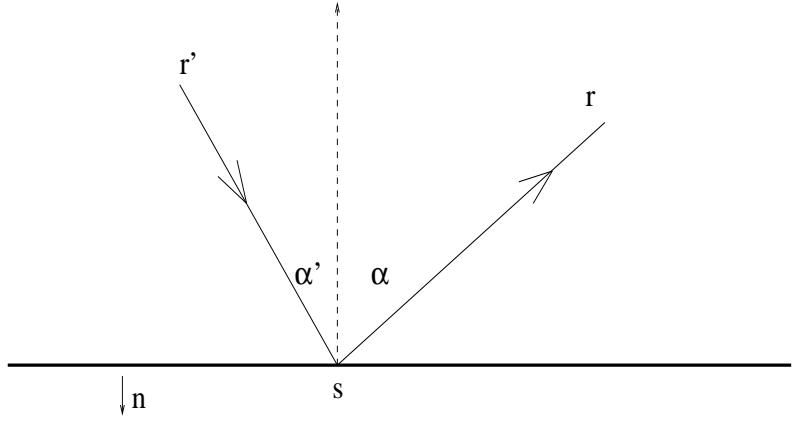,height=5.5cm,width=10.5cm,angle=0}
\centering{\caption{A trajectory from ${\bf r'}=(x',y')$ to ${\bf r}=(x,y)$ with one bounce \label{comp}}}
\end{flushright}
\vspace{0cm}
\end{figure}

The calculation of the boundary integrals is vastly simplified by the composition law. Its semiclassical version is given by
\begin{equation}
\label{complaw}
(-2)^n \oint_{\partial {\cal D}} d s_1 \cdots d s_n \; \partial_{n_n} G_0 (k; {\bf r},{\bf r}_n)\cdots \partial_{n_1} G_0 (k; {\bf r}_2,{\bf r}_1)  G_0 (k; {\bf r}_1, {\bf r'})  \simeq G_{sc}^{(n)} (k; {\bf r}, {\bf r'})
\end{equation}
where the function $G_{sc}^{(n)} (k; {\bf r}, {\bf r'})$ is the part of the semiclassical Green's function which results from the neighborhood of all classical trajectories with $n$ bounces from the walls between ${\bf r}$ and ${\bf r'}$.
Semiclassically the integrals are evaluated using the stationary phase approximation and for this the boundary is assumed to be smooth and all the
Green's function are taken in the semiclassical approximation 
(for a clear discussion see~\cite{sieber97}). Here only the simple
case of one reflection from a straight edge will be treated  to
show how this method works. 
Consider the integral
\begin{equation}
 I(k; {\bf r'},{\bf r})= -2 \int_{-\infty}^{\infty} ds \; G_0 (k; {\bf s},{\bf r'}) \partial_n G_0 (k; {\bf r},{\bf s})
\end{equation} 
where ${\bf s}=(s,0)$ is a parameterization of the boundary. According to the composition law this integral should give the Green's function that describes propagation from ${\bf r'}$ to ${\bf r}$ with one bounce on the boundary. We
will compute this integral using the stationary phase approximation and show that this is indeed the case. The free Green's function is given by the Hankel function of the first kind and therefore $G_0(k; {\bf s},{\bf r'})=-\frac{i}{4} H_0^{(1)} \left( k \sqrt{(x'-s)^2 +{y'}^2} \right) $ and also $\partial_n G_0 (k; {\bf r},{\bf s}) = \frac{i}{4} k \cos (\alpha(s)) H_1^{(1)} \left( k \sqrt{(x-s)^2+y^2}\right)$ where $\alpha(s)$ is the angle between the trajectory from
${\bf s}$ to ${\bf r}$ and the normal to the boundary as is shown in Fig.~\ref{comp},
Thus the integral to be evaluated is given by
\begin{equation}
\label{int}
I(k; {\bf r},{\bf r'})=-\frac{1}{8} \int_{-\infty}^{\infty} ds H_0^{(1)} \left( k l'(s) \right)  H_1^{(1)} \left( k l(s) \right) k \cos (\alpha(s)) 
\end{equation}
where $l(s)=\sqrt{(x-s)^2+y^2}$ and $l'(s)=\sqrt{(x'-s)^2 +{y'}^2}$ are the lengths of the orbits
from ${\bf r}$ and ${\bf r'}$ to the reflection point ${\bf s}$.
Since the arguments of the Hankel functions in~(\ref{int}) are large in the
semiclassical limit the Hankel functions can be replaced by their asymptotic
value for large argument
\begin{equation}
H_{\nu}^{(1)} (z) \sim \sqrt{\frac{2}{\pi z}} e^{i (z-\frac{1}{2} \nu \pi - \frac{1}{4} \pi)}.
\end{equation}
The resulting integral is given by
\begin{equation}
\label{b4st}
I(k; {\bf r},{\bf r'}) \simeq \frac{1}{4 \pi } \int_{-\infty}^{\infty} ds \frac{\cos (\alpha(s))}{\sqrt{l(s) l'(s)}} e^{ik( l(s)+l'(s))}.
\end{equation}

In the semiclassical limit $k \rightarrow \infty$ the phases in the exponent
oscillate rapidly and the dominant contribution is from the region in which the
phase is stationary, that is, near the point $\bar{s}$ which satisfies
\begin{equation}
\sin{\alpha'} = \frac{\bar{s}-x'}{l'(\bar{s})}=\frac{x-\bar{s}}{l(\bar{s})}=\sin{\alpha}.
\end{equation}
The dominant contribution is from the neighborhood of the classical orbit which
performs specular reflection from the hard wall. The slowly varying functions in
the integral~(\ref{b4st}) are replaced by their values at the stationary point ($l=l(\bar{s})$ and similarly for other values). The phase in the exponent is
expanded to second order in the distance from $\bar{s}$ to give $l(s)+l'(s)\simeq l + l' +\frac{\cos^2 (\alpha) (l+l')}{2 l l'} \zeta^2$ where $\zeta=s-\bar{s}$. The resulting Gaussian integral can be easily evaluated leading to
\begin{equation}
I(k; {\bf r},{\bf r'})\simeq \frac{-1}{\sqrt{8\pi k (l+l')}} e^{i(k(l+l') -\frac{3}{4} \pi)}.
\end{equation}
This is just the semiclassical limit of minus of the Green's function from the image
of ${\bf r'}$ with respect to reflection around the boundary to ${\bf r}$.
For the case of a half plane this is a semiclassical version of the known
exact result~\cite{balian72}.
 Since only the neighborhood of the classical reflection point contributes in the semiclassical limit it is not important that the wall of the rectangle 
is not infinite. Multiple reflections are handled by using the same method over and over. We refer the reader to~\cite{sieber97}, where the case of curved boundary is also treated, for details. 

So far the boundary was assumed to be smooth, however the boundary of the
rectangular billiard has corners. In the general case corners can lead to diffraction contributions but when the angle of the wedge is
of the form $\frac{\pi}{q}$ (with $q$ integer) the only contribution is given by using the method
of images~\cite{sieber97}.
 In the case when the classical trajectory hits the corner, the resulting semiclassical Green's function (after integration of~(\ref{complaw})) is 
the free Green's function from the image of ${\bf r'}$ under  reflections 
with respect to both walls,
to ${\bf r}$. This is reasonable when one considers the method of images but 
for the angle $\frac{\pi}{2}$ one
can also show that it is the semiclassical limit of the double integral over
the boundary even when the classical trajectory is reflected from the corner
(the sign of the Green's function is positive since two reflections are involved). 

Using the composition law~(\ref{complaw}) it is now possible to compute the contributions
of periodic and diffracting orbits to the density of states. 

\subsection{Periodic orbits contributions}

Periodic orbits contributions are obtained when the contributions of $D$ in the
Green's function~(\ref{green}) are ignored when substituted in~(\ref{dexp}). 
Periodic orbits can be identified by two integer numbers $(N,M)$ which denote
half of the number of bounces from the horizontal and vertical sides. Therefore the
contribution from a specific periodic orbit results from the boundary integral with $n=2N+2M$ bounces. This contribution is thus
\begin{eqnarray}
\label{po-int}
d_{osc}^{(N,M)} (k) = \frac{1}{\pi} \Im \frac{1}{2N+2M} (-2)^{2N+2M}  \frac{d}{dk} \oint_{\partial {\cal D}} ds_1 \cdots ds_{2N+2M} \partial_{n_{2N+2M}} G_0 (k; {\bf r}_1,{\bf r}_{2N+2M}) \nonumber \\ \partial_{n_{2N+2M-1}} G_0 (k; {\bf r}_{2N+2M},{\bf r}_{2N+2M-1})   \cdots  \partial_{n_2} G_0 (k; {\bf r}_3,{\bf r}_2) \partial_{n_1} G_0 (k; {\bf r}_2,{\bf r}_1).
\end{eqnarray}
The integral~(\ref{po-int}) contains all the contributions of periodic orbits
with $2N+2M$ reflections from the walls. 

\begin{figure}[p]
\begin{flushright}
\vspace{.5cm}
 \leavevmode
\epsfig{file=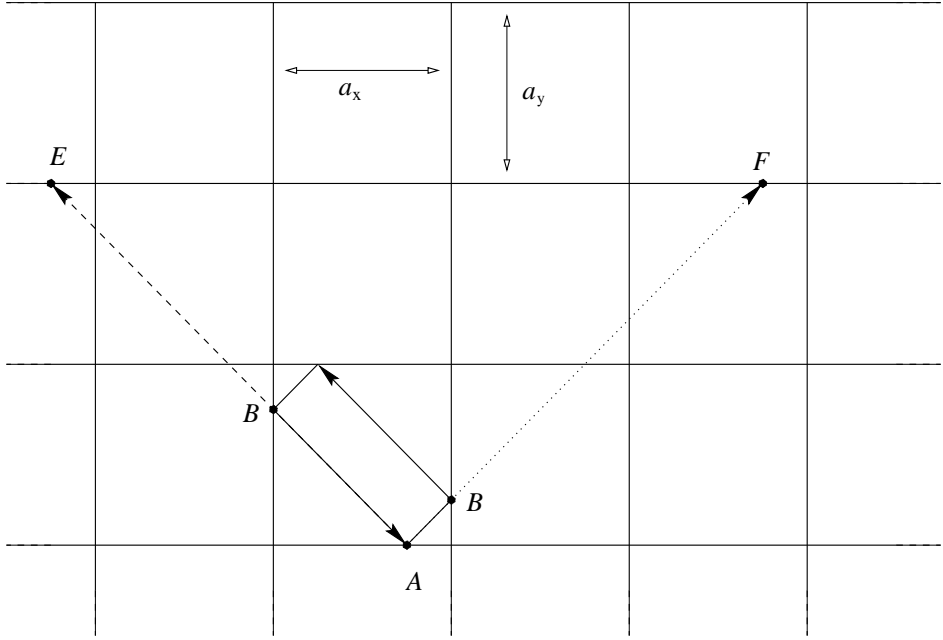,height=8.5cm,width=12.5cm,angle=0}
\centering{\caption{The stationary phase contributions for periodic orbits \label{per}}}
\end{flushright}
\vspace{0cm}
\end{figure}

For the calculation we choose
a point $s_1$ on the boundary, assume, without loss of generality, that it
is on the lower side of the rectangle. The composition law can be used to 
compute the integrals over $s_2,\cdots,s_{2N+2M}$. The contribution of
the stationary phase points which belong to the orbit of interest is given
by the Green's function that describes
 propagation from the image of $s_1$ under
$2N$ and $2M-1$ reflections to
$s_1$ as is seen in Fig.~\ref{per}.
The variable $s_1$ is such that it corresponds to the point $A$ in Fig.~\ref{per}. The integral over the rest of the variables will give contributions corresponding to the orbit of interest when $s_2$ corresponds to either of the two
points marked by $B$. These contributions are contributions of the orbit and
its time reversal, and when either $N=0$ or $M=0$ there will be only one 
contributing point. From the composition law the result of the integral over
$s_2,\cdots,s_n$ (in the semiclassical limit) is the sum of the Green's functions from the image points
$E$ and $F$ to $A$. These contributions (and also their normal derivatives) are equal.
For every point on the lower side there is a stationary 
phase contribution corresponding to an orbit (unless $N=0)$. This demonstrates
that the periodic orbits are arranged in families.
The contribution to the $s_1$ integral (in~(\ref{po-int})) from the lower side is thus
\begin{equation}
-4 \int_{l.s.} ds_1 \; \partial_{n_1} \frac{(-1)^{2N+2M-1}}{\sqrt{8 \pi k l_p}} e^{i(kl_p -\frac{3}{4} \pi )} \simeq 4 \frac{2M a_x a_y i}{\sqrt{8 \pi k l_p}} \frac{k}{l_p} e^{i(kl_p -\frac{3}{4} \pi )}
\end{equation}
where $l_p=\sqrt{(2Na_x)^2+(2Ma_y)^2}$ is the length of the periodic orbit, and
$(N,M)$ is denoted by $p$. The factor of $-4$ is present since there was a factor of $-2$ in front of the integral~(\ref{po-int}) that was not used in the composition law and also there are two equal contributions from the orbit and its time reversed one. In the semiclassical limit the dominant contribution
of the partial derivative in~(\ref{po-int}) is when it is applied to the exponent since it contributes a factor of $k$. The fact that $\partial_{n_1} l_p = \frac{2Ma_y}{l_p}$ was also used. The integrand is constant and thus the integral is just a multiplication by the side length $a_x$. 

The contribution from the top side is equal to the one from the bottom side, and on
the vertical side a similar calculation can be done, and one obtains the same contribution with the roles of $N$ and $M$ interchanged.
The leading contribution (in the semiclassical limit) is obtained when the derivative $\frac{d}{dk}$ is applied to the exponential in~(\ref{po-int}).
 The resulting contribution
of the periodic orbit $p$ is thus
\begin{equation}
\label{perioddosc}
d_{osc}^{(N,M)} (k) = \frac{1}{\pi} \Im \frac{1}{2M+2N} \frac{d}{dk} 
\left( \frac{8{\cal A} \beta_{(N,M)}}{\sqrt{8 \pi k l_p}} \frac{k}{l_p} (2M+2N) e^{i(kl_p -\frac{1}{4} \pi )} \right) \simeq \frac{8 {\cal A} k \beta_{(N,M)}}{\pi \sqrt{8 \pi k l_p}} \cos (k l_p -\frac{\pi}{4}),
\end{equation}
where $\beta (N,M)$ is a counting factor taking the values $1$ and $\frac{1}{2}$.
 Since the orbits with a vanishing index are their own time reversals, there is only one contribution
for the $s_2,\cdots,s_n$ integrals, therefore  $\beta_{(N,M)}$ is equal to $\frac{1}{2}$ when
$N=0$ or $M=0$ and is otherwise $1$.

The density of states measured in units of energy is  $d(E)=\frac{d(k)}{2k}$. 
The contribution of a periodic orbits is thus given by
\begin{equation}
\label{doscofpr}
d_{osc}^{(N,M)} ( E)= \frac{ 4 {\cal A} \beta_{(N,M)}}{\pi \sqrt{8 \pi k l_p}} \cos ( k l_p - \frac{\pi}{4}). 
\end{equation}
This is exactly the contribution that can be obtained using the Berry-Tabor
formula for integrable systems~\cite{BT-I}.

\subsection{Diffracting orbits}

The contribution of diffracting orbits, with one diffraction, to the integrals~(\ref{dexp}) results from terms where for one Green's function
the contribution of the second terms in~(\ref{green}) is used
while for the other $n-1$ terms the first term in~(\ref{green}) is used.
 There are $n$ such terms, which are all equal (they only differ by the labels of the integration variables). 
To compute the contribution of diffracting orbits 
one has to evaluate the integral 
\begin{equation}
\label{tempi}
I_1 = (-2)^n \!  \oint \! ds_1 ds_2 \cdots ds_n \, \partial_{n_1} \! G_0 (k; {\bf r}_2,{\bf r}_{1})
\cdots  \partial_{n_{n-1}} \! G_0 (k; {\bf r}_n,{\bf r}_{n-1}) \partial_{n_n} \! G_0 (k; {\bf r}_0,{\bf r}_n) D (\theta_1,\theta_n)  G_0 (k; {\bf r}_1,{\bf r}_{0}).
\end{equation}
It will be assumed that the diffraction constant $D$ is slowly varying so that
in the region which gives the stationary phase contribution it can be approximated by a constant. 
This means that $D(\theta,\theta ')$ does not change considerably when
the angles change by $\delta \theta \sim \sqrt{\frac{\lambda}{l_\bot}}$ where $l_\bot$ is the distance from the scatterer to the hard wall. 
When $D$ is indeed slowly varying it can be replaced by its value at the stationary points
of the boundary integrals and
the composition law can be used to replace the integral~(\ref{tempi}) by the sum of all Green's functions from images of the scatterer (which cross $n$ walls) to the scatterer. 

\begin{figure}[p]
\begin{flushright}
\vspace{.5cm}
 \leavevmode
\epsfig{file=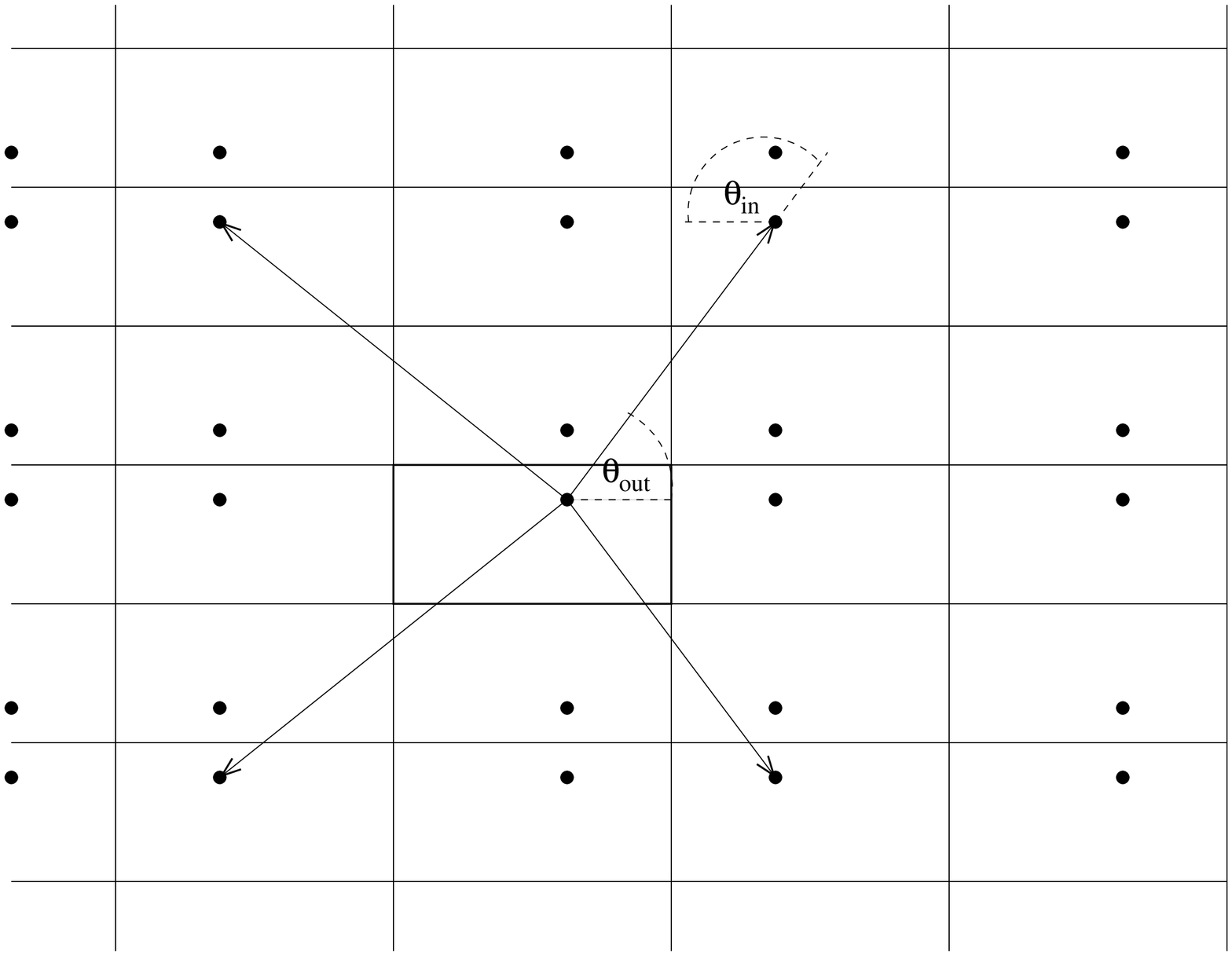,height=8.5cm,width=10.5cm,angle=0}
\centering{\caption{The $(1,2)$ once diffracting orbits \label{difffig}}}
\end{flushright}
\vspace{0cm}
\end{figure}

One can denote the once-diffracting orbits by the number of times they are reflected from the vertical and horizontal sides of the rectangle. For a
pair of indices $(N_1,M_1)$ there are four (two when one of the indices vanishes) such orbits as can be seen from Fig.~\ref{difffig}.
The dots in the figure denote the scatterer and its images, the orbits shown are all $(1,2)$ diffracting orbits. The contributions of these orbits will
result from the integral with $3$ boundary points 
(there are also other contributions with $N_1+M_1=3$). Typically, there are four such orbits, but when one of the indices vanishes there are only two. We will add another index $\mu$ which runs between $1$ and $4$ and counts these orbits (alternatively it is possible
to allow the indices $M_1,N_1$ to be negative).
The direction in which the diffracting orbit returns to the scatterer depends
upon the outgoing direction and upon the parity of $(N_1,M_1)$.
By counting the number of reflections on the boundary one finds
\begin{equation}
\label{inangle}
\theta_{in}=\Theta_{(N_1,M_1)} (\theta_{out}) = \left\{ { \begin{array}{ll} \theta_{out} \;\;  & even-even \\ \pi-\theta_{out} \;\; & odd-even \\ 2\pi-\theta_{out} \;\; & even-odd \\ \pi+\theta_{out} \;\; & odd-odd. \end{array}} \right.  
\end{equation}
In the example of Fig.~\ref{difffig} the parity of $(N_1,M_1)$ is odd-even. The lengths of the orbits and the angles can be calculated easily. If the scatterer is at $(a_x \delta_x,a_y \delta_y)$ from the center of the rectangle then the lengths of odd-odd orbits are given by $l_p^{(o-o)}=\sqrt{a_x^2 (N_1 \pm 2\delta_x)^2 +a_y^2 (M_1 \pm 2\delta_y)^2 }$. The angles can be computed similarly, since $\tan (\theta) =\frac{\pm N_1 a_x + 2a_x\delta_x}{\pm M_1 a_y + 2a_y \delta_y}$ (note that this angle is in the third quadrant if both the numerator and denominator are negative). The lengths and angles of other orbits can be computed too, just by replacing
 $\delta_x$ by $0$ when $N_1$ is even and $\delta_y$ by $0$ for even $M_1$. 

The contribution to the integral~(\ref{tempi}) which corresponds to the $(N_1,M_1)$ once diffracting orbits is thus
\begin{equation}
\label{eqa18}
I_1^{(N_1,M_1)} = \sum_{\mu} (-1)^{N_1+M_1} D (\theta_{{\bf N}_1,\mu},\Theta (\theta_{{\bf N}_1,\mu})) \frac{1}{\sqrt{8 \pi k l_{{\bf N}_1,\mu}}} e^{i(k l_{{\bf N}_1,\mu} -\frac{3}{4} \pi)}
\end{equation}
where $\mu$ runs over the four contributions (two contribution when either $M_1$ or $N_1$ vanish) and ${\bf N}_1=(N_1,M_1)$ is the pair of indices that define the orbits. 
In what follows some of the subscripts ${\bf N}_j$ will be replaced by $j$,
that is $l_{{\bf N}_j}=l_j$, $\theta_{{\bf N}_j}=\theta_j$, etc.
The contribution of each one of these orbits
to the density of states is
\begin{equation}
\tilde{d}_{osc}^{(N_1,M_1),\mu}(k) = \frac{1}{\pi} \Im \frac{d}{dk} \left((-1)^{N_1+M_1} D (\theta_{1,\mu},\Theta (\theta_{1,\mu})) \frac{1}{\sqrt{8 \pi k l_{1,\mu}}} e^{i(k l_{1,\mu} -\frac{3}{4} \pi)} \right)
\end{equation}
 In the semiclassical
limit the main contribution is from the differentiation of the exponent.
Thus the contribution of a once diffracting orbit is given by
\begin{equation}
\tilde{d}_{osc}^{(N_1,M_1),\mu}(k) = (-1)^{N_1+M_1} \frac{\sqrt{l_{1,\mu} } }{\pi \sqrt{8 \pi k }} \Re \left( D (\theta_{1,\mu},\Theta (\theta_{1,\mu})) e^{i(k l_{1,\mu} -\frac{3}{4} \pi)} \right).
\end{equation}
The contribution to $d(E)$ is thus
\begin{equation}
\label{doscof1d}
\tilde{d}_{osc}^{(N_1,M_1),\mu}(E) = (-1)^{N_1+M_1} \frac{\sqrt{l_{1,\mu} } }{2 \pi k \sqrt{8 \pi k }} \Re \left( D (\theta_{1,\mu},\Theta (\theta_{1,\mu})) e^{i(k l_{1,\mu} -\frac{3}{4} \pi)} \right).
\end{equation}

The contribution of orbits with two diffractions is also needed in this work. The computation is quite similar, since these orbits are just a combination of
two once diffracting orbits they can be identified by two
pairs of integer indices ${\bf N}_1=(N_1,M_1)$ and ${\bf N}_2=(M_2,N_2)$ and
by two indices $\mu$ ($\nu$ respectively) which denote
along with the ${\bf N}_1$ (${\bf N}_2$ respectively) orbits
that are parts of the twice diffracting orbit. 
The number of boundary reflections of each segment of the orbit
is denoted by ${\cal N}_i =  N_i+M_i$.
The contribution of such an orbit to the density of states results from terms with two diffraction coefficients in the integral~(\ref{dexp}) with $n={\cal N}_1+{\cal N}_2$. Not all of these terms describe a desired orbit, since only terms
in which the number of reflections between diffractions is ${\cal N}_1$ (and also
${\cal N}_2$ since the orbits is closed) can describe the specified orbit. There are $\frac{n}{2}$ such
contributions if ${\cal N}_1={\cal N}_2$ and $n$ contributions otherwise.
If ${\cal N}_1 \neq {\cal N}_2$ there are $n$ equal contributions
that can be represented by an integral in which the first diffraction is, for instance, between
${\bf r}_1$ and ${\bf r}_2$ and the second one is after ${\cal N}_1$
reflections from the boundary (the orbit is closed after additional ${\cal N}_2$ reflections from the boundary). One can interchange ${\cal N}_1$ and ${\cal N}_2$
without changing the resulting contribution. It is convenient to split this
contribution into two equal integrals in which the roles of ${\cal N}_1$ and ${\cal N}_2$ are interchanged.
Each of these integrals, with the diffractions between ${\bf r}_1$ and ${\bf r}_2$
and between ${\bf r}_{{\cal N}_1+1}$ and ${\bf r}_{{\cal N}_1+2}$,
should be multiplied by $\frac{n}{2}$ to take into account all permutations.
Therefore all of the contributions of orbits with two diffractions are given
by
\begin{eqnarray}
\label{start2dif}
d_{osc}^{(2)} (k) & = & \frac{1}{\pi} \frac{d}{dk} \Im \sum_{n=1}^{\infty} \frac{1}{n} \frac{n}{2} \sum_{ {\cal N}_1,{\cal N}_2=1}^{\infty} \delta_{{\cal N}_1+{\cal N}_2,n} (-2)^{{\cal N}_1+{\cal N}_2} \int_{\partial {\cal D}} ds_1 \cdots ds_{{\cal N}_1+{\cal N}_2} \partial_{{n}_{{\cal N}_1+{\cal N}_2}} G_0 ({\bf r}_1,{\bf r}_{{\cal N}_1+{\cal N}_2}) \nonumber \\
& \times & \partial_{{n}_{{\cal N}_1+{\cal N}_2-1}} G_0 ({\bf r}_{{\cal N}_1+{\cal N}_2},{\bf r}_{{\cal N}_1+{\cal N}_2-1}) \cdots \partial_{{n}_{{\cal N}_1+2}} G_0 ({\bf r}_{{\cal N}_1+3},{\bf r}_{{\cal N}_1+2}) G_0 ({\bf r}_{{\cal N}_1+2},{\bf r}_0) D \nonumber \\ 
& \times & \partial_{{n}_{{\cal N}_1+1}} G_0 ({\bf r}_0,{\bf r}_{{\cal N}_1+1}) \partial_{n_{{\cal N}_1}} G_0 ({\bf r}_{{\cal N}_1+1},{\bf r}_{{\cal N}_1}) \cdots \partial_{n_2} G_0 ({\bf r}_{3},{\bf r}_{2}) G_0 ({\bf r}_{2},{\bf r}_{0}) D \partial_{n_1} G_0 ({\bf r}_{0},{\bf r}_{1}).
\end{eqnarray}
The dependence of $D$ on angles and of $G_0$ on $k$ was omitted in the notations for the sake of brevity. The sum over $n$ is trivial. The terms of this sum (which are products of two integrals) can be calculated using the composition rule. The
result is a sum over all possible classical paths from the scatterer to itself
with ${\cal N}_1$ reflections for the first segment (and ${\cal N}_2$ for the second).
The contribution that corresponds
to all of the twice diffracting orbits is thus 
\begin{eqnarray}
d_{osc}^{(2)} (k) = \sum_{{\bf N}_1 \mu,{\bf N}_2 \nu} \frac{1}{\pi} \frac{d}{dk} \Im (-1)^{N_1+M_1+N_2+M_2}
D (\theta_{1,\mu},\Theta (\theta_{2,\nu})) \times \nonumber \\ D (\theta_{2,\nu},\Theta (\theta_{1,\mu})) \frac{1}{16 \pi k \sqrt{l_{1,\mu}l_{2,\nu}}} e^{i(k(l_{1,\mu}+l_{2,\nu})-\frac{3}{2}\pi)}
\end{eqnarray}
The dependence of the diffraction constants on the angles results from the
fact that
 the in-going direction before the second diffraction is determined by the
out-going direction from the first diffraction and vice versa.
The relation between the out-going and in-going direction is given by~(\ref{inangle}).
In the leading order the derivative in~(\ref{start2dif}) operates only on the exponential, leading to 
the (energy dependent) density
\begin{eqnarray}
\label{doscof2d}
d_{osc}^{(2)} (E) = \sum_{{\bf N}_1 \mu,{\bf N}_2 \nu} 
 (-1)^{N_1+M_1+N_2+M_2} 
\frac{l_{1,\mu}+l_{2,\nu}}{32 \pi^2 k^2 \sqrt{l_{1,\mu}l_{2,\nu}}} \times \nonumber \\
\Re \left( D (\theta_{1,\mu},\Theta (\theta_{2,\nu}))  D (\theta_{2,\nu},\Theta (\theta_{1,\mu})) e^{i(k(l_{1,\mu}+l_{2,\nu})-\frac{3}{2}\pi)} \right).
\end{eqnarray}
Combining~(\ref{doscof2d}) with~(\ref{doscofpr}) and~(\ref{doscof1d})
we find that the total density of states taking into account
terms up to the second order in $D$ is given by
\begin{eqnarray}
\label{finaldos}
d_{osc} (E) & = & \sum_{{\bf N}} \frac{ 2 a b \beta_{(N,M)}}{\pi \sqrt{8 \pi k l_p}} e^{i( k l_p - \frac{\pi}{4})} + 
\sum_{{\bf N}_1,\mu} (-1)^{N_1+M_1} \frac{\sqrt{l_{1,\mu} } }{4 \pi k \sqrt{8 \pi k }}  D (\theta_{1,\mu},\Theta (\theta_{1,\mu}))  e^{i(k l_{1,\mu} -\frac{3}{4} \pi)}  \nonumber \\ & + & 
 \sum_{{\bf N}_1,\mu} \sum_{{\bf N}_2,\nu} 
 (-1)^{N_1 \! +\! M_1\! +\! N_2\! +\! M_2} 
\frac{l_{1,\mu}+l_{2,\nu}}{64 \pi^2 k^2 \sqrt{l_{1,\mu}l_{2,\nu}}} 
\left( D (\theta_{1,\mu},\Theta (\theta_{2,\nu}))  D (\theta_{2,\nu},\Theta (\theta_{1,\mu})) e^{i(k(l_{1,\mu}+l_{2,\nu})-\frac{3}{2}\pi)} \right) \nonumber \\
& + & c.c.
\end{eqnarray}

Note that when one of the indices of ${\bf N}_1$ (${\bf N}_2$ respectively) vanishes the sum over $\mu$ ($\nu$ respectively) includes two contributions
instead of four. Thus for an angle independent diffraction constant such
orbits will have a factor of half similar to the one that appears in
the periodic orbit contributions.
Equation~(\ref{finaldos}) will be used to compute the various contributions to
the spectral from factor.

Similarly one can calculate the contributions of orbits with three diffraction.
The calculation follows the one of two diffractions closely, and the only difference is that one now has a factor of $\frac{n}{3}$ instead of $\frac{n}{2}$
in~(\ref{start2dif}). Since there are three diffractions the sum is over ${\bf N}_1 , \mu_1$, ${\bf N}_2 , \mu_2$ and ${\bf N}_3 , \mu_3$
which describe the three segments of the orbits. 
In this work the contribution to third
order in $D$ is required only for angle independent scattering.
In this case the contribution
of all orbits with three (angle independent) diffractions is found to be
\begin{eqnarray}
\label{3difdos}
d_{osc}^{(3)} (E) & = & \sum_{{\bf N}_1, \mu_1} \sum_{{\bf N}_2, \mu_2} \sum_{{\bf N}_3, \mu_3} (-1)^{N_1+M_1+N_2+M_2+N_3+M_3} \frac{D^3}{12 \pi k (8 \pi k)^{\frac{3}{2}}} \frac{l_{1,\mu_1}+l_{2,\mu_2}+l_{3,\mu_3}}{\sqrt{l_{1,\mu_1} l_{2,\mu_2} l_{3,\mu_3}}} e^{- i \frac{\pi}{4}} \nonumber \\
& \times & e^{i k (l_{1,\mu_1}+l_{2,\mu_2}+l_{3,\mu_3})} + c.c.
\end{eqnarray}

In Sec.~\ref{noangle} it is assumed that the scattering is angle independent,
therefore the function $D(\theta,\theta')$ is replaced by an angle independent
constant $D$. Moreover we are interested in the form factor
where in the calculation energy averages are taken and sums over
periodic orbits are approximated by integrals. Within such approximations the
weight of the orbits with $\beta_{(N,M)} \neq 1$ (one vanishing index) is negligible, therefore in other parts of the paper
these factors were replaced by unity. Within these assumptions~(\ref{finaldos}) and~(\ref{3difdos})
reduce to~(\ref{dosc2}).

\section{The diffraction constant of the point scatterer with the free Green's function $G_0$}
\label{pointd}

The diffraction constant is the on shell transition matrix, $T$, which for 
a point interaction is given by
\begin{equation}
\label{findwhatdis}
D^{-1}(z)=\frac{z-i\Lambda}{1-e^{i\xi_1}} \int d^2 {x} \; G_0 (z; {\bf x},{\bf 0}) G_0 (i\Lambda; {\bf x},{\bf 0}) + \frac{z+i\Lambda}{1-e^{-i\xi_1}}\int d^2 {\bf x} \; G_0 (z; {\bf x},{\bf 0}) G_0 (-i\Lambda; {\bf x},{\bf 0}) 
\end{equation}
where $\Im \sqrt{z} > 0$ and $G_0$ is the Green's function in free space. To compute the integrals we use the following integral representation for the free Green's function in two dimensions
\begin{equation}
G_0 (z; {\bf x},{\bf x}_0) = \frac{1}{(2 \pi)^2} \int d^2 k \;\frac{e^{i {\bf k} \cdot ({\bf x}_0-{\bf x})}}{z-k^2}. 
\end{equation}
Let us compute the integral
\begin{eqnarray}
I_1 & = & \int d^2{x} \; G_0 (z; {\bf x},{\bf 0}) G_0 (i\Lambda; {\bf x},{\bf 0})=\frac{1}{(2 \pi)^4} \int d^2 k_1 \; d^2 k_2 \; d^2 x \; \frac{e^{-i {\bf k}_1\cdot {\bf x}}}{(k_1^2 - z)} \frac{e^{-i {\bf k}_2 \cdot {\bf x}}}{(k_2^2 - i \Lambda)} \nonumber \\ & = &\frac{1}{(2\pi)^2} \int d^2 k \; \frac{1}{(k^2 -z)(k^2 -i \Lambda)}.
\end{eqnarray}
Changing variables to $E=k^2$ one obtains
\begin{equation}
I_1=\frac{1}{4 \pi} \int_0^{\infty} d E \; \frac{1}{(E-z)(E-i\Lambda)}=\frac{1}{4 \pi (z-i\Lambda)} \int_0^{\infty} d E \; \left[ \frac{1}{E-z} - \frac{1}{E-i\Lambda}\right].
\end{equation}
Using the known relation
\begin{equation}
\frac{1}{x-x_0 \pm i \epsilon} = \wp{\frac{1}{x-x_0}} \mp i \pi \delta (x-x_0)
\end{equation}
and noting that we are interested in the case where $\Im z$ tends to zero from above we obtain
\begin{equation}
I_1=\frac{1}{4 \pi (z-i\Lambda)} \lim_{M \rightarrow \infty} \lim_{\beta \rightarrow 0} \left[ \int_0^{z-\beta} d E \; \frac{1}{E-z} + \int_{z+\beta}^M d E \;\frac{1}{E-z} - \int_0^M d E \; \frac{1}{E- i \Lambda} + i \pi\right].
\end{equation}
The integrals are elementary, and taking the limits results in
\begin{equation}
\label{int1}
I_1 = \frac{1}{4 \pi (z-i\Lambda)} \left[ -\ln \left(\frac{z}{\Lambda} \right) +i \frac{\pi}{2} \right].
\end{equation}
The second integral in~(\ref{findwhatdis}) is computed in the same way
\begin{equation}
\label{int2}
I_2=\int d^2 x \; G_0 (z; {\bf x},{\bf 0}) G_0 (-i\Lambda; {\bf x},{\bf 0}) = \frac{1}{4 \pi (z+i\Lambda)} \left[ - \ln\left(\frac{z}{\Lambda} \right) + i \frac{3 \pi}{2} \right].
\end{equation}
Substituting~(\ref{int1}) and~(\ref{int2}) into~(\ref{findwhatdis}) leads to the diffraction constant:
\begin{equation}
D(z)=\frac{2 \pi}{ -\frac{1}{2}\ln\left(\frac{z}{\Lambda} \right) + \frac{\pi}{4} \frac{\sin \xi_1}{1 - \cos \xi_1} +i \frac{\pi}{2}}.
\end{equation}

\section{Asymptotics of scattering from localized potentials}
\label{definitions}

In this Appendix $D_g({\bf q})$ and $h({\bf k},{\bf q})$ are computed up to order $k^2a^2$.
The computation is presented explicitly only to order $ka$. 
The calculation of the second order is straight forward and fairly similar to the 
calculation of the first order, but its details are too long to be included 
in the paper. 

\subsection{Calculation of $D_g({\bf q})$ to the order $(qa)^2$}

In the calculation of $D_g ({\bf q})$ many of the contributions cancel, but equation~(\ref{dg}) does not show this explicitly. A more convenient way to
compute $D_g({\bf q})$ is obtained when~(\ref{noyce2}) is iterated, and all the resolutions of the identity in momentum space which appear there are replaced
by resolutions of identity in coordinate space. Using~(\ref{noyce1})
leads to
\begin{eqnarray}
\label{simpledg}
D_g & = & 1 - g \int d^2 y_1^{(a)} d^2 y_1 \; e^{-i a {\bf q \cdot y}_1^{(a)}} f({\bf y}_1^{(a)}) G_0 (ka; {\bf y}_1^{(a)} ,{\bf y}_1 ) \frac{f( {\bf y}_1)}{V_0} e^{ia {\bf q \cdot y}_1} \nonumber \\ & - & g^2 \int d^2 y_1^{(a)} d^2 y_1 d^2 y_2 \; e^{-i a {\bf q \cdot y}_1^{(a)}} f({\bf y}_1^{(a)}) G_0 (ka; {\bf y}_1^{(a)} ,{\bf y}_1 ) \times \nonumber \\ & \times & \left ( f({\bf y}_1)  G_0 (ka; {\bf y}_1, {\bf y}_2) - \int d^2 y_2^{(a)} \frac{f( {\bf y}_1) f( {\bf y}_2^{(a)})}{V_0} e^{i a {\bf q \cdot} ({\bf y}_1 -{\bf y}_2^{(a)})} G_0 (ka; {\bf y}_2^{(a)}, {\bf y}_2) \right) \frac{f({\bf y}_2)}{V_0} e^{i a {\bf q \cdot y}_2} \nonumber \\ &- & g^3 \int d^2 y_1^{(a)} d^2 y_1 d^2 y_2 d^2 y_3 \; e^{-i a {\bf q \cdot y}_1^{(a)}} f({\bf y}_1^{(a)}) G_0 (ka; {\bf y}_1^{(a)} ,{\bf y}_1 ) \times \nonumber \\ & \times & \! \left ( f({\bf y}_1)  G_0 (ka; {\bf y}_1, {\bf y}_2) - \! \int d^2 y_2^{(a)} \frac{f( {\bf y}_1 ) f( {\bf y}_2^{(a)})}{V_0} e^{i a {\bf q \cdot}({\bf y}_1 -{\bf y}_2^{(a)})} G_0 (ka; {\bf y}_2^{(a)}, {\bf y}_2) \right) \times \nonumber \\ & \times &\! \left ( f({\bf y}_2)  G_0 (ka; {\bf y}_2, {\bf y}_3) - \! \int \! d^2 y_3^{(a)} \frac{f( {\bf y}_2 ) f( {\bf y}_3^{(a)})}{V_0} e^{i a {\bf q \cdot} ({\bf y}_2 -{\bf y}_3^{(a)})} G_0 (ka; {\bf y}_3^{(a)}, {\bf y}_3) \right) \! \frac{f({\bf y}_3)}{V_0} e^{i a {\bf q \cdot y}_3} -\cdots
\end{eqnarray}
where 
\begin{equation}
\label{defv0}
V_0= \langle{\bf q}|{\bf U}|{\bf q}\rangle=\int d^2 y f( {\bf y}).
\end{equation}
The terms of higher order in $g$ have the same structure, but with more of
the factors such as 
\begin{equation}
\label{defbj}
B_j=f({\bf y}_j)  G_0 (ka; {\bf y}_j, {\bf y}_{j+1}) - \int d^2 y_{j+1}^{(a)} \frac{f( {\bf y}_j) f( {\bf y}_{j+1}^{(a)})}{V_0} e^{i a {\bf q \cdot} ({\bf y}_j - {\bf y}_{j+1}^{(a)})} G_0 (ka; {\bf y}_{j+1}^{(a)}, {\bf y}_{j+1}).
\end{equation}
 Equation~(\ref{simpledg}) is a convenient starting point
from which  the asymptotics of $D_g({\bf q})$ in the limit $qa \rightarrow 0$ can be calculated .

To calculate the leading order of $D_g({\bf q})$ (with respect to $qa$) all the exponentials are replaced by unity and the Green's function by the leading order
as given by~(\ref{greenexp}). The leading order of any of 
the brackets is given by
\begin{equation}
\label{leadingbrackets}
B_j \simeq f({\bf y}_j)  \phi ( {\bf y}_j , {\bf y}_{j+1}) - \int d^2 y_{j+1}^{(a)} \frac{f( {\bf y}_j) f( {\bf y}_{j+1}^{(a)})}{V_0}  \phi ( {\bf y}_{j+1}^{(a)}, {\bf y}_{j+1}). 
\end{equation}
Note that the contribution from the constant parts of the Green's functions, $d_1$, cancel each other. This is the advantage of the Noyce method when describing scattering 
at low energies in two dimensions.
Most of the Green's functions in equation~(\ref{simpledg}) appear in brackets 
of the form $B_j$,
and thus, in the leading order, their constant parts cancel. However, in every 
term of some power in $g$ there is one Green's function that does not
appear in some brackets. When this Green's function is multiplied 
by one of the brackets one finds that the contribution of its constant part, $d_1$,
vanishes. This is true since for any functions $g_1({\bf y})$,$g_2({\bf y})$, $G({\bf y}_1,{\bf y}_2)$ one finds 
\begin{eqnarray}
\label{leadinggreen}
\int d^2 y_1^{(a)} d^2 y_1 d^2 y_2 \; g_1( {\bf y}_1^{(a)}) d_1 \left( f({\bf y}_1) G({\bf y}_1,{\bf y}_2) - \int d^2 y_2^{(a)} \frac{f({\bf y}_1) f({\bf y}_2^{(a)})}{V_0} G({\bf y}_2^{(a)},{\bf y}_2) \right) g_2 ({\bf y}_2) & = & \nonumber \\
\int d^2 y_1^{(a)}  d^2 y_2 \; g_1( {\bf y}_1^{(a)}) d_1 \left( \int d^2 y_1 f({\bf y}_1) G({\bf y}_1,{\bf y}_2) - \int d^2 y_2^{(a)}f({\bf y}_2^{(a)}) G({\bf y}_2^{(a)},{\bf y}_2 )\right)  g_2 ({\bf y}_2)  & = & 0.
\end{eqnarray}
Therefore the constant part of the Green's function does not contribute 
to the leading order of $D_g({\bf q})$ except for the term of order $g$. 
Using (\ref{leadingbrackets}) and (\ref{leadinggreen}) the leading order (in $qa$) contribution to $D_g({\bf q})$ is given by
\begin{eqnarray}
\label{leaddg}
D_g & \simeq & 1 - g \int d^2 y_1^{(a)} d^2 y_1 \; f( {\bf y}_1^{(a)}) \left( d_1 + \phi ( {\bf y}_1^{(a)},{\bf y}_1) \right) \frac{f({\bf y}_1)}{V_0} \nonumber \\
& - & \! g^2 \int \! d^2 y_1^{(a)} d^2 y_1 d^2 y_2 \; f( {\bf y}_1^{(a)}) \phi ( {\bf y}_1^{(a)},{\bf y}_1) \left( f( {\bf y}_1) \phi ( {\bf y}_1 ,{\bf y}_2) -\int \! d^2 y_2^{(a)} \frac{f({\bf y}_1)f({\bf y}_2^{(a)})}{V_0} \phi ({\bf y}_2^{(a)},{\bf y}_2 ) \right) \frac{ f( {\bf y}_2)}{V_0} \nonumber \\
& - & \! g^3 \int \! d^2 y_1^{(a)} d^2 y_1 d^2 y_2 d^2 y_3 \; f( {\bf y}_1^{(a)}) \phi ( {\bf y}_1^{(a)},{\bf y}_1) \left( f( {\bf y}_1) \phi ( {\bf y}_1,{\bf y}_2) -\int \! d^2 y_2^{(a)} \frac{f({\bf y}_1)f({\bf y}_2^{(a)})}{V_0} \phi ({\bf y}_2^{(a)},{\bf y}_2) \right) \nonumber \\
& \times & \left( f( {\bf y}_2) \phi ( {\bf y}_2 ,{\bf y}_3 ) -\int \! d^2 y_3^{(a)} \frac{f({\bf y}_2 )f({\bf y}_3^{(a)})}{V_0} \phi ({\bf y}_3^{(a)} ,{\bf y}_3) \right) \frac{ f( {\bf y}_3 )}{V_0} - \cdots.
\end{eqnarray}

Note that the integrals in~(\ref{leaddg}) are angle independent. To simplify
the notations the following definitions are useful
\begin{eqnarray}
L_0 & \equiv & \int d^2 y_1 \; f({\bf y}_1) = V_0 \nonumber \\
L_1 & \equiv & \int d^2 y_1 d^2 y_2 \;  f({\bf y}_1) \phi ({\bf y}_1,{\bf y}_2)f({\bf y}_2) \nonumber \\
L_2 & \equiv &   \int d^2 y_1 d^2 y_2 d^2 y_3 \;  f({\bf y}_1) \phi ({\bf y}_1,{\bf y}_2)f({\bf y}_2) \phi ({\bf y}_2,{\bf y}_3)f({\bf y}_3) \\
 & \vdots & \nonumber 
\end{eqnarray}
Using these definitions~(\ref{leaddg}) can be written as
\begin{equation}
\label{3orderd}
 D_g ({\bf q})\simeq 1 -g d_1 V_0 - g \frac{L_1}{V_0} - g^2 \left( \frac{L_2}{V_0} - \frac{L_1^2}{V_0^2} \right) - g^3 \left( \frac{L_3}{V_0} - 2 \frac{L_2 L_1}{V_0^2} + \frac{L_1^3}{V_0^3} \right) - \cdots
\end{equation}
Note that the contribution of order $g^N$ is a sum where each term is composed of a product of $L_j$'s where the sum of $j$'s is $N$.
It is multiplied by $-\frac{1}{V_0}$ to the power of number of $L_j$'s.
The sum is over all possible partitions of $N$ into positive integer
numbers $m_j$ so that $\sum_j m_j=N$. Such partitions are generated by
the expansion
\begin{equation}
\label{defvg}
V(g) =  \frac{V_0}{\sum_{m=0}^{\infty} g^m L_m}=\frac{V_0}{V_0+ \sum_{m=1}^{\infty} g^m L_m} =  1+ \sum_{k=1}^{\infty} \left[ - \sum_{m=1}^{\infty}g^m \frac{L_m}{V_0}  \right]^k,
\end{equation}
that can be written in terms of partitions as
\begin{equation}
\label{partitution}
V(g) = 1 + \sum_{N=1}^{\infty} g^N \sum_i \prod_{j=1}^{l_i} \left( - \frac{L_{m_j}}{V_0}\right),
\end{equation}
where $i$ is the index that specifies the $i$-th partition
consisting of $l_i$ terms.
The various indices of the $i$-th partition satisfy
$\sum_{j=1}^{l_i} m_j = N$ with the requirement that all $m_j$ are
positive.
Consequently the leading order of $D_g ({\bf q})$ is
\begin{equation}
D_g ({\bf q})= V(g) - g d_1 V_0. 
\end{equation}
Expanding $V(g)$ to the order $g^3$ one recovers~(\ref{3orderd}).

There are no contributions to $D_g({\bf q})$ of order $qa$ and the first correction to $D_g({\bf q})$ scales as $q^2 a^2 $ (times some power of $\ln qa$). 
To see this note that from~(\ref{dg}) it is clear that $D_g({\bf q})$ is built from products
of matrix elements of the form $\langle{\bf q}| {\bf UGUGU \cdots UGU}|{\bf q}\rangle$ with various numbers
of Green's functions. A contribution of the order $qa$ must come from
one of these elements which is expanded to that order. For example,
the contribution of $\langle{\bf q}|{\bf UGU}|{\bf q}\rangle$ to this order is
\begin{equation}
  -  \int d^2 y_1 d^2 y_2 \; i a {\bf q \cdot y}_1 f( {\bf y}_1) G_0 (ka ; {\bf y}_1, {\bf y}_2) f( {\bf y}_2) +  \int d^2 y_1 d^2 y_2 \; i a {\bf q \cdot y}_2 f( {\bf y}_1) G_0 (ka ; {\bf y}_1, {\bf y}_2) f( {\bf y}_2) = 0,
\end{equation}
since $G_0(ka; {\bf y}_1,{\bf y}_2)=G_0(ka; {\bf y}_2,{\bf y}_1)$. This holds if more
Green's functions are used, therefore $D_g ({\bf q})$ cannot
 have contributions of the order $qa$. 

The contribution of higher order in $qa$ to $D_g({\bf q})$ can be computed by expanding
the exponents and the Green's functions in~(\ref{simpledg}) and collecting
all the terms up to the required order of $qa$. Here $D_g({\bf q})$ is calculated to
order $q^2a^2$. The calculation is straightforward
but tedious and only the final result is presented
\begin{eqnarray}
\label{lastdg}
D_g & \simeq & V(g) - g d_1 V_0 + \frac{q^2 a^2}{2} g \left( d_1 - \frac{1}{2 \pi} \right)V_0 V(g) \left[ \sum_{k=0}^{\infty} \frac{\delta_k}{V_0} g^k - V(g) \left| \sum_{k=0}^{\infty} \frac{\alpha_k}{V_0} g^k \right|^2 \right] + \frac{q^2 a^2}{4} V^2(g) \sum_{l=1}^{\infty} \frac{ {{\cal L}_l}}{V_0} g^l \nonumber \\ & & - q^2 a^2 \left( V(g) - g d_1 V_0 \right) \left\{ \left( V(g) - g d_1 V_0 \right) \sum_{l=1}^{\infty} \frac{e^{2i\theta} \mu_l^{(++)} +\mu_l^{(+-)} +\mu_l^{(-+)} +e^{-2 i \theta} \mu_l^{(--)} }{4 V_0} g^l \right. \nonumber \\ & - & \left. \frac{g d_1}{2} \left(\delta_0 + \frac{1}{2} e^{2 i \theta} \sigma_0 + \frac{1}{2} e^{-2i \theta} \sigma_0^* \right) - V(g) \sum_{l=1}^{\infty} \frac{\left( \delta_l + \frac{1}{2} e^{2 i \theta}\sigma_l + \frac{1}{2} e^{-2i \theta} \sigma_l^*\right)}{2V_0} g^l \right. \nonumber \\ & + & \left. V(g) g d_1 V_0 \left( \sum_{k=0}^{\infty} \frac{e^{i \theta} \alpha_k +e^{-i \theta} \alpha_k^* }{2 V_0} g^k\right)^2 \right\},
\end{eqnarray}
where $\theta$ is the direction of the vector ${\bf q}$ in the plane.
The constants ${\cal L}_k$, $\alpha_k$, $\delta_k$, $\sigma_k$ and $\mu_k^{(ab)}$ are independent of $\theta$ and $q$ and are defined
in what follows. First we define 
\begin{equation}
\label{lkdef}
{{\cal L}_k} \equiv \int d^2 y_1 \cdots d^2 y_{k+1} \; \left( \sum_{j=1}^k ( {\bf y}_j - {\bf y}_{j+1})^2 \right) f( {\bf y}_1) \phi ( {\bf y}_1, {\bf y}_2) f( {\bf y}_2) \times \cdots \times \phi ( {\bf y}_{k}, {\bf y}_{k+1}) f( {\bf y}_{k+1})
\end{equation}
and
\begin{eqnarray}
\label{alphadef}
\alpha_0 & \equiv & \int d^2 y_1 \; e^{-i \theta_1 } |{\bf y}_1 | f({\bf y}_1 ) \nonumber \\
\alpha_1 & \equiv & \int d^2 y_1 d^2 y_2 \; e^{-i \theta_1 } |{\bf y}_1 | f({\bf y}_1 ) \phi ( {\bf y}_1 , {\bf y}_2 ) f({\bf y}_2 ) \nonumber \\
\alpha_2 & \equiv & \int d^2 y_1 d^2 y_2 d^2 y_3 \; e^{-i \theta_1 } |{\bf y}_1 | f({\bf y}_1 ) \phi ( {\bf y}_1 , {\bf y}_2 ) f({\bf y}_2 ) \phi ( {\bf y}_2 , {\bf y}_3 ) f({\bf y}_3 ) \\
& \vdots & \nonumber
\end{eqnarray}
where $\theta_1$ denotes the direction of ${\bf y_1}$. The other constants are
\begin{eqnarray}
\label{deltadef}
\delta_0 & \equiv & \int d^2 y_1 \;  |{\bf y}_1 |^2 f({\bf y}_1 ) \nonumber \\
\delta_1 & \equiv & \int d^2 y_1 d^2 y_2 \; |{\bf y}_1 |^2 f({\bf y}_1 ) \phi ( {\bf y}_1 , {\bf y}_2 ) f({\bf y}_2 ) \nonumber \\
\delta_2 & \equiv & \int d^2 y_1 d^2 y_2 d^2 y_3 \; |{\bf y}_1 |^2 f({\bf y}_1 ) \phi ( {\bf y}_1 , {\bf y}_2 ) f({\bf y}_2 ) \phi ( {\bf y}_2 , {\bf y}_3 ) f({\bf y}_3 ) \\
& \vdots & \nonumber
\end{eqnarray}
\begin{eqnarray}
\label{sigmadef}
\sigma_0 & \equiv & \int d^2 y_1 \;  e^{- 2 i \theta_1} |{\bf y}_1 |^2 f({\bf y}_1) \nonumber \\
\sigma_1 & \equiv & \int d^2 y_1 d^2 y_2 \; e^{- 2 i \theta_1} |{\bf y}_1 |^2 f({\bf y}_1 ) \phi ( {\bf y}_1 , {\bf y}_2 ) f({\bf y}_2 ) \nonumber \\
\sigma_2 & \equiv & \int d^2 y_1 d^2 y_2 d^2 y_3 \; e^{- 2 i \theta_1} |{\bf y}_1|^2 f({\bf y}_1 ) \phi ( {\bf y}_1 , {\bf y}_2 ) f({\bf y}_2 ) \phi ( {\bf y}_2 , {\bf y}_3 ) f({\bf y}_3 ) \\
& \vdots & \nonumber
\end{eqnarray}
and
\begin{eqnarray}
\mu_0^{(++)} & \equiv & \int d^2 y_1 \; e^{-i \theta_1 } |{\bf y}_1 | f({\bf y}_1) |{\bf y}_1 |e^{-i \theta_1 } \nonumber \\
\mu_1^{(++)} & \equiv & \int d^2 y_1 d^2 y_2 \; e^{-i \theta_1 } |{\bf y}_1 | f({\bf y}_1 ) \phi ( {\bf y}_1 , {\bf y}_2 ) f({\bf y}_2 ) |{\bf y}_2 |e^{-i \theta_2 }  \nonumber \\
\mu_1^{(+-)} & \equiv & \int d^2 y_1 d^2 y_2 \; e^{-i \theta_1 } |{\bf y}_1 | f({\bf y}_1 ) \phi ( {\bf y}_1 , {\bf y}_2 ) f({\bf y}_2 ) |{\bf y}_2 |e^{+i \theta_2 }  \nonumber \\
\mu_2^{(++)} & \equiv & \int d^2 y_1 d^2 y_2 d^2 y_3 \; e^{-i \theta_1 } |{\bf y}_1 | f({\bf y}_1 ) \phi ( {\bf y}_1 , {\bf y}_2 ) f({\bf y}_2 ) \phi ( {\bf y}_2 , {\bf y}_3 ) f({\bf y}_3 ) |{\bf y}_3 |  e^{-i \theta_3 } \\
\mu_2^{(--)} & \equiv & \int d^2 y_1 d^2 y_2 d^2 y_3 \; e^{i \theta_1 } |{\bf y}_1 | f({\bf y}_1 ) \phi ( {\bf y}_1 , {\bf y}_2 ) f({\bf y}_2 ) \phi ( {\bf y}_2 , {\bf y}_3 ) f({\bf y}_3 ) |{\bf y}_3 |  e^{i \theta_3 } \nonumber \\
& \vdots & \nonumber
\end{eqnarray}
where the lower index specifies the number of $\phi$'s in the integrals and the upper index is the opposite of the sign of the angles in the exponents. Note that
\begin{equation}
\mu_k^{(-+)} = {\mu_k^{(+-)}}^* \;\;\; , \;\;\; \mu_k^{(--)} = {\mu_k^{(++)}}^*
\end{equation}

In addition to these definitions, for future use, it is also convenient to define the sums
\begin{eqnarray}
\label{sumsdef}
\alpha (g) \equiv \sum_{k=0}^{\infty} \alpha_k g^k \nonumber \\
\delta (g) \equiv \sum_{k=0}^{\infty} \delta_k g^k \nonumber \\
\sigma (g) \equiv \sum_{k=0}^{\infty} \sigma_k g^k \\
\mu^{(ab)} (g) \equiv \sum_{k=0}^{\infty} \mu_k^{(ab)} g^k \nonumber .
\end{eqnarray}

\subsection{Expansion of $h({\bf k},{\bf q})$ for $ka$ and $qa \ll 1$}

The most convenient way to obtain the expansion of $h({\bf k},{\bf q})$ for small $ka$ and $qa$
is by iterating~(\ref{noyce2}) and inserting resolutions of
the identity (in coordinate representation) where appropriate.
Note that by definition~(\ref{dunno}), $h({\bf q},{\bf q})=1$.
Changing the integration variables to ${\bf y}_i = \frac{{\bf r}_i}{a}$
results in:
\begin{eqnarray}
\label{simpleh}
h ({\bf k},{\bf q}) & = & \int d^2 y_0 e^{- i a ({\bf k - q}) \cdot {\bf y}_0} \frac{f ( {\bf y}_0)}{V_0} +  g \int d^2 y_0 d^2 y_1 \left( \rule{0mm}{7mm} e^{- i a {\bf k} \cdot {\bf y}_0} f ({\bf y}_0) G_0 (ka ; {\bf y}_0, {\bf y}_1) \right. \nonumber \\ & - & \left. \int d^2 y_1^{(a)} \; \frac{f({\bf y}_0) f({\bf y}_1^{(a)})}{V_0} e^{i a ({\bf q-k}) \cdot {\bf y}_0} e^{- i a {\bf q} \cdot {\bf y}_1^{(a)}} G_0 (ka; {\bf y}_1^{(a)} , {\bf y}_1) \right)  \frac{f ( {\bf y}_1)}{V_0} e^{i a {\bf q} \cdot {\bf y}_1} \nonumber \\ & + & g^2 \int d^2 y_0 d^2 y_1 d^2 y_2 \left( e^{- i a {\bf k} \cdot {\bf y}_0} f ({\bf y}_0) G_0 (ka ; {\bf y}_0, {\bf y}_1) - \int d^2 y_1^{(a)} \frac{f({\bf y}_0) f({\bf y}_1^{(a))}}{V_0} e^{i a ({\bf q-k}) \cdot {\bf y}_0} \right. \nonumber \\ & \times & \left. e^{- i a {\bf q} \cdot {\bf y}_1^{(a)}} G_0 (ka; {\bf y}_1^{(a)} , {\bf y}_1) \rule{0mm}{7mm} \right) \nonumber \\ & \times & \left( f ({\bf y}_1) G_0 (ka; {\bf y}_1,{\bf y}_2) - \int d^2 y_2^{(a)} \frac{f({\bf y}_1 ) f({\bf y}_2^{(a)})}{V_0} e^{i a {\bf q} \cdot ({\bf y}_1 - {\bf y}_2^{(a)})} G_0(ka; {\bf y}_2^{(a)},{\bf y}_2) \right)  \frac{f ( {\bf y}_2)}{V_0} e^{ia {\bf q} \cdot {\bf y}_2} \nonumber \\ & + & \cdots
\end{eqnarray}
The general term in this series is composed of a first factor 
\begin{equation}
A_0 ({\bf y}_0,{\bf y}_1)=e^{- i a {\bf k} \cdot {\bf y}_0} f ({\bf y}_0) G_0 (ka ; {\bf y}_0, {\bf y}_1) - \int d^2 y_1^{(a)} \frac{f({\bf y}_0) f({\bf y}_1^{(a)})}{V_0} e^{i a ({\bf q-k}) \cdot {\bf y}_0} e^{- i a {\bf q} \cdot {\bf y}_1^{(a)}} G_0 (ka; {\bf y}_1^{(a)} , {\bf y}_1)
\end{equation}
 followed by several factors of the type $B_j$ defined by~(\ref{defbj}), and a last factor  
\begin{equation}
A_N=e^{i a {\bf q} \cdot {\bf y}_N} \frac{f({\bf y}_N)}{V_0}
\end{equation}
 (with all variables integrated). 
To compute the $ka$ and $qa$ dependence of $h({\bf k},{\bf q})$ every term in equation~(\ref{simpleh})
is expanded for small $ka$ ($qa$) and the whole series in $g$ is summed.
In what follows $h({\bf k},{\bf q})$ will be computed to order $k^2 a^2$ (and also $q^2 a^2$, $kq a^2$). 

To compute $h({\bf k},{\bf q})$  for small $ka$ the first factor
has to be calculated at least to the first order $ka$. In this order
\begin{eqnarray}
\label{factorh}
I({\bf y}_1) & = & \int d^2 y_0 A_0 ({\bf y}_0,{\bf y}_1) \nonumber \\ & \simeq & \int d^2 y_0 \; i a ({\bf q- k }) \cdot {\bf y}_0 \left( f( {\bf y}_0) \phi ({\bf y}_0,{\bf y}_1) -\int d^2 y_1^{(a)} \frac{ f({ \bf y}_0) f({\bf y}_1^{(a)})}{V_0} \phi ( {\bf y}_1^{(a)},{\bf y}_1) \right).
\end{eqnarray} 
Note that in the leading (zeroth) order in $ka$ this contribution vanishes. This is related to the fact that by definition $h({\bf q},{\bf q}) = 1$, and in the leading order the direction of the momentum is not important.
To compute $h({\bf k},{\bf q})$ up to order $ka$ the factors $B_j$ and $A_N$ can be taken in the leading order (zeroth order in $ka$).
 Using~(\ref{leadingbrackets}) and~(\ref{factorh})
one finds
\begin{eqnarray}
\label{leadh}
h ( {\bf k},{\bf q}) & = & 1 + i a \left\{ \int d^2 y_0 \; ({\bf q- k}) \cdot {\bf y}_0 \frac{f({\bf y}_0)}{V_0} \right. \nonumber \\ 
 & + & g \int d^2 y_0 d^2 y_1 \; ({\bf q- k}) \cdot {\bf y}_0 \left( f({\bf y}_0) \phi ({\bf y}_0,{\bf y}_1) - \int d^2 y_1^{(a)} \frac{f({\bf y}_0) f({\bf y}_1^{(a)})}{V_0} \phi({\bf y}_1^{(a)},{\bf y}_1) \right) \frac{f({\bf y}_1)}{V_0} \nonumber \\
& + & g^2 \int d^2 y_0 d^2 y_1 d^2 y_2 \; ({\bf q- k}) \cdot {\bf y}_0 \left( f({\bf y}_0) \phi ({\bf y}_0 ,{\bf y}_1 ) - \int d^2 y_1^{(a)} \frac{f({\bf y}_0 ) f({\bf y}_1^{(a)})}{V_0} \phi({\bf y}_1^{(a)},{\bf y}_1 ) \right) \nonumber \\
& \times &  \left. \left( f({\bf y}_1 ) \phi ({\bf y}_1,{\bf y}_2 ) - \int d^2 y_2^{(a)} \frac{f({\bf y}_1 ) f({\bf y}_2^{(a)})}{V_0} \phi({\bf y}_2^{(a)},{\bf y}_2) \right) \frac{f({\bf y}_2)}{V_0} + \cdots  \right\}
\end{eqnarray}
Since ${\bf k} \cdot {\bf y}_0 = k |{\bf y}_0| \cos (\theta ' - \theta_0)$ 
(and ${\bf q} \cdot {\bf y}_0 = q |{\bf y}_0| \cos (\theta  - \theta_0)$)
the dependence on the incoming and outgoing momentum can be separated
from the integrals in~(\ref{leadh}). Use of~(\ref{alphadef}) results in
\begin{eqnarray}
h({\bf k},{\bf q}) & \simeq & 1+i\frac{ka}{2} \left\{ \left( e^{i \theta} - e^{i \theta'} \right) \frac{\alpha_0}{V_0} + \left( e^{-i \theta} - e^{-i \theta'} \right) \frac{\alpha_0^*}{V_0} \right. \nonumber \\
& + & g \left[  \left( e^{i \theta} - e^{i \theta'} \right) \frac{\alpha_1}{V_0} + \left( e^{-i \theta} - e^{-i \theta'} \right) \frac{\alpha_1^*}{V_0} -\frac{L_1}{V_0} \left( e^{i \theta} - e^{i \theta'} \right) \frac{\alpha_0}{V_0} -\frac{L_1}{V_0} \left( e^{-i \theta} - e^{-i \theta'} \right) \frac{\alpha_0^*}{V_0} \right] \nonumber \\ 
& + & g^2 \left[ \left( e^{i \theta} - e^{i \theta'} \right) \frac{\alpha_2}{V_0}+\left( e^{-i \theta} - e^{-i \theta'} \right) \frac{\alpha_2^*}{V_0} -\frac{L_1}{V_0} \left( e^{i \theta} - e^{i \theta'} \right) \frac{\alpha_1}{V_0} -\frac{L_1}{V_0} \left( e^{-i \theta} - e^{-i \theta'} \right) \frac{\alpha_1^*}{V_1} \right. \nonumber \\
& -& \left. \left. \left( \frac{L_2}{V_0} - \frac{L_1^2}{V_0^2} \right) \left( e^{i \theta} - e^{i \theta'} \right) \frac{\alpha_0}{V_0} - \left( \frac{L_2}{V_0} - \frac{L_1^2}{V_0^2} \right) \left( e^{-i \theta} - e^{-i \theta'} \right) \frac{\alpha_0^*}{V_0} \right] + \cdots \right\}.
\end{eqnarray} 
First note that $\alpha_0$ appears with the series $1-g\frac{L_1}{V_0} -g^2 \left( \frac{L_2}{V_0} - \frac{L_1^2}{V_0^2} \right) - \cdots = V(g)$. This results from the fact that to get the $\alpha_0$ contribution from the second term in the first
brackets in~(\ref{leadh}) it must be taken in all orders of $g^N$. Then the integrals from ${\bf y_1^{(a)}}$ to ${\bf y_N}$ behave exactly as the leading order (in $qa$) contribution to $D_g({\bf q})$.
Similar considerations can be applied to each of the $\alpha_j$.
This series (in $g$) is summed to give
\begin{eqnarray}
h({\bf k},{\bf q}) & \simeq & 1 + i \frac{ka}{2} V(g) \left\{ \left( e^{i \theta} - e^{i \theta '} \right) \sum_{k=0}^{\infty} \frac{\alpha_k}{V_0} g^k + \left( e^{-i \theta} - e^{-i \theta '} \right) \sum_{k=0}^{\infty} \frac{\alpha_k^*}{V_0} g^k \right\} \nonumber \\
& = & 1 + i \frac{ka}{2} V(g) \left\{ \left( e^{i \theta} - e^{i \theta '} \right) \frac{\alpha (g)}{V_0} + \left( e^{-i \theta} - e^{-i \theta '} \right) \frac{\alpha^* (g)}{V_0} \right\}
\end{eqnarray} 
with $\alpha (g)$ defined by~(\ref{sumsdef}).

The computation of the contribution to $h({\bf k},{\bf q})$ of order $k^2 a^2$ (with additional
powers of $\ln ka$) follows the same lines. However, it is too tedious 
to be included here and only the result is given,
\begin{eqnarray}
\label{thefinalh}
h({\bf k},{\bf q}) & \simeq &1 + i \frac{ka}{2} V(g) \left\{ \left( e^{i \theta} - e^{i \theta '} \right) \frac{\alpha (g)}{V_0} + \left( e^{-i \theta} - e^{-i \theta '} \right) \frac{\alpha^* (g)}{V_0} \right\}  +  k^2 a^2 \left\{ \left( V(g) - g d_1 V_0 \right) \left. \frac{}{} \right. \right. \nonumber \\ 
& \times & \sum_{l=1}^{\infty} \frac{g^l}{4 V_0} \left( e^{i \theta} (e^{i \theta'}-e^{i \theta}) \mu_l^{(++)} + e^{-i \theta} (e^{i \theta'}-e^{i \theta}) \mu_l^{(+-)} + e^{i \theta} (e^{-i \theta'}-e^{-i \theta}) \mu_l^{(-+)} \right. \nonumber \\
& + & \left. e^{-i \theta} (e^{-i \theta'}-e^{-i \theta}) \mu_l^{(--)}\right) + \frac{1}{2 V_0} \left(g V_0 d_1 - V(g) \right) \left[ \delta_0 + \frac{1}{2} e^{2i\theta} \sigma_0+ \frac{1}{2} e^{-2i\theta} \sigma_0^* \right. \nonumber \\
& - & \left. \frac{1}{2} \left[ e^{i(\theta + \theta')} \mu_0^{(++)}+e^{i(-\theta + \theta')} \mu_0^{(+-)}+e^{i(\theta - \theta')} \mu_0^{(-+)}+e^{-i(\theta + \theta')} \mu_0^{(--)}\right] \right] \nonumber \\
& + &  \frac{1}{8} V(g) \left[ \left( e^{2 i \theta} -e^{2 i \theta'} \right) \frac{\sigma (g)}{V_0} +  \left( e^{-2 i \theta} -e^{-2 i \theta'} \right) \frac{\sigma^* (g)}{V_0} \right] \nonumber \\ 
& + & \left. V(g) g d_1 V_0 \left( \left( e^{i \theta'} - e^{i \theta} \right) \frac{\alpha (g)}{2 V_0} + \left( e^{-i \theta'} - e^{-i \theta} \right) \frac{\alpha^* (g)}{2 V_0} \right) \left( e^{i \theta} \frac{\alpha(g)}{2 V_0} + e^{-i \theta} \frac{\alpha^* (g)}{2 V_0} \right) \right\}
\end{eqnarray} 

\subsection{The angular dependence of $D(\theta' , \theta)$}

Finally the diffraction constant of~(\ref{defd}) is computed to the order
$k^2 a^2$ with the help of~(\ref{tnoyce}). It is useful to 
expand~(\ref{lastdg}) as:
\begin{eqnarray}
\label{overdg}
\frac{1}{D_g} & \simeq & \frac{1}{V(g)-g V_0 d_1 + k^2 a^2 Q(g)}
\left[ 1 + k^2 a^2 \left\{ \rule{0mm}{8mm} \left(V(g) - g V_0 d_1 \right) \sum_{l=1}^{\infty} \frac{g^l}{4 V_0} \left( e^{2i\theta} \mu_l^{(++)} +\mu_l^{(+-)} +\mu_l^{(-+)} \right. \right. \right. \nonumber \\ 
& + & \left. \left. e^{-2 i \theta} \mu_l^{(--)} \right) - \frac{g d_1}{2} \left(\delta_0 + \frac{1}{2} e^{2 i \theta} \sigma_0 + \frac{1}{2} e^{-2i \theta} \sigma_0^* \right) - V(g) \sum_{l=1}^{\infty} \frac{\left( \delta_l + \frac{1}{2} e^{2 i \theta}\sigma_l + \frac{1}{2} e^{-2i \theta} \sigma_l^*\right)}{2V_0} g^l \right. \nonumber \\ & + & \left. \left. V(g) g d_1 V_0 \left( \sum_{k=0}^{\infty} \frac{e^{i \theta} \alpha_k +e^{-i \theta} \alpha_k^* }{2 V_0} g^k\right)^2 +\frac{1}{2} V(g) \left( \frac{\delta (g)}{V_0} - V(g) \frac{|\alpha (g)|^2}{V_0^2}\right) \right\} \right]
\end{eqnarray}
where $Q(g)$ is defined by (\ref{defqg}).
Substituting~(\ref{thefinalh}) and (\ref{overdg}) in~(\ref{tnoyce}) leads to~(\ref{desired}).


\section{Calculation of angular dependent contributions of the form factor}
\label{notneeded}

In this Appendix the details of calculations leading to some of the results
of Sec.~\ref{angles} are presented.

\subsection{Diagonal contributions to the form factor for a scatterer at the center \label{appda}} 
For odd-odd orbits the incoming direction
is given by $\pi+\theta$ where $\theta$ is the outgoing direction~(\ref{inangle}).
The contributions from the odd-odd orbits in~(\ref{formparts}) is replaced 
by
\begin{eqnarray}
\label{k1second}
K^{(1,o-o)} (\tau)& = &  \frac{\tau^2}{32 \pi} \int_0^{\frac{\pi}{2}} \! d \theta \; \left| D(\theta,\theta+\pi) + D (\theta+\pi,\theta) +D(\pi-\theta,2\pi-\theta)+D(2\pi-\theta,\pi-\theta) \right|^2 \\
& = &  \frac{\tau^2}{32 \pi} \int_0^{2 \pi} \! d \theta \; D^* (\theta,\theta \!+ \!\pi) \left[ D(\theta,\theta \!+ \! \pi) + D (\theta \! + \! \pi,\theta) +D(\pi \! - \! \theta,2\pi \! - \! \theta)+D(2\pi \! -\! \theta,\pi \! -\! \theta)\right] \nonumber .
\end{eqnarray}
The contribution from the odd-even orbits is given by
\begin{eqnarray}
\label{k1third}
K^{(1,o-e)} (\tau)& = & \frac{\tau^2}{32 \pi} \int_0^{\frac{\pi}{2}} \! d \theta \; \left| D(\theta,\pi-\theta) + D (\pi-\theta,\theta) +D(\pi+\theta,2\pi-\theta)+D(2\pi-\theta,\pi+\theta) \right|^2  \\
& = & \frac{\tau^2}{32 \pi} \int_0^{2 \pi} \! d \theta \, D^* (\theta,\pi-\theta) \left[ D(\theta,\pi \! - \! \theta) + D (\pi \! -\! \theta,\theta) +D(\pi \! +\! \theta,2\pi \! - \! \theta)+D(2\pi \! -\! \theta,\pi \! +\! \theta)\right] \nonumber .
\end{eqnarray}
The contribution from the even-odd orbits is
\begin{eqnarray}
\label{k1forth}
K^{(1,e-o)} (\tau)& = & \frac{\tau^2}{32 \pi} \int_0^{\frac{\pi}{2}} \! d \theta \; \left| D(\theta,2\pi-\theta) + D (2\pi-\theta,\theta) +D(\pi+\theta,\pi-\theta)+D(\pi-\theta,\pi+\theta) \right|^2  \\
& = & \frac{\tau^2}{32 \pi} \int_0^{2 \pi} \! d \theta \, D^* (\theta,2\pi \! -\! \theta) \left[ D(\theta,2\pi \! -\! \theta) + D (2\pi \!- \! \theta,\theta) +D(\pi \! +\! \theta,\pi \! -\! \theta)+D(\pi \! -\! \theta,\pi \! +\! \theta)\right] \nonumber .
\end{eqnarray}
The diagonal contribution of periodic and diffracting orbits to the form factor (up to order $\tau^2$) is obtained by summation of~(\ref{k1first}) and (\ref{k1second}-\ref{k1forth}), resulting in
\begin{eqnarray}
\label{centdiagang}
K^{(1)}(\tau) & = & 1 + \frac{\tau}{2 \pi} \Im \int_0^{2 \pi} d \theta \; D(\theta,\theta) \nonumber \\ & + & \frac{\tau^2}{32 \pi} \int_0^{2 \pi} d \theta \;\left\{ D^{*} (\theta,\theta) \left[ D(\theta,\theta)+ D(\pi-\theta,\pi-\theta)+ D(\pi+\theta,\pi+\theta)+ D(2\pi-\theta,2\pi-\theta) \right] \right. \nonumber \\
& + &  D^* (\theta,\theta+\pi) \left[ D(\theta,\theta+\pi) + D (\theta+\pi,\theta) +D(\pi-\theta,2\pi-\theta)+D(2\pi-\theta,\pi-\theta)\right] \nonumber \\
& + &  D^* (\theta,\pi-\theta) \left[ D(\theta,\pi-\theta) + D (\pi-\theta,\theta) +D(\pi+\theta,2\pi-\theta)+D(2\pi-\theta,\pi+\theta)\right] \nonumber \\
& + & \left.  D^* (\theta,2\pi-\theta) \left[ D(\theta,2\pi-\theta) + D (2\pi-\theta,\theta) +D(\pi+\theta,\pi-\theta)+D(\pi-\theta,\pi+\theta)\right] \right\}.
\end{eqnarray}
Substitution of~(\ref{desired}) yields~(\ref{centshort}).

\subsection{Non diagonal contributions to the form factor for a scatterer at the center \label{appdb}} 

For twice diffracting orbits the diffraction constants appear in the amplitude
in the form
\begin{equation}
\label{d2inamp}
\frac{1}{16} \sum_{\mu \nu} D(\theta_{1,\mu},\Theta(\theta_{2,\nu})) D(\theta_{2,\nu},\Theta(\theta_{1,\mu}))
\end{equation}
instead of $D^2$ that was used for angle independent scattering.
The functions $\Theta$ of (\ref{inangle}) depend on the parity of ${\bf N}_1$ and ${\bf N}_2$
but since on the saddle manifold ${\bf N}_1+{\bf N}_2={\bf N}_p$, where
the indices of ${\bf N}_p$, corresponding to a periodic orbit, are all even. Therefore both ${\bf N}_1$ and ${\bf N}_2$ have the same parity and for orbits that will contribute both the functions
$\Theta$ are identical.

The computation of the contribution of the saddle manifold can be done exactly as in Eq.~(\ref{orbitcomb})-(\ref{form2diff}). Since in the semiclassical limit the angles of orbits which contribute are given by $\theta_p + \delta \theta$ where $\theta_p$ is
the direction of the periodic orbit and $\delta \theta \ll 1$, the expression~(\ref{d2inamp}) can be taken out of the saddle integral because it changes slowly on the saddle manifold. The various non diagonal contributions from any type (even-even, even-odd, etc.) of twice-diffracting orbits and periodic orbits are given by:
\begin{equation}
\label{nondangle}
K^{'(2,x-x)} (\tau) = - \Re \frac{\tau^2}{4} \frac{2}{\pi} \int_0^{\frac{\pi}{2}} d \theta \; \frac{1}{16} \sum_{\mu \nu} D(\theta_{\mu},\Theta(\theta_{\nu})) D(\theta_{\nu},\Theta(\theta_{\mu})),
\end{equation}
where $x-x$ denotes the parity of ${\bf N}_1$ (and also of ${\bf N}_2$), that is, of the segments of the twice diffracting orbits. 

For even-even orbits, from~(\ref{inangle}) one finds $\Theta(\theta)=\theta$. For each segment that starts in the
direction $\theta$ where $0 \le \theta < \frac{\pi}{2}$ (or $\mu=1$) there are also segments of identical length in the directions $\pi - \theta$ ($\mu=2$),
$\pi+\theta$ ($\mu=3$) and $2\pi-\theta$ ($\mu=4$).
Therefore there are $16$ contributions to the sum in~(\ref{nondangle}), 
\begin{eqnarray}
\label{nondee}
K^{'(2,e-e)} (\tau)& = & - \Re \frac{\tau^2}{32 \pi} \int_0^{\frac{\pi}{2}} \left\{ D(\theta,\theta) D(\theta,\theta) + D(\theta,\pi- \theta) D(\pi- \theta,\theta)+D(\theta,\pi+ \theta) D(\pi+\theta,\theta) \right. \nonumber \\
& + & D(\theta,2\pi-\theta) D(2\pi-\theta,\theta)+D(\pi- \theta,\theta) D(\theta,\pi- \theta) +  D(\pi-\theta,\pi-\theta) D(\pi-\theta,\pi-\theta) \nonumber \\
& + &  D(\pi-\theta,\pi+\theta) D(\pi+\theta,\pi-\theta) + D(\pi-\theta,2\pi-\theta) D(2\pi-\theta,\pi-\theta) \nonumber \\
& + &  D(\pi+\theta,\theta) D(\theta,\pi+\theta) +  D(\pi+\theta,\pi-\theta) D(\pi-\theta,\pi+\theta) \nonumber \\
& + & D(\pi+\theta,\pi+\theta) D(\pi+\theta,\pi+\theta) + D(\pi+\theta,2\pi-\theta) D(2\pi-\theta,\pi+\theta) \nonumber \\
& + & D(2\pi-\theta,\theta) D(\theta,2\pi-\theta)+ D(2\pi-\theta,\pi-\theta) D(\pi-\theta,2\pi-\theta) \nonumber \\ 
& + & \left. D(2\pi-\theta,\pi+\theta) D(\pi+\theta,2\pi-\theta) +  D(2\pi-\theta,2\pi-\theta) D(2\pi-\theta,2\pi-\theta) \right\} \nonumber \\
& = & - \frac{\tau^2}{32 \pi} \Re \int_0^{2\pi} d \theta \; \left\{ D(\theta,\theta) D(\theta,\theta) + D(\theta,\pi-\theta) D(\pi-\theta,\theta) + D(\theta,\pi+\theta) D(\pi+\theta,\theta) \right. \nonumber \\
& + & \left.  D(\theta,2\pi-\theta) D(2\pi-\theta,\theta) \right\}.
\end{eqnarray}
The calculation for diffracting segments with other parities is similar.
For the odd-odd orbits $\Theta (\theta)= \pi+\theta$. After a similar calculation
the non diagonal contribution from these orbits to the form factor is found to be:
\begin{eqnarray}
\label{nondoo}
K^{'(2.o-o)} (\tau) & = & - \frac{\tau^2}{32 \pi} \Re \int_0^{2\pi} d \theta \; \left\{ D(\theta,\pi+\theta) D(\theta,\pi+\theta) + D(\theta,2\pi-\theta) D(\pi-\theta,\pi+\theta) \right. \nonumber \\
& + & \left. D(\theta,\theta) D (\pi+\theta, \pi+\theta) + D(\theta,\pi-\theta) D(2\pi-\theta,\pi+\theta) \right\}.
\end{eqnarray}
For orbits of the odd-even type, $\Theta(\theta)=\pi-\theta$ and their contribution is given by
\begin{eqnarray}
\label{nondoe}
K^{'(2,o-e)} (\tau) & = & - \frac{\tau^2}{32 \pi} \Re \int_0^{2\pi} d \theta \; \left\{ D(\theta,\pi-\theta) D(\theta,\pi-\theta) + D(\theta,\theta) D(\pi-\theta,\pi-\theta) \right. \nonumber \\
& + & \left. D(\theta,2\pi-\theta) D(\pi+\theta,\pi-\theta) + D(\theta,\pi+\theta) D(2\pi-\theta,\pi-\theta) \right\}.
\end{eqnarray}
And, finally, for the even-odd orbits, $\Theta(\theta) = 2\pi-\theta$ and
\begin{eqnarray}
\label{nondeo}
K^{'(2,e-o)} (\tau) & = & - \frac{\tau^2}{32 \pi} \Re \int_0^{2\pi} d \theta \; \left\{ D(\theta,2\pi-\theta) D(\theta,2\pi-\theta) + D(\theta,\theta+\pi) D(\pi-\theta,2\pi-\theta) \right. \nonumber \\
& + & \left. D(\theta,\pi-\theta) D(\pi+\theta,2\pi-\theta) + D(\theta,\theta) D(2\pi-\theta,2\pi-\theta) \right\}.
\end{eqnarray}

The total non diagonal contribution to the form factor is obtained when~(\ref{nondee})-(\ref{nondeo}) are summed leading to
\begin{eqnarray}
\label{centnondangle}
K^{'(2)} (\tau) & = &  - \frac{\tau^2}{32 \pi} \Re \int_0^{2\pi} \! d \theta \, \left[ D(\theta,\theta) \left\{ D(\theta,\theta) + D(\pi \!-\!\theta,\pi \! -\! \theta)+   D(\pi \! +\! \theta,\pi \! +\! \theta)+ D(2\pi \! -\! \theta,2\pi \! -\! \theta) \right\} \right. \nonumber \\
& + & D(\theta,\pi-\theta) \left\{  D(\theta,\pi-\theta)+ D(\pi-\theta,\theta)+ D(\pi+\theta,2\pi-\theta)+  D(2\pi-\theta,\pi+\theta) \right\} \nonumber \\
& + & D(\theta,\pi+\theta) \left\{ D(\theta,\pi+\theta)+D(\pi+\theta,\theta)+D(\pi-\theta,2\pi-\theta)+ D(2\pi-\theta,\pi-\theta) \right\} \nonumber \\
& + & \left. D(\theta,2\pi-\theta) \left\{ D(\theta,2\pi-\theta)+D(2\pi-\theta,\theta)+D(\pi+\theta,\pi-\theta)+D(\pi-\theta,\pi+\theta) \right\} \right].
\end{eqnarray}
Substitution of~(\ref{desired}) leads to~(\ref{nondshort}).

\subsection{Diagonal contributions to the form factor for a scatterer at a typical location \label{appdc}} 

Odd-odd orbits with different
indices $\mu$ differ in length and for every $\mu$  
there is a separate contribution to the diagonal part. The sum of these four
contributions is given by
\begin{eqnarray}
\label{noncoo1}
K^{(1,o-o)} (\tau) & = & \frac{\tau^2}{32 \pi} \int_0^{\frac{\pi}{2}} \! d \theta \; \left( |D(\theta,\theta \! +\! \pi)|^2 + |D(\pi\! -\! \theta,2\pi\! -\! \theta)|^2 + |D(\theta\! +\! \pi,\theta)|^2 + |D(2\pi \! -\! \theta,\pi\! -\! \theta)|^2 \right) \nonumber \\
& = &  \frac{\tau^2}{32 \pi} \int_0^{2 \pi} d \theta \; | D(\theta,\theta+\pi)|^2.
\end{eqnarray}
An orbit of the odd-even type which starts at the direction $\theta$ will
return to the scatterer at the direction $\pi-\theta$. Its length is identical
to the length of its time reversal which starts at the direction $2\pi-\theta$
and returns with $\pi+\theta$. Thus orbits of this type with $\mu=1$ and 
$\mu=4$ (with the same index ${\bf N}_j$) have the same length. This is also
true for orbits with $\mu=2$ and $\mu=3$. These length degeneracies lead
to a diagonal contribution to the form factor which is given by
\begin{eqnarray}
\label{noncoe1}
K^{(1,o-e)} (\tau) & = & \frac{\tau^2}{32 \pi} \int_0^{\frac{\pi}{2}} d \theta \; \left( |D(\theta,\pi-\theta)+D(2\pi-\theta,\pi+\theta)|^2 + |D(\pi-\theta,\theta)+ D(\pi+\theta,2\pi-\theta)|^2 \right) \nonumber \\
& = & \frac{\tau^2}{32 \pi} \int_0^{2\pi} d \theta \; D^* (\theta,\pi-\theta) \left[ D(\theta,\pi-\theta) + D(2\pi-\theta, \pi+\theta) \right]. 
\end{eqnarray}
The contribution of the even-odd orbits is very similar to the one of the odd-even ones. The difference is that orbits which start at direction $\theta$ return at the direction $2\pi-\theta$ and their time reversal starts at $\pi-\theta$. Therefore the lengths of orbits with $\mu=1$ and $\mu=2$ (and the same index ${\bf N}_j$) are identical and
it is also true for orbits with $\mu=3$ and $\mu=4$. Their contribution to the
diagonal form factor is given by
\begin{eqnarray}
\label{nonceo1}
K^{(1,e-o)} (\tau) & = & \frac{\tau^2}{32 \pi} \int_0^{\frac{\pi}{2}} d \theta \; \left( |D(\theta,2\pi-\theta)+D(\pi-\theta,\pi+\theta)|^2 + |D(2\pi-\theta,\theta)+ D(\pi+\theta,\pi-\theta)|^2 \right) \nonumber \\
& = & \frac{\tau^2}{32 \pi} \int_0^{2\pi} d \theta \; D^* (\theta,2\pi-\theta) \left[ D(\theta,2\pi-\theta) + D(\pi-\theta, \pi+\theta) \right]. 
\end{eqnarray}
The diagonal contribution to the form factor is obtained when~(\ref{k1first}), (\ref{noncoo1}), (\ref{noncoe1}) and (\ref{nonceo1}) are summed,
resulting in
\begin{eqnarray}
\label{noncdiagand}
K^{(1)} (\tau) & = & 1 + \frac{\tau}{2 \pi} \Im \int_0^{2 \pi} d \theta \; D (\theta,\theta) \nonumber \\
& + &  \frac{\tau^2}{32 \pi} \int_0^{2\pi} d \theta \; \left\{ D^*(\theta,\theta) \left[ D(\theta,\theta) + D(\pi-\theta,\pi-\theta) + D(\pi+\theta,\pi+\theta) + D(2\pi-\theta,2\pi-\theta) \right] \right. \nonumber \\
& + & |D(\theta,\pi+\theta)|^2 + D^* (\theta,\pi-\theta) \left[ D(\theta,\pi-\theta) + D(2\pi-\theta,\pi+\theta) \right] \nonumber \\
& + & \left. D^* (\theta,2\pi-\theta) \left[ D(\theta,2\pi-\theta)+ D(\pi-\theta,\pi+\theta) \right] \right\}.
\end{eqnarray}
Substitution of~(\ref{desired}) leads to (\ref{ncdiagshort}).

\subsection{Non diagonal contributions to the form factor for a scatterer at a typical location \label{appdd}} 

For odd-odd orbits if the first segment has $\mu=1$ then the second
segment must have $\mu=3$ to satisfy~(\ref{nonccond1}). Similarly
if the first segment has $\mu=2$ then only segments with $\mu=4$ can
contribute (and vice versa).
Therefore, instead of the $16$ contributions for~(\ref{nondoo}) when the
scatterer is at the center of the rectangle there are only $4$ contributions
for a scatterer at some typical position.
The resulting contribution to the form factor is given by
\begin{eqnarray}
\label{nondoononc}
K^{'(2,o-o)} (\tau) & = & - \frac{\tau^2}{32 \pi} \Re \int_0^{\frac{\pi}{2}} d \theta \; \left[ D(\theta,\theta) D(\theta+\pi,\theta+\pi) + D(\pi-\theta,\pi-\theta) D(2\pi-\theta,2\pi-\theta) \right. \nonumber \\
& + & \left. D(\theta+\pi,\theta+\pi) D(\theta,\theta) + D(2\pi-\theta,2\pi-\theta) D(\pi-\theta,\pi-\theta) \right] \nonumber \\
& = & - \frac{\tau^2}{32 \pi} \Re \int_0^{2\pi} d \theta \; D(\theta,\theta) D(\theta+\pi,\theta+\pi).
\end{eqnarray}
When both segments are of the odd-even type there are length degeneracies
between a segment and its time reversal. If the first segment has $\mu=1$
then the second segment can either have $\mu=3$ or $\mu=2$ leading to the
combinations of the form $D(\theta,\theta)D(\pi-\theta,\pi-\theta)+D(\theta,2\pi-\theta)D(\pi+\theta,\pi-\theta)$.
Similar contributions are obtained when the first segment has $\mu=2,3,4$.
Thus instead of the $16$ contributions leading to~(\ref{nondoe}) when the scatterer
is at the center there are only $8$ when the scatterer is at a typical
position. These contributions are given by
\begin{eqnarray}
\label{nondoenonc}
K^{'(2,o-e)} (\tau) & = & - \frac{\tau^2}{32 \pi} \Re \int_0^{\frac{\pi}{2}} d \theta \; \left[ D(\theta,\theta)D(\pi-\theta,\pi-\theta)+D(\theta,2\pi-\theta)D(\pi+\theta,\pi-\theta) \right. \nonumber \\
& + & D(\pi-\theta,\pi-\theta) D(\theta,\theta) + D(\pi-\theta,\pi+\theta) D(2\pi-\theta,\theta) \nonumber \\
& + & D(\pi+\theta,\pi+\theta)D(2\pi-\theta,2\pi-\theta) +  D(\pi+\theta,\pi-\theta) D(\theta,2\pi-\theta) \nonumber \\
& + &\left. D (2\pi-\theta,2\pi-\theta) D(\pi+\theta,\pi+\theta) + D(2\pi-\theta,\theta)D(\pi-\theta,\pi+\theta) \right] \nonumber \\
& = &  - \frac{\tau^2}{32 \pi} \Re \int_0^{2\pi} d \theta \; \left( D(\theta,\theta)D(\pi-\theta,\pi-\theta)+D(\theta,2\pi-\theta)D(\pi+\theta,\pi-\theta)\right).
\end{eqnarray} 
The contribution of the even-odd twice diffracting orbits can be computed 
in a similar manner. Here if the first segment has $\mu=1$ the second
is either $\mu=3$ or $\mu=4$, with the same length, resulting in the non diagonal contribution
to the form factor from these orbits, 
\begin{equation}
\label{nondeononc}
K^{'(2,e-o)} (\tau) =  - \frac{\tau^2}{32 \pi} \Re \int_0^{2\pi} d \theta \; \left( D(\theta,\theta)D(2\pi-\theta,2\pi-\theta)+D(\theta,\pi-\theta)D(\pi+\theta,2\pi-\theta)\right).
\end{equation}

The non diagonal contribution is obtained by
summing~(\ref{nondee}) and (\ref{nondoononc}-\ref{nondeononc}) resulting in
\begin{eqnarray}
\label{nonck2tauang}
K^{'(2)} (\tau) & = & - \frac{\tau^2}{32 \pi} \Re \int_0^{2\pi} d \theta \; \left\{ D(\theta,2\pi-\theta) \left[ D(\pi+\theta,\pi-\theta) + D(2\pi-\theta,\theta) \right] \right. \nonumber \\
& + & D(\theta,\theta) \left[  D(\theta,\theta)+D(\pi-\theta,\pi-\theta)+ D(\pi+\theta,\pi+\theta) + D(2\pi-\theta,2\pi-\theta) \right] \nonumber \\
& + & \left. D(\theta,\pi-\theta) \left[ D(\pi+\theta,2\pi-\theta)+D(\pi-\theta,\theta) \right] + D(\theta,\pi+\theta) D(\pi+\theta,\theta) \right\}.
\end{eqnarray}
Substitution of~(\ref{desired}) yields~(\ref{ncndshort}).


\section{Comparison to the paper of  Bogomolny and Giraud~\protect{\cite{BG}}}
\label{appbg}

In this Appendix results of the paper by Bogomolny and Giraud~\cite{BG}
are compared with the results of our
present work where appropriate.
 The notations
in this Appendix are the ones used in \cite{BG}.

Consider first the case of angle independent scattering.
For periodic boundary conditions the result of \cite{BG} is given
by (61) and (62) there. Expanding to order $\tau^3$ and using
\begin{equation}
A_{002} = 1
\end{equation}
leads to
\begin{equation}
\label{kpbc}
K_{pbc} (\tau) = 1 -\frac{1}{4} |D|^2 \tau + \frac{1}{32} |D|^4\tau^2 +\left( \frac{1}{32}|D|^4 - \frac{1}{384}|D|^6 \right) \tau^3.
\end{equation}
The form factor for periodic boundary conditions is related to the form factor
for (an angle independent) scatterer at the center by the relation~(\ref{scale}) of our paper.
Substituting~(\ref{kpbc}) for $K_{per}$ in (\ref{scale}) leads to the result obtained in this work for a 
scatterer at the center~(\ref{centerf}).

When the scatterer is at a typical position the results are given
by Eqs. (83), (84) and (85) of \cite{BG}.
Expanding to order $\tau^3$ one finds
\begin{equation}
\label{ktaur}
K(\tau) = 1 -\frac{1}{4} |D|^2 \langle r \rangle \tau + \frac{1}{32} |D|^4 \langle r^2 \rangle \tau^2 - \frac{1}{384} |D|^6 \langle r^3 \rangle \tau^3 + \frac{1}{32} A_{002} |D|^4 \tau^3. 
\end{equation}
Using
\begin{eqnarray}
\langle r \rangle & = & 1 \nonumber \\
\langle r^2 \rangle & = & \frac{9}{4} \nonumber \\
\langle r^3 \rangle & = & \frac{25}{4}
\end{eqnarray}
and 
\begin{equation}
A_{002} = \frac{81}{16}
\end{equation}
leads to an expression for the form factor
 that is identical to the one computed in our work and is given by~(\ref{typacalf}).

The form factor of an angle dependent scatterer at a typical position
is given by Eqs. (109) and (110) in \cite{BG}. It is sufficient to expand 
it to order $\tau^2$ for comparison with the form factor computed in our
work. 
Eq. (109) in \cite{BG} leads to
\begin{equation}
\label{ktaus}
K(\tau) = \langle \langle 1 + \frac{1}{4} \Im S(\varphi,\varphi) \tau + \frac{1}{64} |S(\varphi,\varphi)|^2 \tau^2 - \frac{1}{64} \Re S^2 (\varphi,\varphi) \tau^2 \rangle \rangle+O(\tau^3),
\end{equation}
where
\begin{eqnarray}
S(\varphi,\varphi) & = & D(\varphi,\varphi)+ D(\pi-\varphi,\pi-\varphi) + D(-\pi+\varphi,-\pi+\varphi)+D(-\varphi,-\varphi) \nonumber \\
 & - & \left[ D(\varphi,-\varphi) + D(\pi-\varphi,-\pi+\varphi) \right] e^{2i\phi_2} - \left[ D(-\varphi,\varphi)+ D(-\pi+\varphi,\pi-\varphi)\right] e^{-2i\phi_2}  \nonumber \\
& - & \left[ D(\varphi,\pi-\varphi) + D(-\varphi,-\pi+\varphi) \right] e^{2 i \phi_1} - \left[ D(-\pi+\varphi,-\varphi) + D(\pi-\varphi,\varphi) \right] e^{- 2 i \phi_1} \nonumber \\
& + & D(\varphi,-\pi+\varphi) e^{2 i (\phi_1+\phi_2)} + D(-\varphi,\pi-\varphi) e^{2i(\phi_1-\phi_2)} + D(-\pi+\varphi,\varphi) e^{-2i(\phi_1+\phi_2)} \nonumber \\
& + & D(\pi-\varphi,-\varphi) e^{2 i (\phi_2-\phi_1)} 
\end{eqnarray}
is obtained with the help of (88), (102) and (105) there.
The brackets
 $\langle \langle \cdots \rangle \rangle$ denote averaging 
over the uniformly distributed random variables $\phi_1,\phi_2$ 
(see the paragraph following (79) of \cite{BG}), and over
the angle $\varphi$ in the range $(0,\frac{\pi}{2})$.
The averaging over $\phi_i$ will leave only the terms without the exponentials.
We show that the terms in~(\ref{ktaus}) are identical to the contributions
to the form factor computed in our work.
For instance, the term linear in $\tau$ is proportional to
\begin{equation}
\label{linears}
\langle \langle S(\varphi,\varphi) \rangle \rangle = \frac{2}{\pi} \int_0^{\frac{\pi}{2}} d \varphi \; \left[ D(\varphi,\varphi)+ D(\pi-\varphi,\pi-\varphi) + D(-\pi+\varphi,-\pi+\varphi)+D(-\varphi,-\varphi) \right]
\end{equation}
which is identical to the contribution linear in $\tau$ in~(\ref{noncdiagand})
of our paper (see also~(\ref{forwardD})).
The term resulting from $|S|^2$ is explicitly given by
\begin{eqnarray}
\frac{1}{64} \langle \langle |S(\varphi,\varphi)|^2 \rangle \rangle & = & \frac{1}{32 \pi} \int_0^{\frac{\pi}{2}} \! d \varphi \, \left( |D(\varphi,\varphi)+ D(\pi-\varphi,\pi-\varphi) + D(-\pi+\varphi,-\pi+\varphi)+D(-\varphi,-\varphi)|^2 \right. \nonumber \\
& + &  |D(\varphi,-\varphi) + D(\pi-\varphi,-\pi+\varphi)|^2 + |D(-\varphi,\varphi)+ D(-\pi+\varphi,\pi-\varphi)|^2 \nonumber \\
& + & |D(\varphi,\pi-\varphi) + D(-\varphi,-\pi+\varphi)|^2 + |D(-\pi+\varphi,-\varphi) + D(\pi-\varphi,\varphi)  |^2 \nonumber \\
& + & \left. |D(\varphi,-\pi+\varphi)|^2 + |D(-\varphi,\pi-\varphi)|^2 + |D(-\pi+\varphi,\varphi)|^2 + | D(\pi-\varphi,-\varphi)|^2 \right) 
\end{eqnarray}
and is identical to the diagonal contribution from once diffracting orbits,
that is, the prefactor of $\tau^2$ in~(\ref{noncdiagand}) as can
be easily be seen examining the preceding equations.
Finally 
\begin{eqnarray}
\label{BGnond}
-\!\frac{1}{64}  \langle \langle \Re S^2(\varphi,\varphi) \rangle \rangle & = & -\frac{1}{32 \pi} \int_0^{\frac{\pi}{2}} \! d \varphi \, \left( \left[ D(\varphi,\varphi)+ D(\pi\! -\! \varphi,\pi\! -\! \varphi) + D(-\pi\! +\! \varphi,-\pi\! +\! \varphi)+D(-\varphi,-\varphi)\right]^2 \right. \nonumber \\
 & + & 2 \left[ D(\varphi,-\varphi) + D(\pi-\varphi,-\pi+\varphi) \right] \left[  D(-\varphi,\varphi) + D(-\pi+\varphi,\pi-\varphi) \right] \nonumber \\
& + & 2 \left[ D(\varphi,\pi-\varphi) + D(-\varphi,-\pi+\varphi) \right] \left[  D(-\pi+\varphi,-\varphi) + D(\pi-\varphi,\varphi) \right] \nonumber \\
 & + & \left. 2 D (\varphi,-\pi+\varphi) D(-\pi+\varphi,\varphi) +2 D(-\varphi,\pi-\varphi) D(\pi-\varphi,-\varphi)\rule{0mm}{4mm} \right). 
\end{eqnarray}
After some manipulations~(\ref{BGnond}) can be shown be identical to~(\ref{nonck2tauang}) of our paper. 

If the scatterer is not at a typical position the calculation can be done 
within the formalism of \cite{BG}, but
the average over $\phi_1$ and $\phi_2$ has to be modified~\cite{bogomolnypr}.
In particular for the scatterer at the center the form factor is obtained from~(\ref{ktaur}) and~(\ref{ktaus}) when the average over $\phi_1$, $\phi_2$ 
(defined by (71) and (72) of \cite{BG}) of a function $f$ is defined by
\begin{equation}
\label{newavg}
\langle f(\phi_1,\phi_2) \rangle = \frac{1}{4} \sum_{\phi_1=0,\frac{\pi}{2}} \sum_{\phi_2=0,\frac{\pi}{2}} f(\phi_1,\phi_2).
\end{equation}
For an angle independent scatterer, with the help of (83) and (85) of \cite{BG},
 it leads to
\begin{equation}
\label{cenetrr}
\langle r^n \rangle =  4^{n-1}
\end{equation}
where $n=1,2, \cdots$ and to
\begin{equation}
\label{centera}
A_{002}=16.
\end{equation}
Substitution in~(\ref{ktaur}) results in the form factor for a scatterer at the center~(\ref{centerf}).

The form factor for an angle dependent scatterer at the center can also be computed
applying the average~(\ref{newavg}) to (\ref{ktaus}) resulting
in (\ref{linears}), that is identical to the prefactor of the linear term in $\tau$
in~(\ref{centdiagang}) (see also~(\ref{forwardD})).
The term resulting from $|S|^2$ is 
\begin{eqnarray}
\frac{1}{64} \langle \langle |S(\varphi,\varphi)|^2 \rangle \rangle & = & \frac{1}{32 \pi} \int_0^{\frac{\pi}{2}} \! d \varphi \, \left( |D(\varphi,\varphi)+ D(\pi-\varphi,\pi-\varphi) + D(-\pi+\varphi,-\pi+\varphi)+D(-\varphi,-\varphi)|^2 \right. \nonumber \\
& + &  |D(\varphi,-\varphi) + D(\pi-\varphi,-\pi+\varphi) + D(-\varphi,\varphi)+ D(-\pi+\varphi,\pi-\varphi)|^2 \nonumber \\
& + & |D(\varphi,\pi-\varphi) + D(-\varphi,-\pi+\varphi) + D(-\pi+\varphi,-\varphi) + D(\pi-\varphi,\varphi)  |^2 \nonumber \\
& + & \left. |D(\varphi,-\pi+\varphi) + D(-\varphi,\pi-\varphi) + D(-\pi+\varphi,\varphi) +  D(\pi-\varphi,-\varphi)|^2 \right) 
\end{eqnarray}
and it is identical to the prefactor of $\tau^2$ in~(\ref{centdiagang}).
The last term is given by
\begin{eqnarray}
\label{BGnondcenter}
-\!\frac{1}{64}  \langle \langle \Re S^2(\varphi,\varphi) \rangle \rangle & = & -\frac{1}{32 \pi} \int_0^{\frac{\pi}{2}} \! d \varphi \, \left( \left[ D(\varphi,\varphi)+ D(\pi\! -\! \varphi,\pi\! -\! \varphi) + D(-\pi\! +\! \varphi,-\pi\! +\! \varphi)+D(-\varphi,-\varphi)\right]^2 \right. \nonumber \\
 & + &  \left[ D(\varphi,-\varphi) + D(\pi-\varphi,-\pi+\varphi)+  D(-\varphi,\varphi) + D(-\pi+\varphi,\pi-\varphi) \right]^2 \nonumber \\
& + &  \left[ D(\varphi,\pi-\varphi) + D(-\varphi,-\pi+\varphi)+  D(-\pi+\varphi,-\varphi) + D(\pi-\varphi,\varphi) \right]^2 \nonumber \\
 & + & \left. \left[ D (\varphi,-\pi+\varphi)+  D(-\pi+\varphi,\varphi) + D(-\varphi,\pi-\varphi)+ D(\pi-\varphi,-\varphi) \right]^2 \rule{0mm}{4mm} \right). 
\end{eqnarray}
It is straightforward to show that~(\ref{BGnondcenter}) is identical to 
the prefactor of $\tau^2$ in~(\ref{centnondangle}). 

We have demonstrated that the results of \cite{BG} are in agreement with our results for all
the cases which we investigated.
\end{appendix}
\typeout{References}

\end{document}